\documentclass[11pt]{article}
\usepackage{amsmath,amssymb,bm,epsfig,color,graphicx,cite,braket,mathrsfs,slashed}
\usepackage{amsmath}
\usepackage{amssymb}
\usepackage{array}
\usepackage{graphicx}
\usepackage{xcolor}
\usepackage{braket}
\usepackage{mathrsfs}
\usepackage{tensor}
\usepackage{tikz-feynman}
\usepackage{dsfont}
\usepackage{calrsfs}
\usepackage{slashed}
\usepackage{hyperref}

\usepackage[margin=1in]{geometry}

\numberwithin{equation}{section}
\hypersetup{
	colorlinks,
	citecolor=black,
	filecolor=black,
	linkcolor=black,
	urlcolor=black}
\hypersetup{linktocpage}

\begin{document}

\begin{figure}
	\centering
	\includegraphics[width=0.7\linewidth]{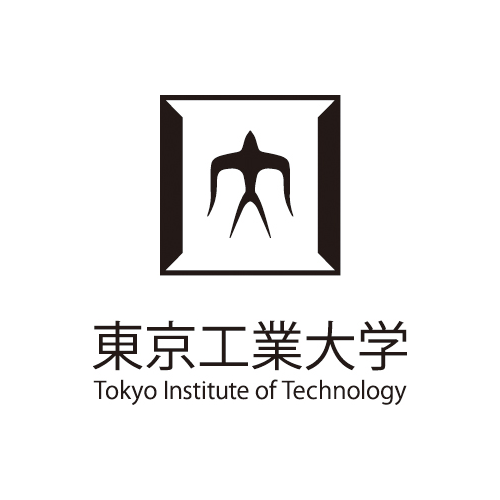}
\end{figure}
	
	\begin{center}
		Presented in Partial Fulfillment of the Author's MSc. in Physics\\
		\vspace{10mm}
		\huge Studying Both the Universe and Neutrinos Themselves Using Cosmic and Astrophysical Neutrinos\\
		\vspace{10mm}
		\Large Saul Isaac Hurwitz\\
		\vspace{10mm}
		\large Supervised by Prof. Masahide Yamaguchi\\
		\vspace{10mm}
		\small The financial assistance of the Japanese Ministry of Education, Culture, Sports, Science and Technology (MEXT) towards this research is hereby acknowledged. Opinions expressed and conclusions arrived at are those of the author and are not necessarily to be attributed to the MEXT.
	\end{center}
	
	\thispagestyle{empty}

	\newpage
	\pagenumbering{roman}

	``No man is an Iland, intire of itselfe; every man is a peece of the Continent, a part of the maine;"
	
	- John Donne, 1624\\
	
	\addcontentsline{toc}{section}{Acknowledgements}
	\section*{Acknowledgements}

	It is with extreme gratitude that first and foremost I need to thank my supervisor, Professor Masahide Yamaguchi - not only for the years of guidance, mentorship, and education, but also for believing in me and affording me the opportunity to learn and grow more than I could have imagined.
	
	I would also like to thank all those with whom I studied and worked over the course of my degree. To my collaborators, mentors, peers and friends: this thesis is built on a body of knowledge that could not have been attained without your generosity and kind help.
	
	To my family: the immense, endless support provided to me over my entire life is what brought me here today. For that, I cannot thank you enough.
	
	Finally, and most importantly, to my partner Nitara, to whom I dedicate this thesis: to put it simply, thank you for absolutely everything - without you, your love, support, encouragement and the million other things you give to me, this would never have been possible.

	\newpage
	
	``There is nothing new to be discovered in physics now. All that remains is more and more precise measurement."
	
	- Lord Kelvin, 1894
	
	\addcontentsline{toc}{section}{Abstract}	
	\section*{Abstract}
	
	With neutrino astronomy just beginning to burgeon, and the prospects of detecting the cosmic neutrino background closer than ever, we live in an era with the unique opportunity not only to investigate the universe with this novel probe, but conversely to utilise the cosmos as a laboratory to study neutrinos themselves. This thesis aims to expound on some of the ways to seize this opportunity, using corollaries of our standard model, gravitational lensing, inverse beta decay, and the neutrino’s spin to achieve our goals. Theoretically, these tools prove insightful and impressive, but difficulties in experimental precision and a low rate of certain astrophysical events hinder the capabilities of these mechanisms. Regardless, the future of neutrino astronomy is certainly bright.

	\newpage
	
	\tableofcontents
	
	\newpage
	\pagenumbering{arabic}
	
	\section{Introduction}
	
	Today, it is well known that our universe is expanding at an ever-increasing pace. Logically, then, at earlier times the universe was smaller. This is the gist of the big bang theory: the further back in time we look, the denser the matter-energy in our universe is; the universe started off very small and hot, and gradually expanded and cooled. Flaunting immense predictive success, the standard big bang theory excels at explaining a plethora of cosmological and astrophysical phenomena, such as the cosmic microwave background (CMB) and the mass abundances of hydrogen and helium in the universe before the first stars began to form.\\
	One important consequence of this model is the idea of decoupling. At very early times, the temperature of the constituents of the universe, which were all in thermal equilibrium owing to their rapid reaction rates, was extremely high. As the universe expanded and cooled, certain reactions became energetically unfavoured. A particle species ``decoupled" when the rate of reactions with other species was low enough such that its mean free path was larger than the size of the universe.\\
    Explored in chapter 4, neutrinos decoupled from the other particles in thermal equilibrium when the universe's temperature was just a few MeV: about 1 second after the universe began. These neutrinos, having decoupled, are expected not to have interacted until today, and can provide information from the time they decoupled. If these relic neutrinos, collectively called the cosmic neutrino background (C$\nu$B), are detected, then we will be able to probe information about our universe from a much earlier stage than we are currently able. For the past few decades, precision cosmology has been based around the CMB: analogous to the C$\nu$B, the CMB describes the relic photons that decoupled when the universe was approximately 380 000 years old. Therefore, the C$\nu$B will act as a probe to observe a much younger universe.\\
    
    Neutrinos are spin-$\frac{1}{2}$ leptons with extremely small masses and no electric charge. First theorised as massless particles in 1930 by Pauli \cite{Bilenky:2012qb}, they were needed to explain issues of energy, angular momentum and momentum conservation during beta decay, and were first seen experimentally in 1956 by Reines and Cowan \cite{Reines:1956rs}. The current standard model predicts three flavours of neutrinos: electron, muon and tau, named according to with which charged lepton they interact.\\
    A particularly elusive particle, the neutrino only interacts via the weak force and gravitationally. Thus, even at high energy, neutrinos are difficult to detect. Cosmic neutrinos from the C$\nu$B, owing to the expansion of the universe for billions of years, have lost most of their momenta and, unless massless, will be non-relativistic. Neutrinos have been detected from astrophysical sources, such as the sun, as well as in laboratories, but these neutrinos had momenta much larger than their masses. Because of this low detection rate, the C$\nu$B has not been directly observed, although indirect evidence, such as the number of relativistic species present during photon decoupling, strongly supports its existence \cite{Mangano:2005cc}.\\
    Though much has been uncovered, there are still many properties of neutrinos that are unknown, such as their masses or whether they are Dirac or Majorana particles (that is, whether their antiparticles are the same as themselves). Through the study of cosmic neutrinos, as well as those from astrophysical sources, we have the unique oppurtunity not only to learn about our universe, but also about neutrinos themselves.\\
    
    Though the C$\nu$B has yet to be directly detected, the near-future holds promise for it. Thus, it is no futile venture to ask ourselves what information we might be able to glean from it once it is detected. This pursuit underlies the main objective of this work: to understand what we expect from the C$\nu$B, how deviations from these expectations can be explained and what they imply, and how we may best detect it. Atop this, neutrinos from astrophysical sources such as supernovae can provide a wealth of information about both the supernovae and neutrinos themselves, and are much easier to detect.\\
    
    The target audience is any individual interested in neutrino astronomy (particularly relic neutrino astronomy) who has covered the basics of both general relativity and quantum field theory. Chapters 2 and 3 are introductory in nature, covering the fundamentals used throughout the rest of our work and ensuring the tools from general relativity and quantum field theory we will utilise are understood. Next we continue by exploring the well-known leading order calculations which follow naturally from the standard big bang theory, which will serve as the starting point for more precise calculations in chapter 5. In the interim, chapter 6 will discuss the current state of different neutrino detectors, and which will best serve our purpose. After ``selecting" our detector, chapter 7 aims to investigate the capture rate of these cosmic neutrinos in our experiment of choice. Changing direction, we then explain how gravitational lensing can be used to our advantage, particularly when it comes to supernova neutrinos and the C$\nu$B. Our penultimate chapter discusses how the known spin of neutrinos might be used to learn about their mass, before concluding.
    
    \newpage
	\section{Crash Course on Cosmology}
	
	The field of cosmology has made absolutely massive strides over the past 50 years. It would of course be impossible to cover all of the basics in one small section of this work, and so only the ideas needed to understand the rest of this work are explored. The derivation of the Friedmann-Lemaitre-Robertson-Walker (FLRW) metric and its corollaries are needed in tracking the evolution of the C$\nu$B and understanding the expansion of the universe quantitatively. A qualitative history of our early universe is also given, in order to provide a context to the significance of relic neutrinos and the time they decoupled. 
	
	\subsection{Homogeneity, Isotropy and Flatness}
	
	About a century ago, it was a philosophical belief that our universe was homogeneous (there is no special point in space), isotropic (there is no preferred direction) and eternal (there is no special moment in time) \cite{Einstein(1917)}. The issue here is that this implies a static, unchanging universe. As observations have shown for decades, the universe is actually expanding at an ever-increasing pace \cite{Riess:1998cb}; our universe certainly is not static, and there was a ``special" moment in time: the big bang, when the universe was at its smallest.\\
	The other two characteristics, homogeneity and isotropy, have stood the test of time. Experimental data supports these assertions\footnote{On large scales: that is, the order of galactic superclusters and larger. Of course, on small scales like our solar system, this is not the case.} and it seems that there is truly no preferred point nor direction in which to look, at least to a very good approximation. The final piece of observational data that is of importance is that our universe appears to be ``flat". In our 3 dimensions of space, it is difficult to visualise the meaning of ``flat", and so we will explain it in 2D:\\
	The intrinsic curvature is the curvature of a manifold described without embedding the manifold into some higher dimensional space. For example, in 2D, consider the surface of a sphere. This surface could of course be embedded into 3D, and we could easily observe its curvature in relation to its surroundings. Alternatively, however, one can describe its curvature without reference to a higher dimension at all. This is necessary when we wish to describe our 3 spatial dimensions, as we do not want to forge some fourth spatial dimension that we can neither visualise nor measure. This analogy comparing 2D and 3D extends into the nomenclature used: when describing the universe as ``flat", we refer to the 3 dimensional equivalent to what flat means in 2 dimensions - having 0 intrinsic curvature.\\
	Using the language of general relativity, let us now translate these words into mathematics and see what we can learn about our metric.\\
	
	Homogeneity states that there is no special point in space - that any point is indistinguishable from another. What this implies then, mathematically, is that there must exist, at each moment in time, a 3 dimensional hypersurface, $\Sigma_t$, and that for any two points on this hypersurface $p$ and $q$, there must be an isometry\footnote{An isometry is a diffeomorphism of the manifold that leaves the metric unaltered.} taking the one into the other. The simplest example would be a spatial translation in Cartesian coordinates $(x,y,z) \rightarrow (x+a,y+b,z+c)$, which moves $p$ to $q$, but the line element
	
	\begin{equation}
	d\ell^2=dx^2+dy^2+dz^2
	\end{equation}
	
	remains unchanged.
	
	Next, isotropy requires that there be no preferred direction in which to look. In fact, isotropy already implies homogeneity, since a preferred point will imply a preferred direction for any point besides for the preferred one. So if a spacetime is spatially isotropic everywhere at each instant in time, it must also be spatially homogeneous everywhere. \\
	Translating this, consider any point $p$ in a spacetime, with a timelike curve with tangent $u^\mu$ at $p$ passing through it and 2 unit spatial vectors, $v_1^\mu$ and $v_2^\mu$, which are both orthogonal to $u^\mu$. This $u^\mu$ is the worldline of an observer at $p$. Isotropy then requires that there be an isometry of the metric that leaves $p$ and $u^\mu$ unchanged, but takes $v_1^\mu$ into $v_2^\mu$ (and vice versa). The simplest example of this would be a spatial rotation in Cartesian coordinates, for example around the $z$-axis $(x,y,z) \rightarrow (x\cos\theta-y\sin\theta , y\cos\theta+x\sin\theta , z)$. This would rotate $v_1^\mu$ into $v_2^\mu$ but leave $u^\mu$ and $p$ unchanged. Clearly, we see that $v_1^\mu$ and $v_2^\mu$ must ``live" in the spatial hypersurfaces $\Sigma_t$, while the worldline with tangent $u^\mu$ must be orthogonal to these hypersurfaces. In fact, we can parametrise the worldline using $t$, such that for every $t$ there is a $\Sigma_t$ orthogonal to it.\\

\begin{figure}
	\centering
	\includegraphics[width=0.7\linewidth]{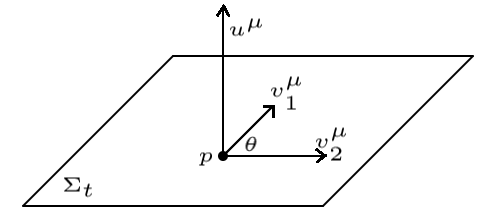}
	\caption{Diagram depicting the setup of the isometry needed for isotropy. Drawn using \cite{Drawingtool}.}
	\label{fig:worldlines}
\end{figure}
	
	Now, to keep our spacetime isotropic, we need to ensure that we are unable to construct any preferred vectors in $\Sigma_t$. Consider then the 3 dimensional Riemann tensor, defined by the spatial metric $\gamma_{ij}$ of $\Sigma_t$, with 2 indices raised: $\tensor{^{(3)}R}{_{ij}^{kl}}$. This tensor can be interpreted as a linear map which maps the space of antisymmetric rank (0,2) tensors (also known as two-forms) into itself. That is, contracting a two-form with this Riemann tensor will provide another two-form. This is analogous to how a matrix maps vectors to vectors in linear algebra. Continuing the analogy, we see that the vector space of two-forms must contain the orthonormal basis of eigenvectors of the 3D Riemann tensor.\\
	Now, if the eigenvalues of these eigenvectors were distinct, one would be able to select a preferred direction! Thus, the 3D Riemann tensor must be a multiple of the identity tensor: $\tensor{^{(3)}R}{_{ij}^{kl}}=KI$. This identity tensor must still satisfy the fact that two-forms are taken to two-forms, and so we have
	
	\begin{equation}
	\tensor{^{(3)}R}{_{ij}^{kl}}=K\tensor{\delta}{^k_{[i}}\tensor{\delta}{^l_{j]}}.
	\end{equation}
	
	Lowering our indices using $\gamma_{ij}$, we obtain $^{(3)}R_{ijkl}=K\gamma_{k[i}\gamma_{j]l}$. Now, homogeneity says that each point in space is indistinguishable. Therefore, their Riemann tensor components cannot be distinguishable, and so clearly $K$ is a constant.\\
	This $K$ actually determines the shape of the hypersurfaces. For any value of $K$, by rescaling of coordinates, $K$ can always be either 1, 0 or -1. $K=1$ is the metric of a 3-sphere, $K=0$ gives that of regular, flat Euclidian space, and $K=-1$ is that of a 3-hyperboloid. As stated above, from observational data, our universe appears to be flat, and so we can take $K=0$.\\
	
	The final piece of the puzzle is to include the scale factor. As we have stated, the universe is expanding. This means that the distance between objects (on very large scales) is not constant, and this can be described using a change of our spatial coordinates over time. Since there is no preferred direction in space, this expansion factor must be the same for all directions, but can evolve in time, and we label it $a(t)$.\\
	Using our tangential velocity at $p$, we can write the metric of our universe as 
	
	\begin{equation}
	ds^2=u^\mu u_\mu dt^2 +a(t)^2\gamma_{ij}dx^idx^j.
	\end{equation}
	
	Now, the norm of a velocity is $u^\mu u_\mu=-1$, and $\gamma_{ij}$ is just the metric of flat Euclidean space. Writing this Euclidean space in spherical coordinates rather than Cartesian, we arrive at our FLRW metric:
	
	\begin{equation}
	ds^2=-dt^2+a(t)^2\left[dr^2+r^2d\Omega^2\right],\label{FLRW}
	\end{equation}
	
	wherein $d\Omega^2=d\theta^2+\sin^2\theta d\phi^2$. Now, the only unknown function is the scale factor $a(t)$, which can be solved for using the Einstein equations.\\
	One important quantity is the Hubble constant\footnote{The word ``constant" is perhaps a misnomer: the parameter is constant throughout space, but changes over time.},
	
	\begin{equation}
	H(t)=\frac{\dot{a}}{a},
	\end{equation}
	
	where $\dot{a}=\frac{da}{dt}$.\\
	We find the non-zero Christoffel symbols to be
	\begin{equation}
	\tensor{\Gamma}{^0_{11}}=a\dot{a} \ \ \ \ \ \ \ \tensor{\Gamma}{^0_{22}}=a\dot{a}r^2 \ \ \ \ \ \ \ \tensor{\Gamma}{^0_{33}}=a\dot{a}r^2\sin^2\theta
	\end{equation}
	\begin{equation}
	\tensor{\Gamma}{^1_{01}}=\tensor{\Gamma}{^1_{10}}=\frac{\dot{a}}{a} \ \ \ \ \ \ \ \tensor{\Gamma}{^1_{22}}=-r \ \ \ \ \ \ \ \tensor{\Gamma}{^1_{33}}=-r\sin^2\theta
	\end{equation}
	\begin{equation}
	\tensor{\Gamma}{^2_{02}}=\tensor{\Gamma}{^2_{20}}=\frac{\dot{a}}{a} \ \ \ \ \ \ \ \tensor{\Gamma}{^2_{12}}=\tensor{\Gamma}{^2_{21}}=\frac{1}{r} \ \ \ \ \ \ \ \tensor{\Gamma}{^2_{33}}=-\sin\theta\cos\theta
	\end{equation}
	\begin{equation}
	\tensor{\Gamma}{^3_{03}}=\tensor{\Gamma}{^3_{30}}=\frac{\dot{a}}{a} \ \ \ \ \ \ \ \tensor{\Gamma}{^3_{13}}=\tensor{\Gamma}{^3_{31}}=\frac{1}{r} \ \ \ \ \ \ \ \tensor{\Gamma}{^3_{23}}=\tensor{\Gamma}{^3_{32}}=\cot\theta.
	\end{equation}.
	
	Next, the non-vanishing components of the Riemann tensor are (along with their permutations):
	\begin{equation}
	\tensor{R}{^0_{101}}=a\ddot{a} \ \ \ \ \ \ \  \tensor{R}{^0_{202}}=a\ddot{a}r^2 \ \ \ \ \ \ \ \tensor{R}{^0_{303}}=a\ddot{a}r^2\sin^2\theta
	\end{equation}
	\begin{equation}
	\tensor{R}{^1_{212}}=\dot{a}^2r^2 \ \ \ \ \ \ \tensor{R}{^1_{313}}=\dot{a}^2r^2\sin^2\theta \ \ \ \ \ \ \ \tensor{R}{^2_{323}}=\dot{a}^2r^2\sin^2\theta.
	\end{equation}

From \eqref{FLRW} and the Riemann tensor elements calculated from it, we can analyse the 00 component of Einstein's equation, $G_{00}=8\pi GT_{00}$. We have from (as we shall see) \eqref{emtensor} that $T_{00}=\rho$, and 

\begin{equation}
	G_{00}=R_{00}-\frac{1}{2}g_{00}R=\tensor{R}{_{0\mu0}^\mu}+\frac{1}{2}g^{\nu\lambda}\tensor{R}{_{\nu\mu\lambda}^\mu} =3\left(\frac{\dot{a}}{a}\right)^2=3H^2.
\end{equation}

And so, Einstein's equation gives us Friedmann's equation (for a flat universe):

\begin{equation}
	H^2=\frac{8\pi G}{3}\rho.\label{Friedmann}
\end{equation}
	
	\subsection{Implications of the Universal Metric}
	Our first corollary is that a particle's physical momentum $p$ is inversely proportional to $a(t)$. For the case of a massive particle, we can see this using the geodesic equation:
	
	\begin{equation}
	\frac{du^\mu}{d\tau}+\tensor{\Gamma}{^\mu_{\rho\sigma}}u^\rho u^\sigma=0.
	\end{equation}
	
	Where $u^\mu=\frac{dx^\mu}{d\tau}$, and $\tau$ is some parameter (in this massive case, it can be chosen to be the proper time). We consider the spatial components of this equation. In this case, the only pertinent Christoffel symbol is $\tensor{\Gamma}{^i_{0j}}=\frac{\dot{a}}{a}\tensor{\delta}{^i_j}$. So we have
	
	\begin{equation}
	\frac{du^i}{d\tau}=-2\frac{\dot{a}}{a}u^0u^i \ \ \ \rightarrow \ \ \ \frac{du^i}{u^i}=-2\frac{da}{a}.
	\end{equation}
	
	Solving this differential equation, we have that $u^i$ is proportional to $a(t)^{-2}$. However, this is not the physical velocity. The physical spatial coordinates are $X^i=a(t)x^i$, and so the physical velocities would be $U^i=a(t)u^i$. Thus, we have that the physical velocity (and in turn, the physical momentum) goes as $a(t)^{-1}$.\\
	
	For massless particles, we know that their momenta are $p^\mu=(E,\bar{p})$. Recall that $\tau$ in the geodesic equation is simply some parameter, and while in the massive case it represented the proper time, in the massless case we can choose it such that $p^\mu=\frac{dx^\mu}{d\tau}$.\\
	Thus, the zeroth component of the geodesic equation reads (with $i$ and $j$ running over spatial components)
	
	\begin{equation}
	\frac{d}{d\tau}\frac{dt}{d\tau}+\frac{\dot{a}}{a}g_{ij}\frac{dx^i}{d\tau}\frac{dx^j}{d\tau}=0 \ \ \ \rightarrow \ \ \ E\frac{dE}{dt} = -\frac{\dot{a}}{a}g_{ij}p^ip^j,\label{EDE}
	\end{equation}
	
	but we know that, for a massless particle, 
	
	\begin{equation}
	g_{\mu\nu}p^\mu p^\nu = 0 = -E^2+g_{ij}p^ip^j
	\end{equation}
	
	and so, using this in \eqref{EDE} we obtain the differential equation for $E$:
	
	\begin{equation}
	\frac{dE}{dt} = -\frac{\dot{a}}{a}E \label{moma}
	\end{equation}
	
	which clearly has the solution that $E$ is inversely proportional to $a$. Finally, since for massless particles $E=|\bar{p}|$, we have that massless particles' momenta $|\bar{p}|$ also go as $a(t)^{-1}$.\\
	
	While the above arguments used single particles to relate momenta to the scale factor, this relation holds generally for a collection of particles as well. Moving now towards large collections of particles, we turn to the conservation of the energy-momentum tensor to prove our next relation:
	
	\begin{equation}
	\nabla_\mu T^{\mu\nu}=0.\label{energycons}
	\end{equation}
	
	In general, we can describe any matter on a universal scale as a perfect fluid with energy density $\rho(t)$ and pressure $P(t)$. Note that, due to our homogeneity, these quantities can only depend on time. This perfect fluid can be used to describe dust (a collection of massive particles with negligible velocity), radiation (any collection of particles with speed approximately that of light) or even more exotic matter like dark energy, using an equation of state relating pressure to density: 
	
	\begin{equation}
	P=\omega\rho.
	\end{equation} 
	
	$\omega$ then takes on different values for different types of matter. The energy-momentum tensor for this perfect fluid is given by
	
	\begin{equation}
	T^{\mu\nu}=(\rho+P)u^\mu u^\nu+Pg^{\mu\nu}.\label{emtensor}
	\end{equation}
	
	Since this fluid fills the entire universe, we can consider things in its rest frame. Thus, only $u^0=1$ and $u^i=0$ for the spatial components and the $\nu=0$ component of \eqref{energycons} gives
	
	\begin{equation}
	\nabla_\mu T^{\mu 0} = \partial_\mu T^{\mu 0}+\tensor{\Gamma}{^\mu_{\mu\sigma}}T^{\sigma 0} +\tensor{\Gamma}{^0_{\mu\sigma}}T^{\mu\sigma} = \partial_0T^{00}+ \tensor{\Gamma}{^\mu_{\mu0}}T^{00} +\tensor{\Gamma}{^0_{\mu\sigma}}T^{\mu\sigma}=0.\label{econs}
	\end{equation}
	
	The necessary Christoffel symbols are that $\tensor{\Gamma}{^\mu_{\mu0}}=3\frac{\dot{a}}{a}$ and $\tensor{\Gamma}{^0_{\mu\sigma}}T^{\mu\sigma} = \frac{\dot{a}}{a}g_{ij}T^{ij}$, where as usual $i$ and $j$ are spatial indices.
	
	Now, from \eqref{emtensor}, we have $T^{00}=\rho(t)$ and $T^{ij}=P(t)g^{ij}$. Plugging this in to \eqref{econs} and using $g_{ij}g^{ij}=3$, we have
	
	\begin{equation}
	\dot{\rho}+3\frac{\dot{a}}{a}(\rho+P)=0.\label{fluideqn}
	\end{equation}
	
	All of this is leading to prove that the effective temperature of some radiation-like species is also inversely proportional to $a(t)$. To this end, we now consider the equation of state for radiation and turn to statistical mechanics. Recall that, for any distribution function $f(t,p)$, the energy density and pressure are defined as 
	
	\begin{equation}
	\rho = g\int\frac{d^3p}{(2\pi)^3}E(p)f(t,p) \ \ \ \ ; \ \ \ \ P= g\int\frac{d^3p}{(2\pi)^3}\frac{|\bar{p}|^2}{3E(p)}f(t,p),\label{statmech}
	\end{equation}
	
	where $g$ is the number of degrees of freedom, $E$ the energy and we integrate over momentum. By inspection then, it is clear that for radiation wherein $E(p)=|\bar{p}|$, we have that, for any distribution function
	
	\begin{equation}
	P= g\int\frac{d^3p}{(2\pi)^3}\frac{|\bar{p}|}{3}f(t,p)=\frac{1}{3}\rho.\label{pressurerho}
	\end{equation}
	
	So, for radiation, our equation of state has $\omega=\frac{1}{3}$, and using $P=\frac{\rho}{3}$ in \eqref{fluideqn}, we have 
	
	\begin{equation}
	\frac{d\rho}{\rho}=-4\frac{da}{a},\label{rhoa}
	\end{equation}
	
	which, of course, shows that for radiation, the energy density decreases with the scale factor as $a(t)^{-4}$.\\
	Going back to \eqref{statmech}, we can find the relationship for radiation between $\rho$ and the temperature $T$:\footnote{In this calculation, the chemical potential is taken to be 0.}
	
	\begin{equation}
	\rho = g\int\frac{d^3p}{(2\pi)^3}|\bar{p}|f(t,p)=g\int\frac{d^3p}{(2\pi)^3}\frac{|\bar{p}|}{e^{|\bar{p}|/T}\pm1}.
	\end{equation}
	
	In momentum-space, we use spherical coordinates - $d^3p=|\bar{p}|^2d|\bar{p}|d\Omega_p$. Integrating trivially over $d\Omega_p$ gives a factor of $4\pi$ as usual, leaving
	
	\begin{equation}
	\rho=\frac{g}{2\pi^2}\int_0^\infty d|\bar{p}|\frac{|\bar{p}|^3}{e^{|\bar{p}|/T}\pm1} = \frac{gT^4}{2\pi^2}\int_0^\infty dx\frac{x^3}{e^{x}\pm1}.\label{rhot}
	\end{equation}
	
	This last integral can be computed, and will be later on, but for now it is not needed. This integral simply gives some numerical value; we were interested in the relationship between $\rho$ and $T$, which we have shown to be quartic. And thus, using \eqref{rhot} and \eqref{rhoa}, we can see that the temperature of a radiative species also goes as $a(t)^{-1}$.\\

	\subsection{Thermal History of the Early Universe}
	
	This section aims to paint a rough picture of the current leading theory of our cosmological origins - the inflationary big bang theory - so as to put neutrino decoupling in context, which we shall discuss more quantitatively in chapter 4.
	As mentioned in the introduction, we know that our universe is expanding, and so tracing this evolution backwards results in a beginning point - the ``big bang" - and a subsequent cooling and expanding of the cosmos.
	While some authors detail the history of our universe starting from the present and working backwards, for simplicity we will work chronologically, despite the fact that in doing so we must begin with the most controversial epochs.\\
	
	The universe began approximately 13.8 billion years ago \cite{Aghanim:2018eyx}, and the very start of the universe is still extremely mysterious, with current physics unable to meaningfully describe it - this is known as the cosmic singularity, and at this time our theories of physics break down. Let us begin our journey then when the universe was approximately $10^{-36}$ seconds old: at this point in time, most theories describe a grand unified force which combine the electromagnetic, weak and strong interactions into one. These grand unified theories (GUT) are still unverifiable, as the energies needed to observe their predictions are of the order $10^{16}$ GeV \cite{Ross:1985ai}, while current experiments like the large hadron collider (LHC) can only reach energies around $10^4$ GeV. Whether GUTs are correct or not, the phase transition wherein the strong interaction becomes distinct is the leading explanation for the conditions needed for baryon asymmetry (discussed slightly more below) and the energy injection into the universe which causes the next controversial era - inflation.\\
	
	Inflation was first theorised in 1981 by Guth \cite{Guth:1980zm} as the next step in our universe's story. Inflation is a transitory period characterised by incredibly rapid expansion, wherein the universe expanded by a factor of around $e^{50\sim60}$ in a period of around $10^{-32}$ seconds. This is the current ``standard" theory, and is needed to explain certain phenomena that without an inflationary period seem either impossible or miraculous \cite{Wang:2013zva}. Inflation is used to solve the horizon problem (the cosmic microwave background's temperature today is very close to homogeneous and isotropic - yet, without inflation, patches of the CMB were never in causal contact. Either there must be some mechanism wherein they were in thermal equilibrium at some point in the past and then expanded rapidly away, or there is a miraculous coincidence), the flatness problem (the universe today is very close to being completely flat. Going back in time without an inflationary phase, this would require that the universe be flat to within $10^{-60}$ \cite{Rydenbook} - either there was an inflationary phase in which the universe went from an arbitrary curvature to being flat, or there was a coincidental finely-tuned perfect initial curvature), the monopole problem (the GUTs described above often result in magnetic monopoles, which is problematic as we don't seem to have any in the universe today. Inflation allows for the few that were created to be diluted to such a low density that they are essentially negligible in our universe today), and issues explaining the anisotropies in the CMB and the formation of large scale structure (with inflation, quantum fluctuations in the extremely early universe can be seen as the cause of these deviations from perfect homogeneity and isotropy). Though there are still many issues with inflationary theory, it is a necessary evil needed to assuage issues of the big bang theory.\\
	
	After inflation, some form of ``reheating" is needed - the energy that caused the rapid expansion is then ``returned" to the components of matter and radiation, usually via the decay of some particle that was responsible for the expansion. At this point, the universe exists in what is known as a quark-gluon plasma: energies are still too high for quarks and gluons to form hadrons, and thus constantly interact with each other and other particles in one big ``soup". Many ``natural" theories of dark matter involve the exotic dark matter particles being created around this era \cite{Gorbunov:2011zz}. At around 100 GeV, the cosmos has cooled to the point of undergoing another phase transition: the electroweak force separates into the weak and electromagnetic interactions, and this transition is discussed in detail in the next chapter. This is the first era wherein the standard model applies, and where speculation is minimal - though we have not directly probed this epoch in our universe, if our universe was at a high enough temperature, there is no reason why the electroweak force would not have been unified.\\
	
	Soon after, when the universe was around 200 MeV, quarks and gluons would have become confined to colourless (a property related to the strong interaction) hadrons - baryons (including protons and neutrons) and mesons. In the quark-gluon plasma, there were almost an equal number of quarks and anti-quarks, with there being only a fractional surplus of quarks at the order of $10^{-10}$ \cite{Gorbunov:2011zz} when compared to the number of photons - meaning there were roughly $10^{10}+1$ quarks for every $10^{10}$ anti-quarks within a certain volume. This baryon asymmetry is the reason the universe is populated today by matter and not anti-matter.\\
	
	At just under 2 MeV, when the universe is only around 1 second old, neutrinos decouple from the other matter in the universe - the subject of chapter 4. These free-streaming neutrinos should still exist today, and detecting them will provide information from this epoch of the universe. Thus, owing to their free-streaming nature, these particles act as the most ancient probe possible in the standard model of particle physics and cosmology. Due to the neutrinos decoupling, protons and neutrons could no longer convert from one to the other, so the amount of protons and neutrons in the universe became fixed with around 15\% of the baryonic component being neutrons \cite{Dodelsonbook}. Since photons are not free-streaming at this time, rather interacting constantly with protons and electrons, the universe at this point is optically ``opaque" - we are unable to observe it directly using electromagnetic radiation: this is why observation of these relic neutrinos is so vital. A very short time after neutrinos decoupled, electrons and positrons annihilated, once again leaving behind only the small surplus of electrons.\\
	
	After around 3 minutes, the universe had cooled enough for nucleosynthesis to occur, wherein the free protons and neutrons began to undergo fusion. Over the next few minutes, protons and neutrons fused to form primarily helium-4 ($^4$He) which made up around a quarter of the baryonic mass \cite{Waldbook}, with small fractions of deuterium, helium-3 and lithium. The remaining protons remained as protons, composing almost three quarters of the baryonic mass. Though this era has yet to be directly probed, the abundances of helium in the early universe are inexplicable via stellar fusion, as the fraction is too high, and thus big bang nucleosynthesis (BBN) has indirect evidence in the observation of helium-4 abundances today \cite{Waldbook}.\\
	
	The universe continued to expand and cool for hundreds of thousands of years, until the temperature of the universe reached approximately the ionisation energy of hydrogen (in actuality it was slightly later \cite{Gorbunov:2011zz}). At this time, it became energetically favourable for electrons to become bound to nuclei and form neutral atoms - an era known as recombination - and it occurred when the universe was around 370 thousand years old \cite{Gorbunov:2011zz}. After recombination, the neutral gas of atoms was now transparent to photons and they were able to free-stream - so this is when photons decoupled, leaving behind the cosmic microwave background. As is clearly apparent, information gathered from the CMB comes from a much later probe than information we could gather using the C$\nu$B. The universe has undergone many changes since the time of photon decoupling, but these changes (for the most part) have not affected relic photons or neutrinos.\\
	
	To conclude this chapter, let us summarise the most vital points: from observations, we can see that we live in a homogeneous, isotropic, flat universe that is expanding at an accelerating rate. Using general relativistic formalism, we were able to see that this expansion is related to the energy density in the universe, as well as the fact that both the momentum and temperature of particles are inversely proportional to the scale factor that characterises the expansion. We have also seen when relic neutrinos were created, and what epoch of the early universe they should probe as opposed to the much later-created relic photons.

	\newpage
	\section{Neutrinos in the Standard Model}
	
	As stated in the introduction, neutrinos have only been theorised since 1930 and detected since 1956. While the full description for neutrinos lies in electroweak theory, the initial heuristic theory devised in 1933 by Fermi \cite{Fermi:1933jpa}, known as the 4-Fermi theory, is adept at explaining phenomena at low energies\footnote{The definition of ``low" energy will become clear in reference to the masses of the gauge bosons.}. While the full electroweak theory will be elaborated in order to fully understand the chiral nature of weak interactions as well as the source for neutrino flavour mixing, we begin by studying the 4-Fermi theory as a simpler tool that we will use to tackle certain interactions.

	\subsection{4-Fermi Theory}
	
	4-Fermi theory is an effective field theory (EFT) which describes weak interactions fairly accurately at low energies. The Lagrangian for this model is given by
	
	\begin{equation}
	\mathcal{L}=\normalfont-\frac{4G_F}{\sqrt{2}}\left[J^\mu_+J^\mu_-+\left(J^\mu_0\right)^2\right],\label{4FermiLag}
	\end{equation}
	
	with each $J^\mu$ a ``V-A" (vector minus axial) current. The idea of a current was to ``turn" one particle into another, with $J^\mu_+$ and $J^\mu_-$ being charged (i.e the end particle's charge was different to that of the initial one) and $J^\mu_0$ being neutral. Given explicitly, we have
	
	\begin{equation}
	J^\mu_+=\bar{\nu}_\alpha\gamma^\mu\left(\frac{1-\gamma^5}{2}\right)e_\alpha+V_{ij}\bar{u}^i\gamma^\mu\left(\frac{1-\gamma^5}{2}\right)d^j,
	\end{equation}
	
	where $\alpha$ runs over electrons $e$, muons $\mu$ and taus $\tau$, while $u^i$ represent the up $u$ or charm $c$ quarks, $d^j$ the down $d$ or strange $s$ quarks, and $V_{ij}$ is the 2-dimensional CKM matrix. The top and bottom quarks are not in this low-energy theory, as the top quark's mass is larger than the cutoff energy of this EFT. All the fermions (charged leptons, neutrinos and quarks) in the above are spinors. The $\bar{\psi}\gamma^\mu\phi$ terms are vector currents while the $\bar{\psi}\gamma^\mu\gamma^5\phi$ terms are the axial currents. $J^\mu_-$ is the Hermitian conjugate of this and the neutral current has a similar structure, but only couples fermions to fermions of the same flavour.\\
	 The original purpose of 4-Fermi theory was to attempt to explain beta decay of neutrons into protons, electrons and (anti-)neutrinos, and at first only contained vector currents. In order to explain both Fermi and Gamow-Teller transitions in more complex systems, the V-A structure was used \cite{Feynman:1958}. As we will see, this was the first inkling that the weak interaction is chiral.\\
	The universal interaction strength is governed by Fermi's constant, $G_F$, and there are no bosons at all in this theory: only a 4-point vertex with 4 fermions (and thus the name 4-Fermi theory).\\
	
	As an example, let us consider the matrix element of the process 
	
	\begin{equation}
	\nu_e+d \ \ \rightarrow \ \ u+e^-.
	\end{equation}
	
	This process is pedagogical, and of course cannot be observed as lone quarks are not stable particles. However, we can read off the matrix element from the Lagrangian. In this example, only the charged current term  will appear, the product wherein the electron and up quark are conjugate spinors as they are the outgoing particles:
	
	\begin{equation}
	\mathcal{L} \supset -\frac{G_F}{\sqrt{2}}V_{ud}\bar{u}\gamma^\mu(1-\gamma^5)d\bar{e}\gamma^\mu(1-\gamma^5)\nu_e.
	\end{equation}
	
\begin{figure}
	\centering
	\includegraphics[width=0.3\linewidth]{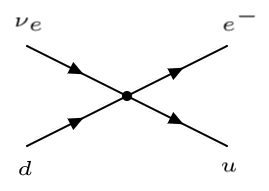}
	\caption{Feynman diagram of electron neutrino and down quark ``turning" into an electron and up quark. Drawn using \cite{Drawingtool}.}
	\label{fig:feynman-diagram}
\end{figure}

	The Feynman diagram for such a process would simply be a 4-point vertex and is depicted in figure \ref{fig:feynman-diagram}.
	
	By inspection, one can read off the matrix element
	
	\begin{equation}
	i\mathcal{M}=-i\frac{G_F}{\sqrt{2}}V_{ud}\bar{u}\gamma^\mu(1-\gamma^5)d\bar{e}\gamma^\mu(1-\gamma^5)\nu_e\label{matrixel}
	\end{equation}

	and the cross section and other quantities calculated. The full, slightly more complex calculation will be done completely in chapter 7, but for now we shall investigate the underlying physics of neutrinos.
	
	\subsection{Spontaneous Symmetry Breaking}
	
	Spontaneous symmetry breaking (SSB) is an interesting phenomenon in quantum field theory, wherein there exists a symmetry in a Lagrangian, but not in the system's ground state. Consider a continuous symmetry characterised by a parameter $\alpha$, and recall that the Noether current that arises from such a symmetry has an associated charge:
	
	\begin{equation}
		Q = \int d^3x J_0(x) = \int d^3x \sum_m \frac{\partial \mathcal{L}}{\partial \dot{\phi}_m}\frac{\delta \phi_m}{\delta \alpha},
	\end{equation}

	which is conserved.
	


	Since the charge is conserved, this means it commutes with the Hamiltonian:
	
	\begin{equation}
		[H,Q]=i\partial_tQ=0
	\end{equation}

	and as a result, there is a degeneracy between the uncharged and charged ground states. Given a vacuum state $|\Omega\rangle$ with energy $E_0$, similarly we have that
	
	\begin{equation}
		HQ|\Omega\rangle = QH|\Omega\rangle = E_0Q|\Omega\rangle,
	\end{equation}
	
	so the state $Q|\Omega\rangle$ is energetically degenerate with the vacuum. Thus, even though altering the Lagrangian via some continuous symmetry left it unchanged, the ground state has been altered: it is possible for it to be either charged or uncharged. \\
	
	From this vacuum, we can construct particle states $|\pi\rangle$ with 3-momentum $\bar{p}$ via 
	
	\begin{equation}
		|\pi(\bar{p})\rangle = \frac{-2i}{F}\int d^3x e^{-i\bar{p}\cdot\bar{x}} J_0(x)|\Omega\rangle,
	\end{equation}

	which have energy $E(\bar{p}) + E_0$, and $F$ is a constant with dimensions of mass. If we consider the particle state with no momentum, we have
	
	\begin{equation}
		|\pi(0)\rangle = \frac{-2i}{F}Q|\Omega\rangle,
	\end{equation}

	which from above has energy $E_0$. Thus, $|\pi\rangle$ must represent a massless particle - these massless particles that arise during spontaneous symmetry breaking are called Goldstone bosons. \\
	
	In order to see some features of this SSB, to begin let us consider a simple complex scalar field Lagrangian:
	
	\begin{equation}
		\mathcal{L} = (\partial_\mu\phi^*)(\partial^\mu\phi)+m^2\phi^*\phi-\frac{\lambda}{4}\phi^{*2}\phi^2,
	\end{equation}

	which contains a U(1) global symmetry under $\phi(x) \rightarrow e^{i\alpha}\phi(x)$.
	
	A theory's ground state is when its potential is minimised. In this case, the potential is given by $V(\phi) = -m^2|\phi^2|+\frac{\lambda}{4}|\phi^4|$, which is obviously minimised when $|\phi|^2=\frac{2m^2}{\lambda}$. Thus, there are an infinite number of equivalent ground states $|\Omega_\alpha\rangle$ wherein
	
	\begin{equation}
		\langle\Omega_\alpha|\phi|\Omega_\alpha\rangle = \sqrt{\frac{2m^2}{\lambda}}e^{i\alpha},
	\end{equation}

	as the exponential phase factor will not appear in the Lagrangian, albeit appearing in the ground states. This is again the signature of SSB - a symmetry in the Lagrangian that does not exist in the vacua. This ground state can be visualised in $\alpha$-space as a ring of degenerate ground states.
	
	To continue, then, we will need to select a vacuum. Of course, the easiest case would be $|\Omega\rangle$ such that $\langle\Omega|\phi|\Omega\rangle = \sqrt{\frac{2m^2}{\lambda}}$. Then, we can write our field as 
		
	\begin{equation}
		\phi(x) = \left(\sqrt{\frac{2m^2}{\lambda}} +\frac{\sigma(x)}{\sqrt{2}}\right)e^{\frac{i}{F_\pi}\pi(x)},\label{phireplace}
	\end{equation}

	with $F_\pi$ a constant. This changes our theory from having 1 complex scalar field to having 2 real scalar fields. Plugging this back in, we get a Lagrangian of 
	
	\begin{equation}
		\mathcal{L} = \frac{1}{2}(\partial_\mu\sigma)^2+\frac{1}{F_\pi^2}\left(\sqrt{\frac{2m^2}{\lambda}} +\frac{\sigma(x)}{\sqrt{2}}\right)^2(\partial_\mu\pi)^2 - \left(-\frac{m^4}{\lambda}+m^2\sigma^2+\frac{1}{2}\sqrt{\lambda}m\sigma^3+\frac{1}{16}\lambda\sigma^4\right).
	\end{equation}

	Expanding this out, we can clearly see that $\pi$ is a massless particle - our Goldstone boson\footnote{We set $F_\pi$ in order to normalise the Goldstone boson's kinetic term properly.}, and $\sigma$ has a mass $\sqrt{2}m$, as opposed to $\phi$'s mass of $m$. This is known as a linear sigma model. In this case, a U(1) symmetry was broken, and 1 massless Goldstone boson entered the theory. As it turns out, for every symmetry that is broken, we get 1 Goldstone boson. For example, if an SU(3) $\times$ SU(3) theory were to undergo spontaneous symmetry breaking and the resultant Lagrangian only admitted one of the SU(3) symmetries, there would be 8 Goldstone bosons. In more complicated situations, some of these may be pseudo-Goldstone bosons, which are massive.\\
	
	The last thing to consider before moving on to the full electroweak theory is what occurs when a gauge boson associated with a symmetry is present. Once again we begin with a simple example - a gauged U(1) theory:
	
	\begin{equation}
		\mathcal{L} = -\frac{1}{4}F_{\mu\nu}^2+(D_\mu\phi^*)(D^\mu\phi)+m^2|\phi|^2-\frac{\lambda}{4}|\phi|^4,
	\end{equation}

	where above $F_{\mu\nu}=\partial_\mu A_\nu - \partial_\nu A_\mu$ and $D_\mu \phi = \partial_\mu \phi +ieA_\mu$. This Lagrangian is called the Abelian-Higgs model, and we once again find that the potential is minimised for (defining $v$) $\braket{\phi} = \sqrt{\frac{2m^2}{\lambda}} = \frac{v}{\sqrt{2}}$. Using our linear sigma model method wherein we replace $\phi$ with \eqref{phireplace}, our Lagrangian now becomes 
	
	\begin{multline}
		\mathcal{L} = -\frac{1}{4}F_{\mu\nu}^2 + \frac{1}{2}\left(v+\sigma\right)^2\left(-\frac{i}{F_\pi}\partial_\mu\pi+\frac{1}{v+\sigma}\partial_\mu\sigma -ieA_\mu\right)\left(\frac{i}{F_\pi}\partial_\mu\pi+\frac{1}{v+\sigma}\partial_\mu\sigma +ieA_\mu\right)\\
		-\left(-\frac{m^4}{\lambda}+m^2\sigma^2+\frac{1}{2}\sqrt{\lambda}m\sigma^3+\frac{1}{16}\lambda\sigma^4\right).
	\end{multline}

	To extract some information from this Lagrangian, it is interesting to take the ``decoupling limit": we want to learn about the theory in the limit that $m$ and $\lambda$ go to $\infty$, while keeping $v$ constant. Since the mass term for the $\sigma$ field is $m_\sigma=\sqrt{2}m$ as before, taking this limit is essentially analysing the theory at energies much smaller than the $\sigma$ field's mass, and so it is a low-energy effective field theory. In this decoupling limit, $\sigma$ is decoupled and we are left with just the gauge boson and the Goldstone boson $\pi$:
	
	\begin{equation}
		\mathcal{L}_{\mathrm{EFT}} = -\frac{1}{4}F_{\mu\nu}^2 + \frac{v^2}{2}\left(\frac{1}{F_\pi}\partial_\mu\pi + eA_\mu\right)^2.
	\end{equation}

	This Lagrangian admits a gauge symmetry of
	
	\begin{equation}
		A_\mu(x) \rightarrow A_\mu(x) + \frac{1}{e}\partial_\mu\theta(x) \ \ \ \ ; \ \ \ \pi(x) \rightarrow \pi(x) - F_\pi\theta(x).
	\end{equation}
	
	 Reading off qualities of these fields is still a bit difficult, and in order to do so we must choose a gauge. The first gauge one can consider is the Lorenz gauge, which similarly to in electromagnetism is $\partial_\mu A_\mu=0$. Expanding the Lagrangian, we have 
	 
	 \begin{equation}
	 	\mathcal{L}_{\mathrm{EFT}} = -\frac{1}{4}F_{\mu\nu}^2 + \frac{v^2}{2}\frac{1}{F_\pi^2}(\partial_\mu\pi)^2 + \frac{v^2e^2}{2}A_\mu^2 +\frac{ev^2}{F_\pi}\partial_\mu\pi A_\mu = -\frac{1}{4}F_{\mu\nu}^2 + \frac{1}{2}(\partial_\mu\pi)^2 + \frac{v^2e^2}{2}A_\mu^2,
	 \end{equation}
 
 	where we used integration by parts and the Lorenz gauge to remove the cross term, and set $F_\pi=v$ to normalise the $\pi$'s kinetic term. This appears to be a theory containing a massive boson $A_\mu$ with mass $ev$, but recall that the gauge removes a degree of freedom and so there are still only 2 degrees of freedom contained in $A_\mu$, not 3, with the 3rd in the massless scalar boson $\pi$.
 	
 	What is more interesting is to consider the unitary gauge, wherein we select $\theta(x)$ such that $\pi(x) \rightarrow 0$. In this case, our Lagrangian collapses to 
 	
 	\begin{equation}
 		\mathcal{L}_{\mathrm{EFT}} = -\frac{1}{4}F_{\mu\nu}^2 + \frac{v^2e^2}{2}A_\mu^2,
 	\end{equation} 
 
 	which is just a Lagrangian for a massive vector boson. Thus, in this gauge, all 3 degrees of freedom are in the vector boson, and it has acquired a mass. This is known as the Higgs mechanism, wherein a vector boson ``eats" a Goldstone boson to acquire a mass. Thus, we have watched as a U(1) symmetry was broken, a massless Goldstone boson field created, and consequently enveloped by the gauge boson to become a massive gauge boson. 
	
	\subsection{Electroweak Theory}
	
	In low energy physics, we know that electromagnetism has a gauge symmetry - specifically a U(1) gauge symmetry, which we will call U(1)$_{\mathrm{EM}}$. In fact, the theory of electromagnetism comes about as a result of a spontaneous symmetry breaking: there is an electroweak (containing both the weak and electromagnetic interactions) theory that originally has an SU(2)$\times$U(1)$_{\mathrm{Y}}$ symmetry (wherein Y is called the hypercharge) but undergoes SSB to be left with U(1)$_{\mathrm{EM}}$.
	The analysis of the electroweak Lagrangian will be done in phases: first seeing how the gauge bosons acquire mass owing to the SSB and the Higgs mechanism, then seeing how fermions couple to these massive bosons, and finally seeing how fermions (particularly neutrinos of course) gain a mass term.\\
	
	To begin, then, let us analyse the section of the electroweak (EW) Lagrangian for the bosons along with a field $H$ known as the Higgs multiplet:
	
	\begin{equation}
		\mathcal{L} = -\frac{1}{4}\left(W^a_{\mu\nu}\right)^2-\frac{1}{4}B_{\mu\nu}^2 +\left(D_\mu H\right)^\dagger\left(D_\mu H\right) +m^2H^\dagger H - \lambda\left(H^\dagger H\right)^2.\label{bosonL}
	\end{equation}

	This Lagrangian has two symmetries. Firstly, the SU(2) symmetry, characterised by the unitary transform $U=e^{i\alpha^a\tau^a}$, with $\tau^a$ the group generators of the SU(2) group. These generators are related to the Pauli spin matrices via $\tau^a=\frac{1}{2}\sigma^a$. In this Lagrangian, $W^a_\mu$ are the three SU(2) gauge bosons (indexed by $a$), with 
	\begin{equation}
			W^a_{\mu\nu} = \partial_\mu W^a_\nu - \partial_\nu W^a_\mu +g\epsilon^{abc}W_\mu^bW_\nu^c
	\end{equation}

	 their field strength. This field strength has the form that it does in order to be gauge invariant with respect to the SU(2) gauge symmetry, with the Levi-Civita symbol being the structure constant of the SU(2) group: these gauge bosons are non-Abelian, meaning that the group generators do not commute with one another, with 
	 \begin{equation}
	 	\left[\tau^a,\tau^b\right] = i\epsilon^{abc}\tau^c,
	 \end{equation}
  which is why the $g\epsilon^{abc}W_\mu^bW_\nu^c$ term appears in the field strength. 
  
  The next symmetry in the Lagrangian is the hypercharge U(1) symmetry, characterised by the unitary transform $U=e^{i\beta}$ - similar to the transformation in electromagnetism. Just as in electromagnetism, particles can be charged relative to the gauge field, and the Higgs multiplet has a hypercharge of $Y=\frac{1}{2}$. The gauge boson in this case is $B_\mu$, which is a regular Abelian gauge boson, such that the field strength is simply $B_{\mu\nu} = \partial_\mu B_\nu - \partial_\nu B_\mu$.
  
  The Higgs multiplet (which has 2 elements) has a kinetic term involving the covariant derivative which accounts for both the SU(2) and U(1)$_{\mathrm{Y}}$ gauge invariances. To ensure gauge invariance, the covariant derivative must be given by 
  
  \begin{equation}
  	D_\mu H = \partial_\mu H - igW^a_\mu\tau^aH - \frac{1}{2}ig'B_\mu H.
  \end{equation}
	
	The factors of $g$ and $g'$ are the coupling constants of the SU(2) and U(1)$_{\mathrm{Y}}$ gauge symmetries, respectively. This is similar to how in electromagnetism, $e$ is the coupling constant. The factor of $\frac{1}{2}$ is owing to the fact that the Higgs multiplet has a hypercharge of $\frac{1}{2}$.\\
	
	Now, let us see how the SSB of SU(2)$\times$U(1)$_{\mathrm{Y}} \rightarrow$ U(1)$_{\mathrm{EM}}$ affects the situation. As we can see in \eqref{bosonL}, the Higgs multiplet has a potential: $V = -m^2|H|^2+\lambda |H|^4$, which is minimised when $|H|^2 = \frac{m^2}{2\lambda} = \frac{v^2}{2}$. Once again, we will use our linear sigma model, similarly to the previous subsection. In this case however, there are two important distinctions: firstly, the multiplet has 2 components, and we only know the norm of it. However, we can make the upper element 0 without any loss of generality \cite{Schwartzbook}. Secondly, the single field appearing in the exponential of \eqref{phireplace} needs to become 3 fields, each coupled to a generator of the SU(2) group. So our Higgs multiplet becomes 
	
	\begin{equation}
		H = \begin{pmatrix}
			0\\
			\frac{v+h}{\sqrt{2}}
		\end{pmatrix} e^{\frac{2i\pi^a\tau^a}{v}}.
	\end{equation}
	
	We can now take a ``shortcut" of sorts - we are interested in the unitary gauge, wherein the gauge bosons are going to ``eat" the Goldstone bosons to acquire a mass. Recall that in this gauge, we simply select our gauge such that the Goldstone bosons ($\pi^a$) vanish, and so we can do so immediately. Thus, plugging this in, we have our Lagrangian as 
	
	\begin{multline}
		\mathcal{L} = -\frac{1}{4}\left(W^a_{\mu\nu}\right)^2-\frac{1}{4}B_{\mu\nu}^2 +\frac{1}{2}\left(\partial_\mu h\right)^2 + \frac{1}{8}g^2(v+h)^2\left((W_\mu^1)^2+(W_\mu^2)^2+(W_\mu^3-\frac{g'}{g}B_\mu)^2\right)\\ +\frac{1}{2}m^2v^2-\frac{1}{4}\lambda v^4+(m^2v-\lambda v^3)h+(\frac{1}{2}m^2-\frac{3}{2}v^2)h^2-\lambda vh^3 -\frac{1}{4}\lambda h^4.
	\end{multline}
	
	The scalar $h$ field that appears in our Lagrangian now is known as the Higgs boson. In order to gain information about the newly-acquired masses of our gauge bosons, we need to diagonalise the mass terms. To do this, we use the relations 
	
	\begin{equation}
		B_\mu = \cos\theta_wA_\mu - \sin\theta_wZ_\mu \ \ \ \ ; \ \ \ \ W^3_\mu = \sin\theta_wA_\mu + \cos\theta_wZ_\mu,\label{rotations}
	\end{equation}
	
	which are simply a 2 dimensional rotation. This $\theta_w$ angle is defined in terms of the coupling constants, with $\tan\theta_w = \frac{g'}{g}$, and this newly defined $A_\mu$ boson will turn out to be our electromagnetic gauge boson. Since the photon is a part of $W^3_\mu$ (as well as $B_\mu$), its couplings to the other SU(2) gauge bosons are contained within the $-\frac{1}{4}\left(W^a_{\mu\nu}\right)^2$ term. Inverting the above relationships, we have that $A_\mu = \sin\theta_wW_\mu^3+\cos\theta_wB_\mu$, such that terms involving the coupling of the EM boson to the $W^a_\mu$ bosons will be related to a coupling constant of (defining $e$) $e=g\sin\theta_w$, which turns out to be the electromagnetic coupling constant, as we shall see.
	
	 Similarly, we can rewrite $W_\mu^1$ and $W_\mu^2$ as 
	
	\begin{equation}
		W_\mu^1 = \frac{1}{\sqrt{2}}\left(W_\mu^+ +W_\mu^-\right) \ \ \ \ ; \ \ \ \ W_\mu^2 = \frac{i}{\sqrt{2}}\left(W_\mu^+ -W_\mu^-\right).
	\end{equation}
	
	As we will see, these $W_\mu^+$ and $W_\mu^-$ are defined as they couple to the electromagnetic gauge boson with charges of $\pm1$ with respect to the coupling constant $e$. Plugging all of this in, and after a lot of calculations, we arrive at the Lagrangian
	
	\begin{multline}
		\mathcal{L} = -\frac{1}{4}F_{\mu\nu}^2 -\frac{1}{4}Z_{\mu\nu}^2 - \frac{1}{2}W_{\mu\nu}^+W_{\mu\nu}^- +\frac{1}{2}m_Z^2Z_\mu^2+m_W^2W_\mu^+W_\mu^- + \frac{e^2}{2\sin^2\theta_w}\left[(W_\mu^+)^2(W_\nu^-)^2-(W_\mu^+ W_\mu^-)^2\right]\\
		+ie\left[F_{\mu\nu}W_\mu^+W_\nu^- + A_\mu W_\nu^+W_{\mu\nu}^--A_\mu W_\nu^-W_{\mu\nu}^+\right] +ie\cot\theta_w\left[Z_{\mu\nu}W_\mu^+W_\nu^- + Z_\mu W_\nu^+W_{\mu\nu}^- - Z_\mu W_\nu^-W_{\mu\nu}^+\right]\\
		+e^2\cot^2\theta_w\left[Z_\mu W_\mu^+Z_\nu W_\nu^- - Z_\mu^2 W_\nu^+ W_\nu^-+W_\mu^+W_\nu^-(A_\mu Z_\nu + Z_\mu A_\nu) -2W_\mu^+W_\mu^-A_\nu Z_\nu\right]\\
		+e^2\left[A_\mu W_\mu^+A_\nu W_\nu^- - A_\mu^2W_\nu^+W_\nu^-\right] +\frac{1}{2}\left(\partial_\mu h\right)^2 -\frac{1}{2}m_h^2h^2 - \frac{em_h^2}{4m_W\sin\theta_w}h^3 - \frac{e^2m_h^2}{32m_W^2\sin^2\theta_w}h^4\\
		+\frac{em_W}{\sin\theta_w}hW_\mu^+W_\mu^-+\frac{em_Z}{2\sin\theta_w\cos\theta_w}hZ_\mu^2 +\frac{e^2}{4\sin^2\theta_w}h^2W_\mu^+W_\mu^- + \frac{e^2}{8\sin^2\theta_w\cos^2\theta_w}h^2Z_\mu^2.
	\end{multline}
	
	In the above, the masses are given by $m_h=\sqrt{2}m$, $m_W = \frac{gv}{2}$, and $m_Z=\frac{gv}{2\cos\theta_w}$. The field strength $F_{\mu\nu}=\partial_\mu A_\nu - \partial_\nu A_\mu$ is the usual electromagnetic field strength, and in the above we have used $W_{\mu\nu}^\pm = \partial_\mu W_\nu^\pm - \partial_\nu W_\mu^\pm$.
	
	Clearly, the 3 symmetries broken from the loss of the SU(2) symmetry have resulted in 3 available degrees of freedom, which instead of becoming Goldstone bosons, have caused the incarnation of 3 massive bosons: $Z_\mu , \ W_\mu^+$ and $W_\mu^-$. The two charged bosons have mass $m_W$, and the electrically neutral boson (electrically neutral as it does not interact with the photon field) has a mass $m_Z$, which we know to be larger than $m_W$ as $\cos\theta_w$ must be between 0 and 1. Finally the Higgs boson has a mass $m_h$, but this last boson did not acquire its mass via the Higgs mechanism - it is the massive degree of freedom which came from the Higgs multiplet.
	
	From this Lagrangian, we can see all the interactions that bosons can have with one another - all mediated by the parameters $e$ and $\theta_w$. There are many different 3- and 4-point vertices, and interesting scattering matrices can be calculated, but this is not necessary for our purposes.\\
	
	Let us now turn our attention to the fermionic part of our Lagrangian. The part of the Lagrangian that details the interactions between fermions and electroweak gauge bosons is given by
	
	\begin{multline}
		\mathcal{L} = i\bar{L}_j\left(\slashed{\partial}-ig\slashed{W^a}\tau^a-ig'Y_L\slashed{B}\right)L_j + i\bar{Q}_j\left(\slashed{\partial}-ig\slashed{W^a}\tau^a-ig'Y_Q\slashed{B}\right)Q_j + i\bar{e}^j_R\left(\slashed{\partial}-ig'Y_e\slashed{B}\right)e_R^j\\ +i\bar{\nu}^j_R\left(\slashed{\partial}-ig'Y_\nu\slashed{B}\right)\nu_R^j +i\bar{u}^j_R\left(\slashed{\partial}-ig'Y_u\slashed{B}\right)u_R^j +i\bar{d}^j_R\left(\slashed{\partial}-ig'Y_d\slashed{B}\right)d_R^j.\label{fermionicL}
	\end{multline}

	In the above, $L_j$ and $Q_j$ are left-chiral leptonic and quarkic doublets, with $j$ running from 1 to 3, that behave as left-handed Weyl spinors. These doublets transform under the SU(2) symmetry and thus are charged with respect to the $W_\mu^a$ bosons, and are given by 
	
	\begin{equation}
		L_j = \begin{pmatrix}
			\nu_{eL}\\e_L
		\end{pmatrix}, \ \begin{pmatrix}
		\nu_{\mu L}\\\mu_L
	\end{pmatrix}, \ \begin{pmatrix}
	\nu_{\tau L}\\\tau_L
\end{pmatrix} \ \ \ \ ; \ \ \ \ Q_j = \begin{pmatrix}
u_L\\d_L
\end{pmatrix} , \ \begin{pmatrix}
c_L\\s_L
\end{pmatrix} , \ \begin{pmatrix}
t_L\\b_L
\end{pmatrix} , \ 
	\end{equation}
	
	with $e$ the electron and $\nu_e$ the electron neutrino, and similarly for the muon $\mu$ and tau $\tau$, and the quarks are the up $u$, down $d$, strange $s$, charm $c$, top(truth) $t$ and bottom(beauty) $b$ quarks. These doublets are also charged with respect to the hypercharge gauge boson, with hypercharges given by $Y_L$ for the leptons and $Y_Q$ for the quarks. 
	
	For right-chiral fermions, though they are charged with respect to the hypercharge boson with their respective charges $Y_n$, they do not couple to the $W_\mu^a$ bosons. These fermions act as right-handed Weyl spinors, with 
	
	\begin{equation}
		e_R^j = e_R,\mu_R,\tau_R \ \ \ \ ; \ \ \ \ v_{R}^j = \nu_{eR} ,\nu_{\mu R} , \nu_{\tau R} \ \ \ \ ; \ \ \ u_R^j = u_R,c_R,t_R \ \ \ \ ; \ \ \ d_R^j = d_R,s_R,b_R.
	\end{equation}

	So the fermionic part of our Lagrangian is entirely chiral - left- and right-handed particles interact very differently. This is the connection we see between the chiral electroweak theory and the V-A current formulation of the 4-Fermi theory. \\
	
	For the purposes of this work, we wish to focus on neutrinos. To this end, ignoring the quarkic contents for now, let us analyse the interactions between leptons and the neutral bosons. These components are given by
	
	\begin{equation}
		\mathcal{L} = \bar{e}_L^j\left(-\frac{1}{2}g\slashed{W}^3+g'Y_L\slashed{B}\right)e_L^j + \bar{\nu}_L^j\left(\frac{1}{2}g\slashed{W}^3+g'Y_L\slashed{B}\right)\nu_L^j + g'Y_e\bar{e}^j_R\slashed{B}e_R^j + g'Y_\nu \bar{\nu}_R^j\slashed{B}\nu_R^j.
	\end{equation}

	Using the rotations in \eqref{rotations}, we obtain 
	
	\begin{multline}
		\mathcal{L} = e\bar{e}_L^j\left((-\frac{1}{2}+Y_L)\slashed{A}-\cos\theta_w(\frac{1}{2}+Y_L)\slashed{Z}\right)e_L^j + e\bar{\nu}_L^j\left((\frac{1}{2}+Y_L)\slashed{A}-\cos\theta_w(-\frac{1}{2}+Y_L)\slashed{Z}\right)\nu_L^j\\
		+ eY_e\bar{e}_R^j\left(\slashed{A}-\tan\theta_w\slashed{Z}\right)e_R^j 
		+ eY_\nu\bar{\nu}_R^j\left(\slashed{A}-\tan\theta_w\slashed{Z}\right)\nu_R^j,\label{neutralints}
	\end{multline}

	where again we have used that $e=g\sin\theta_w$. While before the $e=g\sin\theta_w$ relation may have simply been conventional in order to satisfactorily define charged massive bosons, here we see that it is the same charge as the electromagnetic charge that electrons have. Since we know that electrons must couple to the photonic gauge field with a coupling of $-e$, we can immediately read off that $Y_L=-\frac{1}{2}$ and $Y_e=-1$. This also gives us the correct prediction that left-handed neutrinos do not couple to the electromagnetic force - they are electrically neutral. In order for right-handed neutrinos to be the same, we need $Y_\nu=0$. Going through the same process, one can calculate the hypercharges of the quarks as well.\\
	
	The first important result is already apparent: looking at \eqref{fermionicL}, as well as \eqref{neutralints}, we see that while left-handed neutrinos do not interact with the electromagnetic gauge bosons, they do interact with the $Z$-bosons and charged $W^\pm$-bosons. On the other hand, the right-handed neutrinos do not couple with anything! Thus, the weak interaction is perfectly chiral: only left-chiral neutrinos may be involved in interactions. 
	
	As an aside, to make the connection to the 4-Fermi theory more apparent, recall that left-handed Weyl spinors can be written in terms of Dirac spinors by the left projection operator $P_L = \frac{1}{2}(1-\gamma^5)$, such that $P_L\psi=\psi_L$ for some spinor. Using this fact, we can write the left-handed Weyl spinors in the above Lagrangians as Dirac spinors but with an added factor of $\frac{1}{2}(1-\gamma^5)$. Combined with the gamma matrices inherent in the Feynman slash notation, the emergence of the vector minus axial (V-A) form of the fermionic interactions becomes apparent, once the bosonic masses are taken to be extremely large. \\
	
	The final task is to analyse where neutrinos' masses came from. In fact, initially the electroweak theory assumed neutrinos to be massless. Their oscillations - discussed next subsection - proved that this is impossible, and there must be 3 distinct masses for the 3 species of neutrinos\footnote{Though one species may still be massless, the other 2 cannot be.}. The most general renormalisable mass terms that involve neutrinos are given by the Lagrangian
	
	\begin{equation}
		\mathcal{L} = -Y_{ij}^e\bar{L}^iHe_R^j -Y_{ij}^\nu\bar{L}^iH\nu_R^j - iM_{ij} \left(\nu_R^i\right)^c\nu_R^j + \mathrm{Hermitian \ conjugate}.
	\end{equation}

	In this mass Lagrangian, the first two terms allow a mass to arise from the SSB of the Higgs multiplet, with the mass of each different particle related to the components of a matrix of Yukawa coupling constants. Note that this is not a case of a Higgs mechanism - those degrees of freedom were already used to generate the masses of the weak gauge bosons. These terms involving the Higgs multiplet are simply invariant under the SU(2)$\times$U(1)$_{\mathrm{Y}}$ symmetries, and once SSB occurs, the constant $v$ terms will multiply terms quadratic in fields and thus will act as mass terms. 
	
	The third term is more interesting, and is called a Majorana mass term. The superscript $c$ denotes a charge conjugation - that is, $\nu_R^c = \nu_R^T\sigma_2$. This Majorana mass term is only allowed since the question of whether neutrinos are Dirac or Majorana particles is still unanswered. Recall that Majorana fermions are fermions that are their own antiparticle: there is no distinction at all between them. We have already seen that neutrinos are uncharged electromagnetically, and so cannot rely on electric charge to distinguish between particle and antiparticle. There may be other quantum numbers that neutrinos carry - the most promising being lepton number - that will solidify their position as Dirac fermions. If neutrinos do not carry lepton number, then phenomena like neutrino-less double-beta decay can occur, which would require a neutrino to annihilate with another neutrino. This would only be possible if their lepton number was 0. For now, we shall keep both mass terms, but it is important to keep in mind that if neutrinos are Dirac particles, this final term would no longer be allowed.\\
	
	Let us now analyse just one species of neutrino. Writing these mass terms more simply, the Lagrangian would look something like 
	
	\begin{equation}
		\mathcal{L} = -m\bar{\nu}_L\nu_R -\frac{1}{2}M\bar{\nu}_R\nu_R + H.c \propto \bar{\nu}\begin{pmatrix}
			0 & m \\ m & M
		\end{pmatrix}\nu
	\end{equation}
	
	with $m$ related to the vacuum expectation value $v$ that arises from SSB and the Yukawa coupling, and $M$ the Majorana mass. Thus, this mass matrix describes the 2 possible masses of neutrinos. If only Dirac mass terms are allowed, then we would have $M=0$, and both left- and right-handed neutrinos would have degenerate mass $m$. If, on the other hand, they are Majorana particles, then $M\neq0$ and we must diagonalise this mass matrix. This ends up giving two possible masses: $m_{1,2} = \sqrt{m^2+\frac{1}{4}M^2}\pm\frac{1}{2}M$. If we consider a scenario wherein $M$ is extremely large, $M\gg m$, using binomial approximations we get that $m_1\approxeq M$, a very heavy particle, and $m_2\approxeq \frac{m^2}{M}$, a very light particle. This is known as the see-saw mechanism: the larger $M$ is, the smaller one of the masses are. This is a more natural way to account for the very tiny masses of left-handed neutrinos. In the Dirac case, a fine-tuning of $m$ is needed for the neutrino masses to all be extremely tiny. In the case of Majorana neutrinos, the smallness of left-handed neutrino masses is directly related to the largeness of sterile (thus named as they do not interact with anything) right-handed neutrino masses. In any case, some condition is required: either the finely-tuned smallness of $m$, or the large difference between $m$ and $M$. 
	
	There are still many mysteries surrounding neutrinos and the electroweak theory in general: the reason behind the chirality of the theory, the reason behind neutrinos' tiny masses, and whether neutrinos are Majorana fermions are still unsolved. However, for the purposes of phenomenology, we know that it is chiral and the masses are minute - and thus, we can still perform useful calculations, especially using the simple 4-Fermi theory above.
	
	\subsection{Neutrino Oscillations}
	
	In the electroweak theory, neutrinos exist in 3 species, and there are 2 bases in which to describe them: the flavour basis and the mass basis. The flavour basis describes the states whose couplings to the charged $W^\pm$ bosons are diagonalised:
	
	\begin{equation}
		\mathcal{L} = -\frac{e}{\sqrt{2}\sin\theta_w}\left(\bar{e}_L\slashed{W}\nu_{Le} + \bar{\mu}_L\slashed{W}\nu_{L\mu} + \bar{\tau}_L\slashed{W}\nu_{L\tau} \right) + H.c,
	\end{equation}
	
	while the mass basis describes the states wherein the mass matrix is diagonalised. Using diagonalisation from linear algreba, the mass eigenstates can be related to the flavour eigenstates using a unitary matrix $U^{ij}$: that is, $\nu_{L\alpha} = U^{\alpha n}\nu_{Ln}$, with $\alpha = e,\mu,\tau$ and $n=1,2,3$. Thus, the interactions above in the mass basis become 
	
	\begin{equation}
		\mathcal{L} = -\frac{e}{\sqrt{2}\sin\theta_w}U^{ij}\bar{e}_{Li}\slashed{W}\nu_{Lj} + H.c.
	\end{equation}

	This unitary matrix $U^{ij}$ is called the Pontecorvo-Maki-Nakagawa-Sakata (PMNS) matrix, and is the leptonic equivalent of the Cabibbo-Kobayashi-Maskawa (CKM) matrix for quarks. 
	
	This PMNS matrix can be categorised using 3 mixing angles $\theta_{12}, \theta_{13}$ and $\theta_{23}$, as well as a Dirac phase $\delta$. If neutrinos are Majorana particles, there are 2 additional phases, $\alpha_{12}$ and $\alpha_{31}$. This is because if neutrinos were Dirac particles, they would have 3 U(1) symmetries under $\nu_L^i \rightarrow e^{i\alpha_i}\nu_L^i$, as the antiparticles would have a $e^{-i\alpha_i}$ factor. If they are Majorana particles, these symmetries are lost as antiparticles are particles, so instead of a factor of $e^{i\alpha_i}e^{-i\alpha_i}$, terms would have a factor of $e^{2i\alpha_i}$. These 3 symmetries in the Dirac case can then be used to remove 2 phases (not all 3, as there is still an overall phase - the Dirac phase - that cannot be removed). Writing the PMNS matrix explicitly, it is equal to a rotation in the 1-2 plane, then a rotation in the 1-3 plane including the Dirac phase, and finally a rotation in the 2-3 plane. The Majorana phases can also be accounted for in a separate matrix:
	
	\begin{multline}
		U = \begin{pmatrix}
			1 & 0 & 0 \\ 0 & e^{i\alpha_{12}/2} & 0 \\ 0 & 0 & e^{i\alpha_{31}/2}
		\end{pmatrix}
	\begin{pmatrix}
		1 & 0 & 0 \\ 0 & \cos\theta_{23} & \sin\theta_{23} \\ 0 & -\sin\theta_{23} & \cos\theta_{23}
	\end{pmatrix}
	\begin{pmatrix}
		\cos\theta_{13} & 0 & \sin\theta_{13}e^{i\delta} \\ 0 & 1 & 0 \\ -\sin\theta_{13}e^{i\delta} & 0 & \cos\theta_{13}
	\end{pmatrix}
\begin{pmatrix}
	\cos\theta_{12} & \sin\theta_{12} & 0 \\ -\sin\theta_{12} & \cos\theta_{12} & 0 \\ 0 & 0 & 1
\end{pmatrix} \\
= \begin{pmatrix}
	c_{12}c_{13} & s_{12}c_{13} & s_{13}e^{-i\delta} \\ -s_{12}c_{23}-c_{12}s_{23}s_{13}e^{i\delta} & c_{12}c_{23} - s_{12}s_{23}s_{13}e^{i\delta} & s_{23}c_{13} \\ s_{12}s_{23} - c_{12}c_{23}s_{13}e^{i\delta} & -c_{12}s_{23} - s_{12}c_{23}s_{13}e^{i\delta} & c_{23}c_{13}
\end{pmatrix}\begin{pmatrix}
1 & 0 & 0 \\ 0 & e^{i\alpha_{12}/2} & 0 \\ 0 & 0 & e^{i\alpha_{31}/2}
\end{pmatrix},
	\end{multline}
	
	where we have written for shorthand $c_{ij} = \cos\theta_{ij}$ and similarly for the sine functions. This is just one parameterisation - for example, the overall phase could have been included in a different rotation. The actual values for the components of this matrix have not yet been measured extremely precisely - the errors in its components are fairly large. Regardless, we will use some values in our calculations in later chapters - but it is important to remember that the inability to measure these mixing angles and phases accurately is a huge source of uncertainty in all calculations. \\
	
	The effect of neutrino oscillations, also known as neutrino mixing, comes into play when we consider the time evolution of a neutrino. When a neutrino is created (or scattered), it does so in a specific flavour state - so that at our initial time $t=0$, our neutrino exists purely in one flavour eigenstate. This means that there is a probability of it being in each of the different mass eigenstates. 
	
	While during interactions the flavour basis is important, during free streaming it is the mass basis that matters, as time evolution of neutrinos is described using mass eigenstates. To see this, consider a neutrino in its rest frame. Then its energy is simply its mass, and its evolution is 
	
	\begin{equation}
		|\nu_j(t)\rangle = e^{-im_jt}|\nu_j(0)\rangle.
	\end{equation}
	
	In flavour eigenstates this evolution does not make sense, as flavour eigenstates do not have distinct masses. Generalising to the lab frame wherein the particle has energy $E$ and momentum $p$, after a time $t$ and travelling a distance $L$, we would have 
	
	\begin{equation}
		|\nu_j(t,L)\rangle = e^{-iEt+ipL}|\nu_j(0)\rangle.\label{oscilprop}
	\end{equation}

	For illustrative purposes, let us consider an extremely relativistic case where $v_\nu \approx1$. Then $t\approxeq L$ and $p=\sqrt{E^2-m_j^2} \approxeq E(1-\frac{m_j^2}{2E^2})$, and we have 
	
	\begin{equation}
		|\nu_j(L)\rangle \approxeq e^{-im_j^2L/2E}|\nu_j(0)\rangle.
	\end{equation}

	So now let us consider the case wherein our neutrino is created in some flavour state $\alpha$ and after some distance $L$, we wish to know the probability that it will be measured in the flavour state $\beta$. We begin with the transition amplitude:
	
	\begin{equation}
		A_{\alpha\beta} = \sum_{j=1}^3\braket{\nu_\beta|\nu_j(L)}\braket{\nu_j(0)|\nu_\alpha} = \sum_{j=1}^3\braket{\nu_\beta|\nu_j}e^{-im_j^2L/2E}\braket{\nu_j|\nu_\alpha} = \sum_{j=1}^3U_{\beta j}e^{-im_j^2L/2E}U^*_{\alpha j},
	\end{equation}

	and then square it to obtain (after some lengthy calculation) the probability:
	
	\begin{multline}
		P(\alpha \rightarrow \beta) = |A_{\alpha\beta}|^2 \\= \delta_{\alpha\beta} - 4\sum_{j>i}\mathrm{Re}\left[U^*_{\alpha j}U_{\beta j}U_{\alpha i}U^*_{\beta i}\right]\sin^2\left(\frac{\Delta m_{ji}^2}{4E}L\right) + 2\sum_{j>i} \mathrm{Im}\left[U^*_{\alpha j}U_{\beta j}U_{\alpha i}U^*_{\beta i}\right]\sin\left(\frac{\Delta m_{ji}^2}{2E}L\right), \label{oscillationProb}
	\end{multline}

	where $\Delta m_{ji}^2 = m_j^2 -m_i^2$ is the difference between neutrino masses squared. This is one of the quantities that can be measured from experimental data, and their values are discussed in the next section. Thus, we have seen that a relativistic neutrino that is created in some flavour state has a probability to, at a later time, exist in a different flavour state entirely. This effect of neutrino mixing was vital in explaining why neutrinos from the sun - which should all be created as electron neutrinos - seemed to have a deficit. In actuality, they had oscillated into other flavours, and this experiment will be discussed in the next subsection.
	
	Though this derivation was done for neutrinos propagating in vacuum, there is an additional effect known as the Mikheyev-Smirnov-Wolfenstein (MSW) effect which describes how neutrinos are affected when travelling in matter. This effect causes a slight change to the effective Lagrangian and the effective masses of the different neutrinos, and is owing to scattering off electrons \cite{Gorbunov:2011zz}.\\

	While this oscillation occurs for relativistic neutrinos, for neutrinos that are non-relativistic, the coherent nature of this mixing is lost\footnote{This is not true in the case of degenerate or quasi-degenerate mass spectra, but in reality we know there are no degenerate or quasi-degenerate species of neutrinos \cite{Akhmedov:2017xxm}.}. Without quantum coherence, interference cannot occur between the wave functions of different mass eigenstates. This decoherence can also occur when the neutrinos are created or detected \cite{Akhmedov:2017xxm}, but for the case of the C$\nu$B it is important to consider decoherence that occurs during propagation. As we will see next chapter, the relic neutrinos created in the early universe were relativistic upon creation, but owing to the expansion of spacetime over the course of billions of years, some (if not all) of these neutrinos will be non-relativistic today.
	
	Decoherence occurs during propagation owing to the differences in group velocities of different wave packets, corresponding to the different mass species. After a certain amount of time called the decoherence time $t_{\mathrm{decoh}}$, the wave packets - each moving with a different group velocity $v_g^i$ - will have separated by a distance exceeding the size of the wave packets, $x_{wp}$. Thus, the wave packets of different mass eigenstates will no longer overlap, and therefore interference - mixing - cannot occur. This will happen when 
	\begin{equation}
		|\Delta v_g|t_{\mathrm{decoh}} = x_{wp},
	\end{equation}

	or similarly at the distance $L_{\mathrm{decoh}} = \bar{v}_gt_{\mathrm{decoh}} = \frac{\bar{v}_g x_{wp}}{|\Delta v_g|}$, with $|\Delta v_g|$ the difference between group velocities of mass species and $\bar{v}_g$ the average group velocity. As we can see, for very relativistic neutrinos with $\bar{v}_g\approxeq1$ and $|\Delta v_g|$ very close to 0, provided that $|\Delta v_g|$ is much smaller than $x_{wp}$, neutrino oscillations can continue to occur for a large part of the neutrinos' journey. For the case of relic neutrinos however, wherein (as we shall see for the inverted hierarchy case) $|\Delta v_g|$'s smallest value is $\approx 10^{-4}$, the decoherence length is only a couple thousand times the size of the wave packet itself: clearly, after having travelled for billions of years, neutrinos from the C$\nu$B are no longer oscillating today, and their composition is essentially fixed.

	\subsection{Parameters Gleaned From Experimental Data}
	
	As we have seen up until now, there are a few parameters in the electroweak theory that are still unknown - namely the masses of the different neutrino mass eigenstates and the elements of the PMNS matrix - which need to be extracted from experimental data. Some pertinent moments from the past five decades will be explored here.
	
	 Though neutrino masses are not yet known exactly, the most stringent bounds are imposed by neutrino oscillation experiments and indirect information from the CMB. The lower limit comes from the fact that we have deduced the absolute values of the differences between the squared masses of different mass eigenstates of neutrinos, $\Delta m^2_{ij}=m_i^2-m_j^2$, by experimental observations and using \eqref{oscillationProb}. \\
	 
	 The first observation that provided some information came from solar neutrinos. In the core of the sun, many nuclear fusion reactions are occurring. In these reactions, which involve the fusion of nuclei and occasionally electrons, neutrinos are constantly created and emitted but only in the electron neutrino flavour eigenstate. Thus, before neutrino oscillation was understood, one would expect that all neutrinos arriving from the sun on earth would be electron neutrinos. However, the first experiment to detect solar neutrinos at the Homestake Solar Neutrino Detector found that the experimental flux of these electron neutrinos from the sun was only around 33.5\% $\pm$ 5\% of the expected theoretical flux \cite{Cleveland:1998nv}.
	 
	 Many experiments were performed around this time, and the results seemed to be consistent. An important experiment that began in 1999 was the Sudbury Neutrino Observatory, which utilised heavy water to detect solar neutrinos. The reason this experiment was interesting is that the deuterium in the heavy water could undergo two different weak reactions: a charged current reaction
	 
	 \begin{equation}
	 	\nu_e + ^2\mathrm{H} \rightarrow p + p + e^-, \label{chargedflux}
	 \end{equation} 
 
 	which, as we have seen in the previous subsections, can only involve electron neutrinos, and a neutral current reaction
 	
 	\begin{equation}
 		\nu + ^2\mathrm{H} \rightarrow p + n + \nu, \label{neutralflux}
 	\end{equation}
 
 	which can occur for any flavour of neutrino. Therefore, by measuring the rate of \eqref{chargedflux} one can obtain the electron neutrino flux, while measuring \eqref{neutralflux} will provide the total neutrino flux. The experiment found that the neutral current flux was 87\% $\pm$ 19\% of the theoretical value, with the electron neutrino flux only around 30\% $\pm$ 5\% of the expected theoretical flux \cite{Jelley:2009zz} (in agreement with the Homestake experiment as well). What this showed is that there was no error in the solar model of nuclear reactions, and it must be that around two thirds of neutrinos created as electron neutrinos converted into other flavours.\\
 	
 	The oscillation of solar neutrinos can be simplified and modelled fairly accurately using only 2 neutrino species. The flavours will be $\nu_e$ and $\nu_h$, with the latter representing a combination of muon- and tau-neutrinos. In this case, the PMNS mixing matrix is simplified immensely to become a simple 2D rotation matrix:
 	
 	\begin{equation}
 		U_{2\mathrm{D}} = \begin{pmatrix}
 			\cos\theta_{sol} & \sin\theta_{sol} \\ -\sin\theta_{sol} & \cos\theta_{sol}
 		\end{pmatrix},\label{2DU}
 	\end{equation}
 
 	and using \eqref{oscillationProb}, we have in vacuum 
 	
 	\begin{equation}
 		P(\alpha \rightarrow \beta) = \delta_{\alpha\beta} + (-1)^{\delta_{\alpha\beta}}\sin^2(2\theta_{sol})\sin^2\left(\frac{\Delta m_{12}^2}{4E}L\right).\label{2dangle}
 	\end{equation}
 	
 	Na\"ively, the amount of electron neutrinos that survived should then be quite high, but the particularly low retention rate of electron neutrinos is owing to the MSW effect described previously \cite{Jelley:2009zz}, wherein the effective potential owing to the dense population of electrons in the sun causes the electron neutrino upon creation to be much more closely related to the heavier mass eigenstate \cite{Gorbunov:2011zz} rather than the lighter one as in vacuum. This MSW effect is energy dependent, and only comes into effect at energies above 3 MeV \cite{Gorbunov:2011zz}. The reactions used for detection at the Sudbury Neutrino Observatory had energy thresholds greater than 5 MeV, and so only neutrinos that were affected by the MSW effect were detected. Therefore, in this case, owing to the fact that neutrinos were almost all created in the $|\nu_2\rangle$ state, from \eqref{2DU} we would have 
 	
 	\begin{equation}
 		P(\nu_e \rightarrow \nu_e) = |\braket{\nu_e|\nu_2}|^2 = \sin^2\theta_{sol},
 	\end{equation}
 
 	a very different case to when the MSW effects are negligible. Using these formulae, and the approximation that $\theta_{13}$ is small (which we will see shortly), one can obtain approximately $\Delta m_{12}^2$ and $\theta_{12}\approx\theta_{sol}$. Of course, the accuracy of these values has been improved since then.\\
 	
 	The next piece of information comes from atmospheric neutrinos arising from cosmic rays. As super high energy particles such as protons and other nuclei arrive from space, their interaction with the atmosphere causes a shower of particles, including pions, muons and neutrinos. The charged pions then decay like 
 	
 	\begin{equation}
 		\pi^+ \rightarrow \mu^+ + \nu_\mu \ \ \ \ ; \ \ \ \pi^- \rightarrow \mu^- + \bar{\nu}_\mu,
 	\end{equation}
 
 	with a branching ratio of over 99\%. Muons also decay via 
 	
 	\begin{equation}
 		\mu^+ \rightarrow e^+ + \nu_e + \bar{\nu}_\mu \ \ \ \ ; \ \ \ \mu^- \rightarrow e^- + \nu_\mu + \bar{\nu}_e.
 	\end{equation}
 
 	Clearly, there are multiple electron and muon neutrinos being showered onto earth from every cosmic ray. In fact, owing to their small interaction strength, these neutrinos can pass through the earth with very little chance of being stopped. Thus, albeit being stationed on land, we would expect a detector to observe a near-isotropic distribution of these neutrinos. 
 	
 	In reality, this was not the case. While electron neutrinos seemed reasonably isotropic, muon neutrinos showed a strong dipole in their distribution: there were more coming from above than from below through the earth \cite{Ashie:2005ik}. This was not owing to interactions with matter in the earth, as in that case the electron neutrinos would have experienced a similar dipole distribution. It must be that the increased travel time for neutrinos coming from below has allowed for more time for muon neutrinos to oscillate into a third flavour, one that does not have a strong mixing with electron neutrinos (this is the first sign that $\theta_{13}$ is small). Thus, we may once again model this using 2 flavours, but in this case instead of electron- and heavy-neutrinos ($\nu_e$ and $\nu_h$), we have muon- and tau-neutrinos ($\nu_\mu$ and $\nu_\tau$). Using then \eqref{2dangle}, we can find approximately $\Delta m_{23}^2$ and $\theta_{23}\approx\theta_{atm}$.\\
 	
 	The final parameter we can analyse is $\theta_{13}$. Experimentally, what is actually measured is $|U_{e3}|^2 = \sin^2\theta_{13}$, and so the Dirac phase unfortunately cannot be measured using this mixing probability. There are a few ways that $|U_{e3}|^2$ has been measured, one of which is the survival probability of electron anti-neutrinos. Under the approximation $\frac{\Delta m_{12}^2}{2E}L\ll1$, and using the unitarity of $U$, \eqref{oscillationProb} becomes 
 	
 	\begin{equation}
 		P(\bar{\nu}_e \rightarrow \bar{\nu}_e) = 1 - 4|U_{e3}|^2(1-|U_{e3}|^2)\sin^2\left(\frac{\Delta m_{13}^2}{4E}L\right).
 	\end{equation}
 
 	Clearly, the amount of electron anti-neutrinos with some energy $E$ will oscillate over distance $L$ based on $|U_{e3}|^2$. By measuring the flux of electron anti-neutrinos at different distances away from their point of creation, the final mixing angle can be deduced.
 	
 	It is important to note here that since the mass differences appear in squared sinusoidal arguments, their sign is not apparent: we are actually measuring the absolute value between mass eigenstates. Thanks to the MSW effect in the sun, we do know that $m_2 > m_1$ (as $m_2$ was the heavier state), but whether $m_3$ is smaller or larger than $m_1$ and $m_2$ is still unknown. \\
 	
 	The presence of the Dirac phase would be an indication of charge-parity (CP) violation: that is, if we take a situation and mirror all the parities and charges of the particles involved, the system should remain the same, but it does not. 
 	The full charge-parity-time (CPT) symmetry must be retained in order for the theory to be Lorentz invariant and unitary \cite{Schwartzbook}. Time reversal involves taking $i\rightarrow-i$ in the Lagrangian \cite{Schwartzbook}, so clearly any term that contains both real and imaginary parts must violate CP invariance through either its real or imaginary part. Thus, if the Dirac phase (also called the CP phase owing to the phenomenon just described) is anything other than a multiple of $\pi$, there is CP violation in the leptonic fermions.\\

	 The current values for mass differences, taking into account current experiments including T2K, NOvA, Super-Kamiokande and many others are \cite{Esteban:2020cvm}
	
	\begin{equation}
		\Delta m^2_{21} \approxeq (8.6 \ \mathrm{meV})^2 \ \ \ \ \ ; \ \ \ \ \ |\Delta m^2_{3i}| \approxeq (50 \ \mathrm{meV})^2,\label{massdiff}
	\end{equation}
	
	wherein $i$ can be 1 or 2, depending on the hierarchy. Since the sign of $\Delta m^2_{3i}$ is still unknown, there are 2 hierarchies or orderings: the normal hierarchy, which has $m_1$ as the smallest mass, and the inverted hierarchy, which has $m_3$ as the smallest. Thus, we know that at least 2 neutrino species are not massless. Setting the smallest neutrino's mass to 0, we get the lower limit for each ordering. For the normal hierarchy, we have
	
	\begin{equation}
		m_1\approxeq0 \ \mathrm{meV} \ \ \ ; \ \ \ m_2\approxeq 8.6 \ \mathrm{meV} \ \ \ ; \ \ \ m_3\approxeq 50 \ \mathrm{meV} \ \ \ ; \ \ \ \sum_{j=1}^{3}m_j\approxeq 58.6 \ \mathrm{meV}, \label{normalorder}
	\end{equation}
	
	in which we took $i=1$ in \eqref{massdiff}.\\
	In the inverted hierarchy then, taking $i=2$ in \eqref{massdiff}, we get the lower limit
	
	\begin{equation}
		m_1\approxeq49.3 \ \mathrm{meV} \ \ \ ; \ \ \ m_2\approxeq50 \ \mathrm{meV} \ \ \ ; \ \ \ m_3\approxeq0 \ \mathrm{meV} \ \ \ ; \ \ \ \sum_{j=1}^{3}m_j\approxeq 99.3 \ \mathrm{meV}. \label{invertedorder}
	\end{equation}

	From \cite{Esteban:2020cvm} we also have the most recently updated experimental values for the mixing angles and Dirac phase, and as a result we have for the elements of the PMNS matrix (listing the 3$\sigma$ ranges):
	
	\begin{equation}
		U_{\mathrm{PMNS}} = \begin{pmatrix}
			0.801 \rightarrow 0.845 & 0.513 \rightarrow 0.579 & 0.143 \rightarrow 0.155 \\ 
			0.234 \rightarrow 0.500 & 0.471 \rightarrow 0.689 & 0.637 \rightarrow 0.776 \\
			0.271 \rightarrow 0.525 & 0.477 \rightarrow 0.694 & 0.613 \rightarrow 0.756
		\end{pmatrix}.
	\end{equation}

	It is vital to note that these values are not all independent, as the unitarity of the matrix must be ensured. For the normal hierarchy, the CP phase is given by a best fit of $\delta=197^\circ$, with a huge 3$\sigma$ range of $\delta \in [120 ; 369]$ \cite{Esteban:2020cvm}, indicating that there is a chance that CP symmetry is conserved in the leptonic sector but that there is most likely a slight violation. In the inverted hierarchy on the other hand, the CP phase is given by a best fit of $\delta=282^\circ$, with a slightly smaller 3$\sigma$ range of $\delta \in [193 ; 352]$ \cite{Esteban:2020cvm}, indicating that in the inverted ordering case CP symmetry is almost certainly violated, and is most likely maximally violated (that is, $\delta$ is an odd multiple of $\frac{\pi}{2}$). \\
	
	From direct experimental evidence the current upper limit on neutrino masses is (for electron neutrinos) $m_\nu < 1.1$ eV \cite{Aker:2019uuj}. If neutrinos are Majorana particles, from experiments searching for neutrino-less double beta decay, the current upper limit on the lightest neutrino species is $m_\nu < 0.16$ eV, with their sum constrained by $\sum_{j=1}^{3}m_j< 1.3 \ \mathrm{eV}$ \cite{Agostini:2019hzm}.
	
	The most stringent constraint on the upper limit of neutrino masses comes indirectly from data collected by the Planck satellite and takes into account baryon acoustic oscillations, CMB lensing, temperature fluctuations and polarisation of the CMB, and gives (with a 95\% confidence level) \cite{Aghanim:2018eyx}
	\begin{equation}
		\sum_{j=1}^{3}m_j< 120 \ \mathrm{meV}.\label{massupperlim}
	\end{equation}
	
	It should be noted that this value assumes all neutrino masses are degenerate, which we know is not true. Also, including the value of the Hubble constant $H_0$ from early universe measurements would decrease \eqref{massupperlim} even further, but owing to the current tension in $H_0$ measurement, this parameter is not yet taken into consideration. It is also important to remember that this constraint depends on which other cosmological parameters are fixed in the model.\\

	In this chapter, we have seen that neutrinos are a vital component of the electroweak theory - a chiral theory that describes both electromagnetic and weak interactions, and which arises from a spontaneous symmetry breaking. We have seen how neutrinos' tiny masses either arose from this symmetry breaking, or that combined with the see-saw mechanism, if they are Majorana fermions. From the analysis of the fermionic interactions in this theory, we see how the 4-Fermi theory can act as an effective field theory for energies much smaller than the masses of the weak gauge bosons. We have also seen how neutrino oscillation occurs, and that a neutrino created in one flavour state may not be in that state at a later time, but once the relic neutrinos slowed down sufficiently they were ``locked" into specific eigenstates. Finally we reviewed some of the latest data surrounding neutrino masses and mixing angles, which we shall use later on in our calculations.
	
	\newpage
	\section{The Instantaneous Decoupling Limit}
	
	Now that we have covered the basics of both neutrino physics and our cosmological model, the next logical step would be to consider how the population of neutrinos in our universe - particularly during the early epochs - should have evolved. This chapter aims to cover the most important leading-order aspects, with smaller corrections being considered in the following chapter.
	
	 \subsection{Decoupling Time}
	 
	 At early times in the universe, when temperatures were much higher, neutrinos reacted constantly in thermal equilibrium with themselves and charged
	 leptons via the weak interaction. The time of decoupling is the time when the mean free travel time of neutrinos would be comparable to the Hubble time (essentially, when a neutrino can travel, on average, through the whole universe before it interacts with another particle).\\
	 We can estimate a ball-park figure for this time using the 4-Fermi theory discussed above. 
	 
	 We wish to obtain the mean free time $\tau$ between neutrino interactions. This is the inverse of the reaction rate $\Gamma=\braket{n\sigma v}$, where $v$ is the velocity ($\approxeq1$ before and during decoupling), $n$ is the number density and $\sigma$ the interaction cross section.\\
	 The number density, since the neutrinos are still relativistic, is given as usual by
	 
	 \begin{equation}
	 	n=\int\frac{d^3\bar{p}}{(2\pi)^3}\frac{1}{e^{|\bar{p}|/T}+1}=\frac{T^3}{2\pi^2}\int_0^\infty dx\frac{x^2}{e^{x}+1},
	 \end{equation}
	 
	 where we have integrated over the angular coordinates in momentum space. By the use of \cite{Gradshteyn}, integrals in the form $I(n)=\int_0^\infty dx\frac{x^n}{e^{x}\pm1}$ are given by
	 
	 \begin{equation}
	 	\int_0^\infty dx\frac{x^n}{e^{x}-1}=\Gamma(n+1)\zeta(n+1) \ \ \ \forall \ \ n>0,\label{BE}
	 \end{equation}
	 
	 \begin{equation}
	 	\int_0^\infty dx\frac{x^n}{e^{x}+1}=(1-2^{-n})\Gamma(n+1)\zeta(n+1) \ \ \ \forall \ \ n>-1, \label{FD}
	 \end{equation}
	 
	 where $\Gamma(n+1)$ is the usual Gamma function of $n+1$, while $\zeta(n+1)$ is the Riemann zeta function\footnote{We will see later that this is actually a special case of the polylogarithm function.}. So, for this case wherein $n=2$, we have $\Gamma(3)=2$ and $\zeta(3)\approxeq1.20206$. Also note that for fermionic species, there is the added factor of (in the case of $n=2$) $\frac{3}{4}$, and so 
	 
	 \begin{equation}
	 	n=\frac{3\zeta(3)}{4\pi^2}T^3.
	 \end{equation}
	 
	 The important thing to take away from this is that $n\approx T^3$. Next, we need to consider the cross section $\sigma$. The cross section is related to the matrix element by $\sigma\propto|\mathcal{M}|^2$. As can be seen from \eqref{matrixel}, the cross section goes as $G_F^2$ times some masses and energies. By dimensional analysis, since $G_F^2$ has units of energy$^{-4}$ and a cross section must have units of energy$^{-2}$ (when $c=\hbar=1$), we need a factor of mass-energy squared. At energies larger than the electron and neutrino masses then, we have 
	 
	 \begin{equation}
	 	\sigma\approx G_F^2E^2.\label{4Fermiapprox}
	 \end{equation}
	 
	 Since all the neutrinos and electrons (and positrons) are still in thermal equilibirum, treating them as ideal, we know that $E \sim T$ (taking also $k_B=1$). Putting this all together, the mean free time goes like 
	 
	 \begin{equation}
	 \tau=\frac{1}{\braket{n\sigma v}}\approx\frac{1}{G_F^2T^5}.
	 \end{equation}
	 
	 Next, the Hubble time is 
	 
	 \begin{equation}
	 t_H=\frac{1}{H}.
	 \end{equation}

 	So, since we are interested in relativistic particles, from \eqref{Friedmann} and \eqref{rhot}, we have using \eqref{FD} that $\rho=\frac{\pi^2}{30}g_*T^4$ and defining $M_P^*=\sqrt{\frac{90}{8\pi^3g_*}}M_P$ with $G=\frac{1}{M_P^2}$ and $M_P$ the Planck mass, we find that
 	
 	\begin{equation}
 		t_H=\frac{M_P^*}{T^2}.
 	\end{equation}

	 Equating $\tau$ and $t_H$, and using that the Fermi constant is (when $\hbar=c=1$) $G_F=1.1664\times10^{-5} \ \mathrm{GeV}^{-2}$ \cite{Griffithsbook} and the reduced Planck mass at the time of decoupling is $M_P^*=\frac{M_P}{1.66\sqrt{g_*}}=2.242 \times 10^{18} \ \mathrm{GeV}$,\footnote{$g_*$ represents the number of effective degrees of freedom at any given temperature. At neutrino decoupling, we get 2 from photons, $\frac{7}{8}\times4$ from electrons and positrons and $\frac{7}{8}\times6$ from neutrinos.} we find the approximate decoupling temperature to be around 1.5MeV, which corresponds to when the universe was just under 1 second old.

	 \subsection{Present Properties}
	 Clearly, before neutrinos decoupled, the universe's temperature would have been a lot larger than the neutrinos' masses (by at least 4 orders of magnitude). Thus, at the time of decoupling, the distribution function for neutrinos is approximately the massless Fermi-Dirac distribution:
	 
	 \begin{equation}
	 f_j(\bar{p},T)=\frac{1}{e^{(E-\mu)/T}+1}\approxeq\frac{1}{e^{|\bar{p}|/T}+1},
	 \end{equation}
	
	where we have used that for each species, the energy is $E=\sqrt{\bar{p}^2+m_j^2}\approxeq|\bar{p}|$ and that the chemical potential is negligible. This chemical potential is discussed further in chapter 5, and is related to the lepton asymmetry in the universe.\\
	Following this distribution function over the history of the universe, we appeal to Liouville's theorem: because the neutrinos are free-streaming, they do not interact with anything and so Boltzmann's equation for it reads
	
	\begin{equation}
	\frac{df_j}{dt}=0 \ \ \ \rightarrow \ \ \ f_j(\bar{p},T) = f_j(\bar{p}_0,T_0),
	\end{equation}
	
	where the 0 subscript indicates the value a quantity has at present day. Thus, we see that today, albeit no longer being relativistic, even massive neutrinos still obey the massless Fermi-Dirac distribution function. Since the relic neutrinos are not in thermal equilibrium with the relic photons, this $T_0$ is an effective temperature, defined by
	
	\begin{equation}
	T_0a_0=T_{dec}a_{dec},\label{effT}
	\end{equation}
	
	with $a$ being the scale factor. This is simply a rephrasing of the $T\propto a(t)^{-1}$ relation we derived using \eqref{rhot} above. Clearly, using \eqref{moma} and \eqref{rhot} and their results, we can verify Liouville's theorem as $\frac{p(t)}{T(t)}$ would be constant over time, and this is the only variable function in the Fermi-Dirac massless distribution function. \\
	At leading order, one can calculate the number density of these relic neutrinos today using an instantaneous decoupling model, wherein all neutrinos instantly stopped interacting at some precise temperature, $T_{dec}$. Of course, in reality, there exists a range of temperatures wherein each individual neutrino interacted for the last time,\footnote{An important example is that around 2 MeV, electrons (and positrons) were still in thermal equilibrium, while muons and taus were not. Thus, electron neutrinos could interact through both charged and neutral currents, while for muon and tau neutrinos, there was only the neutral current at their disposal near the time of decoupling.} and this is explored further in chapter 5.\\
	In this instantaneous decoupling model, it is useful to note the relationship between the temperatures of the C$\nu$B and the CMB. This is done by considering entropy conservation of electrons, photons and positrons. To this end, we begin with the 1st law of thermodynamics (with 0 chemical potential):
	
	\begin{equation}
	dE=TdS-PdV,\label{thermo1}
	\end{equation}
	
	where we have $E$ as energy, $S$ as entropy, $P$ as pressure and $V$ as volume. Using the energy density $\rho$ and defining the entropy density $s=\frac{S}{V}$, we have 
	
	\begin{equation}
	TVds+TsdV=Vd\rho+\rho dV+PdV.
	\end{equation}
	
	Since this relation must be valid for both the entire system (universe) and a small part, when we consider some region of constant volume such that $dV=0$, we have the relation
	
	\begin{equation}
	Tds=d\rho,
	\end{equation}
	
	and plugging this back in, we have in general $(Ts-\rho-P)dV=0$ or
	
	\begin{equation}
	s=\frac{\rho+P}{T}.
	\end{equation}
	
	Our interests lie in relativistic matter, as both our neutrinos and photons are relativistic during the timeframes of their decouplings. Thus, as was shown in \eqref{pressurerho}, we have $P=\frac{\rho}{3}$ and so for each type of particle $i$, we have from \eqref{rhot}
	
	\begin{equation}
	s_i=\frac{4}{3T}\rho_i=\frac{4}{3T}\frac{g_iT^4}{2\pi^2}\int_0^\infty dx\frac{x^3}{e^{x}\pm1}.
	\end{equation}
	
	We can define 
	
	\begin{equation}
	g_*=\sum_{boson} g_i +\frac{7}{8}\sum_{fermion} g_i,
	\end{equation}
	
	such that, by use of \eqref{FD} and \eqref{BE}
	
	\begin{equation}
	s=\sum s_i = \frac{4}{3T}\frac{g_*T^4}{2\pi^2}\frac{6\pi^4}{90}=\frac{4\pi^2g_*T^3}{90}.\label{entropyT}
	\end{equation}
	
	Next, we again use \eqref{thermo1} and this time, we shall consider a region of space with a volume of $V=a^3$. Differentiating with respect to time, this results in
	
	\begin{equation}
	T\frac{dS}{dt}=\frac{dE}{dt}+P\frac{dV}{dt} \ \ \ \rightarrow \ \ \ T\frac{d(sa^3)}{dt}=(\rho+P)\frac{dV}{dt}+V\frac{d\rho}{dt}.
	\end{equation}
	
	Then, using \eqref{fluideqn} and $\frac{dV}{dt}=3a^2\dot{a}$, we have
	
	\begin{equation}
	T\frac{d(sa^3)}{dt}=3a^2\dot{a}(\rho+P)+a^3\left(-3\frac{\dot{a}}{a}(\rho+P)\right)=0,
	\end{equation}
	
	and thus $sa^3$ is a conserved quantity, and is constant. Putting this together with \eqref{entropyT}, we have that
	
	\begin{equation}
	g_*a^3T^3=C,\label{entropycons}
	\end{equation}
	
	where $C$ is a constant. At the time of neutrino decoupling, electrons and positrons were still relativistic, and thus we have $g_*=2+\frac{7}{8}\times4=\frac{11}{2}$, accounting for the 2 polarisations of the bosonic photons and the 2 spin states of the fermionic electrons and positrons. Very soon after, (at an energy of around 0.5 MeV), electrons and positrons annihilated away, leaving a very small remainder of electrons and injecting the remaining energy into the photons. After this, $g_*=2$, and so using \eqref{entropycons}, we have 
	
	\begin{equation}
	\frac{T_{\gamma,0}}{T_{\gamma,dec}}=\frac{a_{dec}}{a_0}\left(\frac{g_*(T_{dec})}{g_*(T_0)}\right)^{1/3}=\frac{a_{dec}}{a_0}\left(\frac{11}{4}\right)^{1/3}.
	\end{equation}
	
	Then, since at neutrino decoupling, $T_\nu=T_\gamma$ and using \eqref{effT}, we have that
	
	\begin{equation}
	\frac{T_{\gamma,0}}{T_{\nu,0}}=\left(\frac{11}{4}\right)^{1/3} \approxeq 1.40102.\label{Trelation}
	\end{equation}
	
	So, to leading order, and based on the latest Planck satellite data wherein on average $T_{\gamma,0}\approxeq2.7255$ K \cite{Fixsen:2009ug}, we have at present day $T_{\nu,0}\approxeq1.945$ K.\\
	The present day number density is given in terms of the distribution function by
	
	\begin{equation}
	n_0=g\int\frac{d^3\bar{p}}{(2\pi)^3}f(\bar{p},T_0),
	\end{equation}
	
	where $g$ is the number of degrees of freedom. So, per degree of freedom, we have using \eqref{FD} and the present value of $T_\nu$
	
	\begin{equation}
	n_0=\int\frac{d^3\bar{p}}{(2\pi)^3}\frac{1}{e^{|\bar{p}|/T_{\nu,0}}+1}=\frac{3\zeta(3)}{4\pi^2}T_{\nu,0}^3\approxeq 56.01 \ \mathrm{cm}^{-3}.\label{n0}
	\end{equation}
	
	$n_0$ will be useful in calculating the expected capture rate of neutrinos. Another quantity of interest is the average momentum, which after again using \eqref{FD} is given by
	
	\begin{equation}
	\braket{p_0}=\frac{1}{n_0}\int\frac{d^3\bar{p}}{(2\pi)^3}|\bar{p}|f(\bar{p},T_0)\approxeq 0.53 \ \mathrm{meV}.\label{lowenergy}
	\end{equation}
	
	The reason this is interesting is because, when compared to \eqref{massdiff}, we can see that in any hierarchy, at least 2 mass species of neutrinos will be non-relativistic today.
	
	\subsection{Helicity States}
	
	The helicity of a particle is defined as the projection of a particle's spin onto its momentum's direction, that is $h=\frac{\bar{s}\cdot\bar{p}}{|\bar{p}|}$. For massless particles, helicity and chirality coincide, and what we describe as ``left-handed" or ``right-handed" can refer to both left-(right-)chiral and left-(right-) helical states. However, when mass cannot be ignored, chirality is no longer conserved, while helicity is \cite{Schwartzbook}. Thus, while initially left- or right-chiral states can become right- or left-chiral, initially left- or right-helical states remain so.\\
	In the early universe, when all neutrinos were extremely relativistic, their masses could be safely neglected, and chirality and helicity coincide. During this time, before decoupling, the neutrinos were in thermal equilibrium and interacting weakly with themselves and charged leptons. As we know, electroweak theory is chiral. For the remainder of this section, we must consider two possibilities: that neutrinos are Dirac particles, or that they are Majorana particles.\\
	In the case that they are Dirac particles, only left-chiral neutrinos $\nu_L$ and right-chiral antineutrinos $\bar{\nu}_R$ were in this equilibrium, with right-chiral neutrinos $\nu_R$ and left-chiral antineutrinos $\bar{\nu}_L$ being ``sterile" - that is, not interacting with anything other than gravitational forces. Since these sterile neutrinos could not be produced via the weak interaction, barring some miraculous initial quantity, we can safely assume that there were never any sterile neutrinos. Now, because during this epoch neutrinos were essentially massless, chirality and helicity coincided, and so only left-helical neutrinos $\nu_L$ and right-helical antineutrinos $\bar{\nu}_R$\footnote{Though we have used the same notation to denote the chirality and helicity of neutrinos, when the difference is important it will be clarified.} were created. So today, since helicity is conserved, we have for each massive species that 
	
	\begin{equation}
			f_{\nu_L}=f_{\bar{\nu}_R}=f(\bar{p}_0, T_0),
	\end{equation}
	\begin{equation}
		f_{\nu_R}=f_{\bar{\nu}_L}=0.
	\end{equation}
	
	Integrating, we of course obtain that
	
	\begin{equation}
		n_{\nu_L}=n_{\bar{\nu}_R}=n_0,
	\end{equation}
	\begin{equation}
	n_{\nu_R}=n_{\bar{\nu}_L}=0,
	\end{equation}

	and owing to the fact that there are (in the standard model) 3 massive species, we have a total number density of the C$\nu$B 
	
	\begin{equation}
		n_T=3\times n_{\nu_L}+3\times n_{\bar{\nu}_R}=6n_0 \approxeq 336.06 \  \mathrm{cm}^{-3},
	\end{equation}
	
	with half being neutrinos and half anti-neutrinos. Note that in the above, as already stated, we have neglected the possibility of an appreciable lepton asymmetry (that is, we have taken the approximation that the chemical potential is essentially 0), but will discuss this with other ideas that go beyond the standard model next chapter.\\
	
	If, however, neutrinos are Majorana particles, then their anti-particles are themselves, and lepton number loses meaning. In this case, sterile neutrinos are slightly different. The ``active" neutrinos (those that interact weakly) $\nu_L$ and $\nu_R$ have extremely small masses owing to the extremely large masses of the sterile neutrinos $N_L$ and $N_R$ via the see-saw mechanism, discussed prior. In this case then, active neutrinos were created via the weak interaction, while sterile neutrinos were not. Once again, unless there was some initial quantity of sterile neutrinos, they could not be produced. In this case, even in the case where there was some initial population of sterile neutrinos, their extremely large masses would mean that they would have decayed long ago, and there would be none remaining today. So, the distribution functions today would be
	
	\begin{equation}
		f_{\nu_L}=f_{\nu_R}=f(\bar{p}_0, T_0),
	\end{equation}
	\begin{equation}
		f_{N_R}=f_{N_L}=0.
	\end{equation}
	
	Once again upon integrating, we of course obtain that
	
	\begin{equation}
		n_{\nu_L}=n_{\nu_R}=n_0,
	\end{equation}
	\begin{equation}
		n_{N_R}=n_{N_L}=0,
	\end{equation}
	
	and owing to the fact that there are (in the standard model) 3 massive active species, we have a total number density of the C$\nu$B as 
	
	\begin{equation}
		n_T=3\times n_{\nu_L}+3\times n_{\nu_R}=6n_0 \approxeq 336.06 \  \mathrm{cm}^{-3},
	\end{equation}
	
	the exact same as when we considered Dirac neutrinos. However, the main difference is that in this case, there is no distinction between neutrinos and anti-neutrinos. This fact will make the most significant difference to the capture rate, as we will see below in chapter 7.\\
	
	The goal of this chapter was to elucidate the general qualities that we expect relic neutrinos to have: an effective temperature of almost 2 K, masses in the range of tens of meV with momenta in the range of tenths of meV, and helicity components based on whether they are Dirac or Majorana particles. Besides for the small corrections discussed next chapter, any significant deviation from these predictions would necessitate a reformation of either our cosmological model, our particle physics model, or both.

	\newpage
	\section{Higher Order Effects}
	
	As we showed in chapter 4, the leading order approximation for the C$\nu$B number density on earth today is roughly around $56\ \mathrm{cm}^{-3}$ per degree of freedom, with an average momentum of the order $10^{-4}$ eV. For the purposes of observing the effects of higher order corrections to this number density, we will focus on the (formerly known as the Princeton Tritium Observatory for Light-Early Universe Massive-neutrino Yield) PTOLEMY (Pontecorvo Tritium
	Observatory for Light, Early-Universe, Massive-Neutrino Yield) project, which utilises the method of neutrino capture by beta-decaying nuclei described next chapter.
	
	Keeping this experiment in mind, we begin by analysing the standard physics affecting the current neutrino number density: non-instantaneous decoupling, gravitational clustering and annual modulation, and end off by considering the effects of some proposed new physics. By calculating the expected capture rate to a very high precision according to the standard model of particle physics and cosmology, any experimental observation that deviates from these precise quantities may provide a signature of new physics.
	
	\subsection{Spectral Distortion During Decoupling}
	
	The first correction to these values comes from the fact that neutrinos did not all instantly decouple, and some interactions continued to occur during the decoupling period between themselves and also with electrons and positrons\footnote{Note that long before decoupling, neutrinos interacted with all charged leptons, but during decoupling we need only consider the interactions between the neutrinos and electrons and positrons. This means that the distortions to $n_0$ will be most significant for the electron-neutrino and, via the PMNS matrix, the 1st species of massive neutrino. This is of course because muons and taus would already have annihilated.}. This is because not all neutrinos had the same energy; their momenta were distributed according to the Fermi-Dirac distribution. Along with this, when electrons and positrons annihiliated (very soon after the time of instantaneous decoupling), not all neutrinos would have decoupled, and so some of the energy from this annihilation would have entered into the neutrino component, changing $\frac{T_\nu}{T_\gamma}$, the ratio of the C$\nu$B and CMB temperatures.\\
	
	The injection of energy into the neutrino component will make the ratio smaller, as the neutrinos' effective temperature will be slightly higher and the photons' lower. The latest calculation finds \cite{Akita:2020szl}
	
	\begin{equation}
		\frac{T_{\gamma,0}}{\bar{T}_{\nu,0}}\approxeq 1.39797,
	\end{equation}

	a change of around 0.218\%. Since the F-D distribution is itself now an approximation, this $\bar{T}_{\nu,0}$ is the \itshape effective \normalfont temperature of the C$\nu$B, defined as the temperature found in the distribution function (that is, it is the temperature that is inversely proportionate to the scale factor $a$). This distribution function is distorted from the Fermi-Dirac case above, and is written as
	
	\begin{equation}
		f_{\nu_i}^d(\bar{p},t)=\frac{1}{e^{|\bar{p}|/\bar{T}_{\nu}(t)}+1}\bigl(1+\delta f^d_{\nu_i}(\bar{p}, t) \bigl).
	\end{equation}

	The subscript $i$ refers to each species in the mass basis, and the $\delta f^d_{\nu_i}(\bar{p}, t)$ represents the distortion to the F-D distribution. The 3 massive species need to be seperated as they are each affected differently by the interactions occurring around the time of decoupling. The main reason is that electron neutrinos can interact with the electrons and positrons via both neutral and charged currents, while the muon and tau neutrinos can only do so via the neutral current. Going on from that, experimental PMNS matrix element values show that the electron neutrino is most closely related to the 1st massive species of neutrino, $\nu_1$, with the other massive species being affected to a lesser degree.\\
	
	Another quantity affected by these interactions is what is known as the effective number of neutrino species, $N_{\mathrm{eff}}$. In the standard model, there are exactly 3 species of neutrinos. However, the value that can be observed is in the relation
	
	\begin{equation}
		\rho_r=\rho_\gamma\left[1+\frac{7}{8}\left(\frac{4}{11}\right)^{4/3}N_{\mathrm{eff}}\right],
	\end{equation}
	
	wherein $\rho_r$ is the energy density of radiation and $\rho_\gamma$ that of photons. The factor of $\frac{7}{8}$ comes from the fact that neutrinos are fermions while photons are bosons, and as such, when integrating over the distribution functions to calculate energy density, we gain this factor, as seen in \eqref{FD}. The factor of $\left(\frac{4}{11}\right)^{4/3}$ is owing to the fact that the energy density is related to the temperature by $\rho\propto T^4$, and from \eqref{Trelation}, we see the $\frac{4}{11}$ factor. However, we already know that the neutrino effective temperature is not related to that of photons by exactly a factor of $\frac{4}{11}$. To account for this, and for the other distortions to the energy density of each flavour, we can write $N_{\mathrm{eff}}$ as \cite{Akita:2020szl}
	
	\begin{equation}
		N_{\mathrm{eff}}=\left(\frac{1.40102}{1.39797}\right)^4 \left(3+\frac{\delta\rho_{\nu_e}}{\rho_{\nu_0}}+\frac{\delta\rho_{\nu_\mu}}{\rho_{\nu_0}}+\frac{\delta\rho_{\nu_\tau}}{\rho_{\nu_0}}\right),\label{Neff}
	\end{equation}

	where $\delta\rho_{\nu_\alpha}$ is the small deviation to the energy density of each flavour from the instantaneous decoupling energy density, $\rho_{\nu_0}$. Note that this calculation assumes only the 3 standard flavours of neutrinos.\\
	
	Before accounting for the flavour- or mass-specific corrections to the number and energy densities, owing to the change in the effective temperature, the average number density $\bar{n}_0$ is now given numerically by
	
	\begin{equation}
		\bar{n}_0=\frac{3\zeta(3)}{4\pi^2}\bar{T}_{\nu,0}^3\approxeq 56.376\  \mathrm{cm}^{-3},\label{newnumdens}
	\end{equation}

	approximately 0.65\% larger than in the instantaneous decoupling limit. The number density for each mass species is then given by 
	
	\begin{equation}
		n^d_{\nu_i}=\bar{n}_0(1+\delta\bar{n}^d_{\nu_i}) = n_0(1+\delta n^d_{\nu_i}).\label{noninstant}
	\end{equation}
	
	Taken from \cite{Akita:2020szl}, the values for the distortions to the energy densities in the flavour basis and to the number densities in the mass basis are shown in tables \ref{energy densities} and \ref{number densities}. This then allows us to calculate $N_{\mathrm{eff}}\approxeq3.044$. This value falls within the current experimental bounds. To calculate these density distortions, $N_{\mathrm{eff}}$ and the effective temperature, finite temperature quantum field theory calculations were performed, including loop corrections as well as accounting for neutrino oscillations, and then completed numerically in \cite{Akita:2020szl}.

		\begin{table}
		\centering
		\begin{tabular}{|c|c|c|}
			\hline
			$\delta\bar{\rho}_{\nu_e}$ & $\delta\bar{\rho}_{\nu_\mu}$ & $\delta\bar{\rho}_{\nu_\tau}$ \\ 
			\hline
			$7.12\times10^{-3}$ & $5.11\times10^{-3}$ & $5.23\times10^{-3}$\\
			\hline
		\end{tabular}
	\caption{Distortions to the energy densities of each neutrino flavour owing to non-instantaneous decoupling, with $\delta\bar{\rho}_{\nu_\alpha}=\delta\rho_{\nu_\alpha}/\rho_{\nu_0}$. These values account for neutrino oscillations and neutrino-neutrino and neutrino-electron/positron interactions in finite temperature field theory up to order $e^3$, taken from \cite{Akita:2020szl}. }
	\label{energy densities}
	
	\end{table}

	\begin{table}
		\centering
		\begin{tabular}{|c|c|c|c|c|c|}
			\hline
			$\delta\bar{n}^d_{\nu_1}$ & $\delta\bar{n}^d_{\nu_2}$ & $\delta\bar{n}^d_{\nu_3}$ & $\delta n^d_{\nu_1}$ & $\delta n^d_{\nu_2}$ & $\delta n^d_{\nu_3}$ \\ 
			\hline
			$4.68\times10^{-3}$ & $3.50\times10^{-3}$ & $2.48\times10^{-3}$ & 0.0113 & 0.0101 & $9.10\times10^{-3}$\\
			\hline
		\end{tabular}
		\caption{Distortions to the number densities of each neutrino mass species owing to non-instantaneous decoupling. These values account for neutrino-neutrino and neutrino-electron/positron interactions in finite temperature field theory up to order $e^3$, taken from \cite{Akita:2020szl}.}
		\label{number densities}
		
	\end{table}

	\subsection{Gravitational Clustering}
	
	Though we have never directly observed neutrinos interact gravitationally, being massive particles, there is no reason for them not to do so. A direct experiment, analogous to the Pound-Rebka experiment \cite{Pound1959} done for light would be the most direct evidence possible. In the mean time, one effect of this gravitational interaction would be the clustering of non-relativistic neutrinos caught in the potential near earth, while another would be the effects of lensing, discussed in chapter 8, which would also affect relativistic neutrinos.\\
	
	For high energy astrophysical neutrinos from sources such as supernovae, gravitational clustering is not a factor that needs to be considered as the particles' velocities are too high to be captured by the gravitational field near earth. For some C$\nu$B neutrinos, this is no longer the case. If there exists a massless species of neutrino\footnote{More accurately, any species with mass much less than the average momentum today, of order $10^{-4}$eV.}, then no matter how low its momentum, it will not cluster. However, we know that at least 2 species of neutrinos are massive and have mass much greater than the average momentum today, seen in \eqref{lowenergy}. Thus, it is natural to consider how the Milky Way, along with other nearby objects, might affect the relic neutrino number density on earth today. 
	
	The latest work in \cite{Mertsch:2019qjv} shows that besides for the matter (both dark and baryonic) in the Milky Way, the second most important effect comes from the gravitational potential induced by the Virgo cluster, with the nearby Andromeda galaxy playing an almost negligible role. In fact, the Virgo cluster's effect is quite complex, as it may even divert some neutrinos away from us. In general however, it greatly increases the clustering effect for smaller masses (less than 100\ meV) and has little effect on larger masses, although for very large masses (those actually disallowed by the constraints in \eqref{massupperlim}) it decreases the clustering effect on earth, attracting the neutrinos more toward its centre. Though calculated numerically, 2 different models were used to describe the dark matter distribution of the Milky Way and Virgo cluster (the NFW and Einasto models) and the results were consistent with one another. The numerical calculation involved an N-body simulation, utilising a back-tracking method, which uses final rather than initial conditions to calculate the trajectories of neutrinos. This method helps reduce computation time such that more accurate results could be obtained. Another interesting feature of the numerical calculation was that it showed that nearly all the clustering takes place at very small redshift: this makes sense, as the neutrinos would have ``slowed down", their momentum decreasing via \eqref{moma} as the universe expanded and they free-streamed. Thus, only at later times (small redshift) were they non-relativistic and able to cluster.\\
	
	Accounting for clustering in the number density, we can write our number density as 
	
	\begin{equation}
		n_{\nu_i}=n_{\nu_i}^d(1+\delta n^c_{\nu_i}),
	\end{equation}

where we have retained the effects of non-instantaneous decoupling in $n_{\nu_i}^d$. However, from \eqref{noninstant}, we see that we can write this as 

\begin{equation}
	n_{\nu_i}=n_0(1+\delta n^d_{\nu_i})(1+\delta n^c_{\nu_i}) \approxeq n_0(1+\delta n^d_{\nu_i}+\delta n^c_{\nu_i}),\label{numdens}
\end{equation}

where we have taken our answer to linear order, as the $\delta n^{d,c}$ values are small compared to 1.
	
	\begin{table}
		\centering
		\begin{tabular}{|c|c|}
			\hline
			$m_\nu$ (meV) & $\delta n^c$\\
			\hline
			10 & $5.3\times 10^{-3}$ \\ 
			20 &  0.02\\
			30 &  0.04\\ 
			40 &  0.07\\
			50 &  0.12\\ 
			\hline
		\end{tabular}
		\caption{Distortions to the number densities of a neutrino species with various masses owing to gravitational clustering, taken from \cite{Mertsch:2019qjv}.}
		\label{clustering table}
		
	\end{table}

	Numerically, it is found that the neutrino's mass plays a huge role in the clustering effect. Listed in table \ref{clustering table}, we see that the effect on the present number density changes by almost 2 orders while the mass only changes by a factor of 5. More importantly, however, we see that for masses of less than 10\ meV that the distortive effect of the interactions during decoupling are more prominent than those of gravitational clustering. Conversely, for masses of 50\ meV or greater, the effect of the distortion owing to non-instantaneous decoupling would be masked completely, as the error margin of clustering and the distortive effects are indistinguishable, and for masses around 10 meV the two effects are comparable. Thus, for us to probe the era of neutrino decoupling precisely, we would need to work with the lightest neutrino species, and it would need to have a mass much less than 10\ meV.\\
	
	The final points we may need to consider related to gravitational clustering is the effect it may have on the helicity of neutrinos and the isotropy of their incoming angles. Recall that the helicity is given by the projection of the spin of the particle onto its momentum, and while the action of clustering will not flip any particle's spin, it can change the direction of its momentum, and thus change its helicity. This helicity flipping can be anywhere between slight and complete, where ``complete" helicity flipping would mean in the Dirac case that initially unpopulated helicities and those that interact via the weak force would now all have the same number density: $n_{\nu_L}=n_{\nu_R}=n_{\bar{\nu}_L}=n_{\bar{\nu}_R}=\frac{n_{\nu_i}}{2}$, for each mass species $i$. As we will see in chapter 7, at leading order this does not make any difference, as the capture rate is summed over the helicities. However, at the next order, the capture rate will be altered slightly, with complete helicity flipping of all neutrinos changing the capture rate maximally at the order of $10^{-3}$yr$^{-1}$, as will be shown. This effect has not been calculated exactly yet, but since only clustered neutrinos' helicities would be flipped, as well as not all clustered neutrinos' helicities being flipped, this effect would be more on the order of $10^{-4}$yr$^{-1}$.
	
	The isotropy of the detected C$\nu$B at earth may also be changed. Already the gravitational wells of the Milky Way and other large bodies have broken the perfect homogeneity of the C$\nu$B, and the effect of clustering may cause more relic neutrinos to be incident from the direction of the centre of the Milky Way than from outside it. Once again, this effect will only change the capture rate by a tiny amount
	. Thus, for the numerical calculations, we will neglect the effects of helicity flipping and anisotropy, and leave the exact calculations of their effects on the distribution function and number density of cosmic neutrinos to future works.

	\subsection{Annual Modulation}
	
	As we know, the universe is expanding at an ever-increasing rate, and owing to homogeneity and isotropy, on large scales every point is a ``centre" of the universe. Thus, it is very difficult to define a ``static" object, against which the peculiar velocities of objects can be measured. This is usually chosen to be the CMB, and it is often assumed that the C$\nu$B's rest frame is the same as that of the CMB, as there is no reason for the C$\nu$B to have some preferred direction. From this, from observations relative to the CMB, we can see how our solar system moves relative to the C$\nu$B. The current observation is that the sun is moving at $v_{sun}\approx369\ \mathrm{km.s}^{-1}$ relative to the CMB \cite{Aghanim:2013suk}, and the earth moves with velocity $v_{earth}\approx29.79\ \mathrm{km.s}^{-1}$ around the sun \cite{McCabe:2013kea}.\\
	
	Annual modulation is a natural, expected feature of the presence of a C$\nu$B, and can also aid in measuring the exact number of clustered versus unclustered relic neutrinos. The gravitational potential of the sun causes focusing of neutrinos at earth when the earth is ``downwind" of the sun (that is, behind the sun relative to the incoming neutrinos). For clustered and unclustered neutrinos, the earth is downwind of the sun at opposite times of the year. Therefore, if there are equally as many incoming neutrinos that have gravitationally clustered in the Milky Way as those that have not, the annual modulation would not be visible.\\
	
	The latest numerical calculations in \cite{Safdi:2014rza} show that for larger masses, the effect of annual modulation is far more visible. The smallest mass considered was for $m_\nu=150$meV, a mass now already excluded by Planck constraints \cite{Aghanim:2018eyx}. However, even for this (relatively) large mass, the effect of annual modulation was only a change of $1.6\times10^{-3}$ to the number density for unclustered neutrinos and between $10^{-3}$ and $10^{-2}$ for those coming from inside the Milky Way, depending on the model used. The effect of lessening the mass was to lessen the annual modulation, and so for even smaller masses that fit the latest data, the effect of annual modulation would be even smaller than the $10^{-3}$ change expected for 150 meV neutrinos. For example, for a 50 meV neutrino species, the annual modulation would be approximately $10^{-4}$, as the neutrino would both cluster less effectively and deflect less as they are moving faster and thus spend less time near the sun's gravitational potential. Also, assuming neutrinos would come both from within and outside the Milky Way, the two annual modulations would cancel out to a large degree, leaving an even smaller and less regular signature. Considering this, albeit falling under standard physics, in the calculation of the capture rate in chapter 7 below, we can safely neglect the effects of annual modulation.

	\subsection{Beyond Standard Physics}
	
	In this section, we consider various theories that, as a result, would affect the number densities of the different mass species of neutrinos. These include the presence of sterile neutrinos, a lepton asymmetry, decay patterns for neutrinos and other non-standard thermal effects.\\
	
	In the standard model, there are only 3 flavours of neutrinos (and thus, 3 massive species) known as ``active" neutrinos. However, there may be other flavours that can be oscillated into, but do not interact with the weak force. These ``sterile" neutrinos are not the same as those discussed above, as these neutrinos are chiral in the same way as the active species, and do not have extremely large masses. There may be a fourth, fifth, sixth, etc. flavour, but for simplicity we will consider the case wherein there is one extra flavour, denoted $\nu_s$, and as a result, one extra mass species $\nu_4$.
	
	In this scenario, the PMNS mixing matrix then becomes a 4 $\times$ 4 matrix, with the $U_{\alpha4}$ elements describing the flavour composition of the $\nu_4$ mass eigenstate. With a mass in the eV range, some experimental data can be interpreted as the presence of a fourth, sterile neutrino. Some excesses in neutrino beam experiments can be explained by a $\nu_4$ with $\Delta m_{41}^2 \approx (0.1 - 10) \mathrm{eV}^2$ and $|U_{e4}|^2\approx|U_{\mu4}|^2\approx 0.03$ \cite{Aguilar-Arevalo:2013pmq}. These PMNS elements may seem miniscule, but $|U_{e3}|^2\approx 0.02$ is even smaller, and so the effect would not be negligible. 
	
	The inclusion of a fourth, sterile neutrino can actually alleviate tension in many areas. As already mentioned, the inclusion of a fourth species can help explain the excess of $\nu_e$ and $\bar{\nu}_e$ flavour (anti-)neutrinos in beams of $\nu_\mu$ and $\bar{\nu}_\mu$ flavour (anti-)neutrinos, as more muon-neutrinos would oscillate into electron neutrinos via the sterile neutrino. It can also explain the flux deficit compared to theory in both reactor \cite{Mention:2011rk} and solar neutrino experiments \cite{Hampel:1997fc}, as some of the neutrinos would be sterile. 
	
	Cosmologically, including a sterile neutrino in models can help explain why the observed $N_{\mathrm{eff}}$ can be larger than 3, why early-universe and local measurements of the Hubble constant $H_0$ are in unresolvable tension, and why the BICEP2 tensor perturbations measurement is inconsistent with Planck's CMB data \cite{Zhang:2014dxk}. This is owing to the fact that if we include a sterile species, gravitational clumping in the early universe would occur more slowly (as more energy density would be relativistic than in current models), suppressing the growth of structure and bringing the Planck CMB scalar-to-tensor ratio constraints more in line with data from other sources. Similarly, inclusion of sterile neutrinos would increase the early-time Hubble expansion rate, as recall $H\propto\sqrt{\rho}$, so a larger energy density would imply a larger Hubble constant.\\
	
	Let us consider how this possibility would be realised quantitatively. If the entire excess in the observed $N_{\mathrm{eff}}$ is owing to one extra species, we would have 
	
	\begin{equation}
		n_{\nu_s} = \Delta N_{\mathrm{eff}}n_0(1+\delta n^c_{\nu_s}), \label{sterilen}
	\end{equation}

	where $\Delta N_{\mathrm{eff}}$ is the difference between the observed value and the theoretical value, calculated in \eqref{Neff}. The effect that these sterile neutrinos would have on the expected capture rates and energy spectra will be discussed in chapter 7.\\
	
	The final note on sterile neutrinos is that they could be much more massive, such as the keV range: this is one of the candidates for dark matter. Recently, a 3.5 keV X-ray line has been observed \cite{Boyarsky:2014jta}, and if this is explained by a 7 keV sterile neutrino decaying into 2 photons, the PMNS elements would be of the order $|U_{\alpha4}|^2\approx 10^{-11}$. This may seem minute, but owing to their extremely large mass, the enhancement owing to clustering would be very large. If they account for all of dark matter, then their enhancement factor would be given by $\frac{n_{\mathrm{local}}}{n_0} \approxeq 8 \times 10^2$, where $n_{\mathrm{local}}$ would be $\frac{\rho_{DM}}{m_{\nu_4}}\approxeq \frac{0.3 \mathrm{GeV cm}^{-3}}{7 \mathrm{keV}}$. As can be seen, this clustering enhancement is not nearly enough to offset the very tiny proportion that the sterile neutrino would compose of the detected electron neutrino, and the capture rate (as we will see) would be proportional to $|U_{\alpha4}|^2 \frac{n_{\mathrm{local}}}{n_0} \approx 8\times 10^{-9}$ times that of active neutrinos - essentially impossible to observe.\\
	
	It has been observed that the baryon asymmetry in the universe is extremely small: that is, the number density of baryons and anti-baryons is very similar, with the difference compared to the photon density being \cite{Cooke:2013cba}
	
	\begin{equation}
		\eta_b=\frac{n_b-\bar{n}_b}{n_\gamma} \approx 10^{-10}.
	\end{equation}
	
	At extremely high energies, it is predicted that baryon $B$ and lepton $L$ number are not individually conserved, and that rather the difference $B-L$ is. This idea leads to sphaleron processes, wherein baryons can become anti-leptons and vice versa \cite{Gorbunov:2011zz}. If this is the case, then the lepton asymmetry is as tiny as the baryon asymmetry, and can be safely neglected.
	
	If, as is the case in some models, some sizable lepton asymmetry is created, then this will affect the number density today. If there is an asymmetry, there is an associated chemical potential in each species. This means the distribution function would be altered to be as such:
	
	\begin{equation}
		f_{\nu}(p,T_\nu) = \frac{1}{e^{(p-\mu)/T_\nu}+1},
	\end{equation}
	
	and where the distribution function for the anti-neutrinos would be the same, but with the opposite sign of the chemical potential $\mu$. In the case where there is a difference between the number densities, the chemical potential would exist in a way such that it will make the species with more particles (either neutrinos or anti-neutrinos) change into the other, and so its sign is unknown as well. Calculating the number density, we have 
	
	\begin{equation}
		n_{v}= \int \frac{d^3p}{(2\pi)^3}\frac{1}{e^{(p-\mu)/T_\nu}+1} = \frac{T_\nu}{2\pi^2}\int_0^\infty\frac{dx(T_\nu x+\mu)^2}{e^x+1}.
	\end{equation}
		
	This integral was arrived at using $x=\frac{p-\mu}{T_\nu}$, and expanding the brackets gives 3 complete Fermi-Dirac integrals, the solution to each is a polylogarithm \cite{Gradshteyn2}: 
	
	\begin{equation}
		n_{v} = \frac{T_\nu^3}{2\pi^2}\left(-\mathrm{Li}_3(-1)\Gamma(3) -2\xi\mathrm{Li}_2(-1)\Gamma(2) -\xi^2\mathrm{Li}_1(-1)\Gamma(1)\right),
	\end{equation}
    \begin{equation}
	n_{v} = T_\nu^3\left(\frac{3\zeta(3)}{4\pi^2} +\frac{\xi}{12} +\frac{\ln(2)}{2\pi^2}\xi^2\right). \label{asym}
	\end{equation}

	In the above, $\xi=\frac{\mu}{T_\nu}$. The first term is the same as above, and we recover \eqref{n0} in the limit $\xi\rightarrow0$. \\
	
	Using the above calculation, the consequences for Dirac and Majorana particles are different. In the Dirac case, this asymmetry manifests as $n_{\nu_L} \neq n_{\bar{\nu}_R}$, while in the Majorana case it is $n_{\nu_L} \neq n_{\nu_R}$. This difference is pertinent because in the Dirac case, the $\xi$ values for neutrinos and anti-neutrinos have the same value but opposite sign, whereas for the Majorana case the $\xi$ value is the same for both left- and right-handed neutrinos\footnote{We must also keep in mind that helicity flipping owing to gravitational clustering will alter the results in the Majorana case.}. The effect that this asymmetry would have on the capture rate for the two cases will be explored in the respective chapter.\\
	
	Next, we consider the possibility that neutrinos (or perhaps, some neutrinos) decay: either into photons (radiative decay), lighter species of neutrinos (weak decay), or other, exotic particles (invisible decay).
	
	The bounds on radiative decays are very strict coming from solar neutrino and photon flux observations. The decay rate bounds are usually quoted as $\frac{\tau_\nu}{m_\nu}$, with $\tau_\nu$ the neutrino's lifetime and of course $m_\nu$ its mass. For radiative decays, the bound is \cite{Raffelt:1985rj}
	
	\begin{equation}
		\frac{\tau_\nu}{m_\nu}\bigg|_{\nu\rightarrow\gamma} \geq 7 \times 10^9 \  \mathrm{s\cdot eV}^{-1}.
	\end{equation}

	For masses around 50 meV then, this gives a lifetime of $3.5 \times 10^8$ seconds or around 11 years. This is more than enough time for a relic neutrino to decay since the universe was 1 second old.\\
	
	In general, the model-independent contraint on neutrino decays is \cite{Hirata:1988ad}
	
	\begin{equation}
		\frac{\tau_\nu}{m_\nu} \geq  10^5 \  \mathrm{s\cdot eV}^{-1}.
	\end{equation}

	If neutrinos did decay into photons, or into some exotic matter that we can't detect, then the number of C$\nu$B neutrinos remaining to be detected would be 
	
	\begin{equation}
		N=N_0e^{-\lambda_\nu},
	\end{equation}

	where $N_0$ is the number after decoupling. This $\lambda_\nu$ is given by \cite{Long:2014zva}
	
	\begin{equation}
		\lambda_\nu = \int \frac{dt}{\tau_\nu} = \int_{0}^{z_{dec}} \frac{dz}{(1+z)H(z)\gamma(z)\tau_\nu^0},
	\end{equation}

	where $\tau_\nu^0$ is the proper lifetime of neutrinos, and $H(z)$ is the Hubble constant at some redshift $z$. This integral may be difficult to perform, but if we make the approximation that most neutrinos have decayed at later times (as $z$ goes to 0), we can insert a Dirac delta function and obtain
	
	\begin{equation}
		\lambda_\nu\approx \frac{1}{H_0\tau_\nu^0} = \frac{t_0}{\tau_\nu^0}.
	\end{equation}

	So, the approximate number of relic neutrinos left today would be given by 
	
	\begin{equation}
		N=N_0e^{-t_0/\tau_\nu^0}.
	\end{equation}

	If $\tau_\nu^0$ is much smaller than the age of the universe, then we would not be able to detect the C$\nu$B at all - a very lamentable thought. If, as we hope, relic neutrinos are observed, this will place an extremely strong constraint on neutrino decays, as their lifetime would have to be at least larger than the age of the universe.\\
	
	If neutrinos all decayed into the lightest species, this could either amplify or diminish the detection rate. If neutrinos are normally ordered, then all neutrinos would decay into $\nu_1$ neutrinos, and instead of summing over all species, we would simply have thrice the amount of $\nu_1$ neutrinos. Similarly for the inverted hierarchy, except with $\nu_3$ neutrinos. As we will see in the capture rate chapter, the ordering will result in either doubling the capture rate, or diminishing it almost completely, if neutrinos do indeed decay into other neutrinos.\\
	
	The final consideration made in this subsection is what the consequences of a non-standard thermal history would be. Recall from \eqref{n0} that the number density is dependent on the effective neutrino temperature cubed - a very sensitive quantity. If some ``dark radiation" (some species of matter that behaved as radiation but which we have not yet detected) was present, and it too annihilated between the decoupling of neutrinos and photons (similarly to the electrons and positrons), then the ratio calculated in \eqref{Trelation} would be altered. Owing to the cubic nature of the relationship, a doubling of the neutrino effective temperature would result in the number density today being 8 times larger. Similarly, a colder temperature would diminish the number density on earth. This method of entropy injection from the annihilation of some species would actually result in a colder C$\nu$B, and thus a smaller number density. 
	
	Recall that 
	
	\begin{equation}
		\frac{T_{\gamma,0}}{T_{\nu,0}}\propto\left(\frac{g_*(T_{dec})}{g_*(T_0)}\right)^{1/3}=\left(\frac{11/2+\Delta g_*}{2}\right)^{1/3},
	\end{equation}

	so any dark radiation would increase this ratio. Since $T_{\gamma,0}$, the present-day temperature of the CMB, is a known observable, this would mean the neutrino effective temperature would decrease, and in turn so would the number density. This $\Delta g_*$ is quite well constrained, as $N_{\mathrm{eff}}$ would of course be affected if there was any other radiative species present before recombination. \\
	
	This chapter has aided us in achieving a more accurate prediction of what the cosmic neutrino background may look like around earth today, and which effects would be most noticeable. While annual modulation is one of the most natural characteristics of a C$\nu$B, its effect is too small to be measurable in the near future. Distortions to the number density from non-instantaneous decoupling and gravitational clustering are competing effects, with the dominant one decided by the mass of the neutrino: with neutrino masses above approximately 10 meV, clustering effects overshadow the spectral distortion, while relativistic neutrinos would only exhibit the effects of spectral distortion. Other non-standard physics may explain any deviations from these predictions, but as we shall discuss later, it may be difficult to distinguish between models.

	\newpage
	\section{Types of Neutrino Detectors}
	
	For the purposes of this text, supernovae are the main astrophysical source we will consider. The reason for this is that supernovae are short events, and thus data from both electromagnetic observations and neutrinos (and perhaps in future gravitational waves) can be compared to gain insights. Though much remains to be learned about supernovae, and data extracted is often model-dependent, the addition of precision neutrino astronomy will assist in making great strides in the field.\\
	The main source of information regarding these events is SN1987A, a supernova that occurred in the Large Magellanic Cloud near the Milky Way \cite{Bionta:1987qt}. From this, we learned that supernova neutrinos have energies in the range of tens of MeV - many orders of magnitude larger than the relic neutrinos, as seen in \eqref{lowenergy}.\\
	Owing to this energy difference, the detectors used for supernova neutrinos and those for the C$\nu$B utilise different capture mechanisms. This chapter does not pretend to act as a review of all detection methods, but simply attempts to elucidate some of the most popular methods used.
	
	\subsection{Astrophysical Neutrino Detectors}
	
	Owing to the small masses seen in \eqref{massupperlim}, supernova neutrinos are extremely relativistic, with speeds close to that of light. Their high energy allows for certain interactions that relic neutrinos are not energetic enough to achieve.\\
	
	\underline{Cherenkov Radiation}\\
	
	Cherenkov radiation is a phenomenon wherein charged particles travelling faster than light in a medium (i.e $v>\frac{1}{n}$ when $c=1$) emit radiation in a characteristic conal shape, analogous to the sonic boom achieved by objects moving faster than the speed of sound in a medium. All decelerating charged particles emit radiation, but the conal structure is unique to Cherenkov radiation. One way that Cherenkov radiation can be used to detect neutrinos is if a neutrino transfers energy to a charged particle in a detector.\\
	Consider some charged particle moving with a speed $v$, emitting radiation at some angle $\theta$ from its axis of propagation, symmetrically around the axis.
	
	\begin{figure}[!h]
		\centering
		\includegraphics[width=0.3\linewidth]{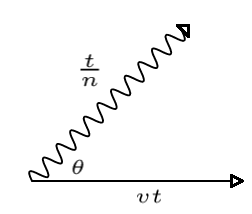}
		\caption{As the particle travels $vt$, the emitted light travels $\frac{t}{n}$. Drawn using \cite{Drawingtool}.}
	\end{figure}
	
	In some time $t$ (short enough such that its velocity does not change considerably in this time),the particle would have moved $vt$ while the radiation would have moved $\frac{t}{n}$. Thus, we find 
	
	\begin{equation}
		\cos\theta=\frac{t/n}{vt}=\frac{1}{nv},
	\end{equation}
	
	and since $nv>1$ for all times of emission, this angle is always acute, and thus as the particle travels we see the characteristic cone of light. For example, for a very relativistic particle $v\approxeq1$ in water with $n\approxeq1.33$, the cone will make an angle of $\approxeq41.25^{\circ}$.\\
	In order for this to be energetically possible, we need the kinetic energy, given by
	
	\begin{equation}
		E_K=(\gamma-1)m=\left(\frac{1}{\sqrt{1-v^2}}-1\right)m
	\end{equation}
	
	to obey $v>\frac{1}{n}$. So the condition for Cherenkov radiation is that the charged particle have energy larger than
	
	\begin{equation}
		E_K^{thr}=\left(\frac{1}{\sqrt{1-\frac{1}{n^2}}}-1\right)m.\label{thresh}
	\end{equation}
	
	As an illustrative example, for an electron with mass 0.511\ MeV in water with refractive index 1.333, we find $E_K^{thr}=264.06\ \mathrm{keV}$.\\
	Remember that this is the energy that the electron (or any other charged particle) must have, not the neutrino we wish to detect. Neutrinos can scatter off electrons via
	
	\begin{equation}
		\nu_\alpha + e^- \ \ \rightarrow \ \ \nu_\alpha + e^-.
	\end{equation}
	
	This technique is extremely useful as we gain information not only about the neutrino's energy, but also about its direction of incidence as charged particles (in this case, the electrons) paths can be tracked and the Cherenkov cones seen. The interaction cross section can be calculated using the electroweak theory above. For electron neutrinos, there will be both charged and neutral current interactions, but for muon and tau neutrinos there will of course only be the neutral current. Thus, this method favours the detection of electron neutrinos. Though the threshhold energy does not translate to exactly the required energy of the neutrino, in elastic scattering with electrons, it is pretty close. If we consider an initially stationary electron hit by a neutrino which afterwards comes to rest, we would have
	
	\begin{equation}
		E_e=m_e+E_\nu-m_\nu
	\end{equation}
	
	or, even more simply, $E_{K,e}=E_{K,\nu}$. Thus, in this limiting case, the kinetic energy of the detected electron is the exact same as that of the incoming neutrino. Of course, if the neutrino does not come to rest, or if the electron has some initial velocity, there can be slight deviations. However, we can take the threshhold energy for Cherenkov radiation as a good estimate for the kinetic energy needed in the incident neutrino. Thus, for even the lightest charged particle (the electron) in any realistic material, relic neutrinos would not be able to trigger Cherenkov radiation, while supernova neutrinos can.\\
	The other way that a neutrino can create a Cherenkov cone is via weak interactions with nuclei in the medium,
	
	\begin{equation}
		\nu+X(A,Z) \ \ \rightarrow \ \ Y(A,Z+1) + e^-,
	\end{equation}
	
	and similarly for the production of positrons by the incidence of an anti-neutrino. In this case, the incident (anti-)neutrino must have enough energy to overcome the binding energy of the $X$ nucleus as well as provide enough kinetic energy to the (positron) electron in order for it to be above $E_K^{thr}$. Once again, taking water as an example, the anti-neutrino could interact with the hydrogen nucleus (proton) or the neutrino with the oxygen nucleus, turning them into a neutron or fluorine nucleus, respectively. The former needs $\approx1.29$\ MeV while the latter needs around 16\ MeV. For supernova neutrinos these values are certainly attainable, even if the latter is only for some, but for cosmic neutrinos it simply is not energetically possible. Finally, note that only the ejected electron might create Cherenkov light, as the energy threshhold for nuclei, as seen in \eqref{thresh} is around $m_{nuc}$, while our supernova neutrinos are of the order of tens of MeV.\\
	
	Another variation of these detectors are ``Long-String" Cherenkov detectors, named for the long strings of photomultiplier tubes (PMTs) placed in the medium and used to detect the Cherenkov radiation. The advantage of this variation is the increased time sensitivity, owing to the array of PMTs, but the drawback is the loss of individual interaction information - only statistical information is gathered \cite{Scholberg:2012id}.\\
	
	\underline{Scintillation Detectors}\\
	
	As we have seen, anti-neutrinos can interact with nuclei via the weak interaction. One of these interactions is inverse beta decay,
	
	\begin{equation}
		\bar{\nu}_\alpha + p \ \rightarrow \ n + e^+.
	\end{equation}
	
	The energy threshold for this interaction is $m_n+m_e-m_p\approxeq1.8 \ \mathrm{MeV}$, and so while essentially impossible for cosmic neutrinos, this reaction is well within the capabilities of supernova neutrinos. Given a detector full of free electrons, the positrons then go on to undergo electron-positron annihilation,
	
	\begin{equation}
		e^- + e^+ \ \rightarrow \ 2\gamma_e
	\end{equation} 
	
	and, after a short delay, the neutrons undergo neutron capture:
	
	\begin{equation}
		p + n \ \rightarrow \ d + \gamma_d.
	\end{equation}
	
	At the energies of supernova neutrinos, the electron-positron annihilation will only produce photons, and not other massive bosons. These $\gamma_e$ photons will have energies around $m_e\approxeq 0.511\ \mathrm{MeV}$, while the energy released by the fusion of $p$ and $n$ means that the $\gamma_d$ photons will have energies around $m_p+m_n-m_d\approxeq2.22\ \mathrm{MeV}$.\\
	
	Detectors that use the detection of these photons to observe neutrinos are known as scintillation detectors, and consist of large volumes of the interacting material, usually in liquid form, surrounded by photomultiplier tubes in order to enhance and capture the photonic signal \cite{Scholberg:2012id}. The clear signal of a detection is the characteristic chain of events: a pair of $\approxeq 0.511\ \mathrm{MeV}$ photons, followed after a few hundred microseconds by the $\approxeq2.22\ \mathrm{MeV}$ signal. This technique was the first to successfully detect neutrinos back in 1956.\\
	
	Because the photons are emitted isotropically, they cannot directly give information about the direction of the incoming anti-neutrino. However, with very small segmentation, information from both $\gamma_e$ and $\gamma_d$ may give some idea of the incident direction \cite{Scholberg:2012id}.\\
	
	There are many other variants of detectors, but they all function using similar methods: either produced photons are measured or other particles produced. For example, liquid argon time-projection chambers use the interaction 
	
	\begin{equation}
		\nu_e +\ ^{40}\mathrm{Ar} \ \rightarrow \ ^{40}\mathrm{K}^* + e^- \ \rightarrow \ ^{40}\mathrm{K} + e^- + \gamma_K,
	\end{equation}
	
	and detect the $\gamma_K$ arising from the de-excitation of the krypton produced during the charged current interaction.  In the same experimental setup, anti-neutrinos can be detected using 
	
	\begin{equation}
		\bar{\nu}_e +\ ^{40}\mathrm{Ar} \ \rightarrow \ ^{40}\mathrm{Cl}^* + e^- \ \rightarrow \ ^{40}\mathrm{Cl} + e^- + \gamma_{Cl},
	\end{equation}

	and the characteristic de-excitation photons detected once again. This same design is used in so-called heavy nuclei detectors, which use substances such as iron and lead.

	\subsection{Mechanisms for Detecting Relic Neutrinos}
	
	\underline{Elastic Scattering}\\
	
	As we have seen, the factor of $G_F=1.1664\times10^{-5} \ \mathrm{GeV}^{-2}$ which dictates weak interactions (in the 4-Fermi approximation, which is appropriate for the low energies at which we are working) is a small quantity: that is, terms including higher orders of $G_F$ are smaller. Thus, it seems to be in our best interest to analyse interactions that are of the order $G_F$.
	
	The first mechanism we can consider is ``neutrino optics" - if the relic neutrinos are travelling through a medium with inter-atomic spacing smaller than their de Broglie wavelength, this medium can be characterised by a refractive index $n$, with $n=\frac{p'}{p}$, $p$ being the neutrino's original momentum and $p'$ its altered form. This $n$ is found to be $n=1+\mathcal{O}\left(G_F\right)$ \cite{Gelmini:2004hg}, and so at first this seems promising. However, it was shown that the force of this refraction would be proportional to $\bar{\nabla}n_\nu$, and in the case of an experiment on earth, the neutrino density is negligibly different from constant, and so this force vanishes \cite{Cabibbo:1982bb}. \\
	
	The second option available at this order of $G_F$ is known as the Stodolsky effect. Consider a relic neutrino scattering off an electron. At low energy, we can use our 4-Fermi approximate Lagrangian, and from \eqref{4FermiLag} the Hamiltonian (density) is clearly just the negative of this. Only considering electrons and neutrinos, we have 
	
	\begin{equation}
		\mathcal{H} = \frac{G_F}{\sqrt{2}}\bar{e}\gamma^\mu(1-\gamma^5)e\bar{\nu_e}\gamma^\mu(1-\gamma^5)\nu_e.
	\end{equation}

	In theory, there will be an induced energy difference $\Delta E$ owing to the spin up and down electrons \cite{Gelmini:2004hg}. In the non-relativistic limit, the electron current yields a factor of $\bar{\sigma_e}\cdot \bar{v_e}$, while the neutrino current is found to be proportional to $n_\nu-n_{\bar{\nu}}$. Thus, not only is the interaction dependent on the alignment of the electrons' spins, but a lepton asymmetry in the neutrino sector is needed. Thus, for the standard model wherein the lepton asymmetry is negligible, this mode too seems unrealistic. \\

	At the next order of $G_F^2$, perhaps our luck will be better. In this case, we consider elastic scattering off of some large target whose nuclei have a mass number $A$, following \cite{Duda:2001hd}. As we discussed in the section above on annual modulation, the earth can be thought of as moving through the sea of neutrinos that is the C$\nu$B. If the earth was stationary relative to the C$\nu$B, the average momentum exchange between relic neutrinos and our target would be 0, as they would all cancel out. The motion of earth induces a dipole distortion in the momentum of the cosmic neutrinos. If the relic neutrino loses all of its momentum in the interaction, then the momentum exchange in earth's frame is 
	
	\begin{equation}
		\braket{p}_{\mathrm{rel}} = v_{\mathrm{earth}}p_\nu \ \ \ \ ; \ \ \ \ \braket{p}_{\mathrm{nonrel}} = v_{\mathrm{earth}}m_\nu,
	\end{equation}

	for massless and massive species, respectively. For the massive case, it is assumed that the neutrino that interacts with the target has clustered in the Milky Way.
	
	This imparting of momentum onto the target would cause a small acceleration:
	
	\begin{equation}
		a=\frac{F}{m_T}=\frac{\braket{p}}{m_T\Delta t}.
	\end{equation}

	This $\frac{1}{\Delta t}$ is the collision rate. Let us consider a target of mass 1 gram, such that $m_T=1$ and $N_T=\frac{N^A}{A}$, with the numerator being Avogadro's constant. Since the collision rate is $\frac{1}{\Delta t} = \phi_\nu N_T \sigma_{\nu N}$ with $\phi_\nu=n_\nu v_\nu$ the neutrino flux, we have 
	
	\begin{equation}
		a = \frac{N^Av_{\mathrm{earth}}E_\nu n_\nu v_\nu \sigma_{\nu N}}{A},
	\end{equation}

	with $E_\nu=p_\nu$ for the relativistic neutrinos and $m_\nu$ for the non-relativistic case. The acceleration of this 1g mass target will be the sum of the effects from the relativistic and non-relativistic neutrino species, as well as from both neutrinos and anti-neutrinos. Recall also from \eqref{4Fermiapprox} that the cross section is given approximately by 
	
	\begin{equation}
		\sigma_{\mathrm{rel}} \approx G_F^2 p_\nu^2 \ \ \ \ ; \ \ \ \ \sigma_{\mathrm{nonrel}} \approx G_F^2 m_\nu^2.
	\end{equation}
	
	At leading order in the instantaneous decoupling limit, the number density for each species (adding both neutrinos and anti-neutrinos) is $\approx 112\mathrm{cm}^{-3}$. 
	
	For the sake of numerical calculation, let us consider the minimal inverted hierarchy case, wherein we have 1 relativistic species and 2 massive species with mass around 50 meV and velocity approximately 0.01, using \eqref{lowenergy}. Then we have
	
	\begin{equation}
		a = 2a_{\mathrm{nonrel}} + a_{\mathrm{rel}} = \frac{2N^AG_F^2v_{\mathrm{earth}}n_0}{A} \left[2\times 0.01\times m_\nu^3+\braket{p_\nu}^3\right],
	\end{equation}

	where $\braket{p_\nu}$ is the average momentum, calculated in \eqref{lowenergy}. We will take $A$ to be around 56, to correspond with using iron as a target. Using $N^A=6.022\times10^{23}$ \cite{Pavese:2015cog} and $v_{\mathrm{earth}}\approx400$ km.s$^{-1} = 1.33\times10^{-3}$,\footnote{This is a maximum, using the fact that the sun is moving at around 369 km.s$^{-1}$ relative to the C$\nu$B \cite{Aghanim:2013suk} and the earth is moving with a velocity of around 30 km.s$^{-1}$ around the sun \cite{McCabe:2013kea}, $v_{\mathrm{earth}}$ can range between $\approx$ 330 and 400 km.s$^{-1}$.} this leads to a numerical acceleration of $a = 6.815 \times 10^{-52}$ cm.s$^{-2}$: our target will move roughly $1\times10^{-16}$ cm over the age of the universe. According to \cite{Duda:2001hd}, a more precise calculation leads to $a \approx 10^{-27}f_c\rho_\nu$ cm.s$^{-2}$, with $f_c$ the amplification owing to clustering (around 1.12 for our 50 meV neutrinos) and $\rho_\nu$ the energy density of neutrinos in grams per cm$^3$.
	
	Clearly, elastic scattering is not our best bet for detecting relic neutrinos.\\

	\underline{High Energy Cosmic Rays}\\
	
	Cosmic ``rays" are in fact particles such as protons and nuclei that come from distant sources and have extremely high energies. Owing to the presence of the CMB, there is an upper limit on the possible energies that these cosmic rays can have. As an illustrative example, let's consider very high energy protons.
	
	As these protons are travelling towards us, they are moving through a sea of relic photons and there is a high chance they will undergo some kind of interaction. For example, 
	
	\begin{equation}
		p \ + \ \gamma_{\mathrm{CMB}} \ \rightarrow \ p \ + \ \pi^0,
	\end{equation}
	
	where $\pi^0$ is the neutral pion. Let us consider 4-momentum conservation. Then we have
	
	\begin{equation}
		(p_i+p_\gamma)^2 = (p_f+p_\pi)^2 \ \ \rightarrow -m_p^2 + 2p_i\cdot p_\gamma = -(m_p+m_\pi)^2,
	\end{equation}

	where in the second part we are interested in the threshold, wherein the outgoing particles are approximately at rest. For simplicity, let us consider that the proton and CMB photon were moving towards each other along the z-axis. Then, since the proton is extremely relativistic, we can take their 4-momenta to be
	
	\begin{equation}
		p_i = (E_i,0,0,E_i) \ \ \ \ ; \ \ \ \ p_\gamma=(E_\gamma,0,0,-E_\gamma).
	\end{equation}

	Plugging this in, we have 
	
	\begin{equation}
		-m_p^2 - 4E_iE_\gamma = -(m_p+m_\pi)^2 \ \ \rightarrow \ \ E_i=\frac{(m_p+m_\pi)^2-m_p^2}{4E_\gamma}.
	\end{equation}

	Using that the mass of the proton is 938.2721 MeV \cite{Mohr:2018hvt}, the mass of the neutral pion is 134.9768 MeV \cite{Zyla:2020zbs}, and the energy of the CMB photon is (on average) $E_\gamma=T_\gamma=T_{\mathrm{CMB}}=0.2348$ meV \cite{Fixsen:2009ug}, we have the cutoff energy of the cosmic ray as around $E_i=2.89\times 10^{20}$ eV. More precise calculations give the GZK (Greisen-Zatsepin-Kuzmin) cutoff of $E_{\mathrm{GZK}}\approxeq 5 \times 10^{19}$ eV \cite{Greisen:1966jv}.\\
	
	Experimentally, however, cosmic rays with energies higher than this cutoff (so called ``Oh-My-God" particles) have been observed, which seems extremely unlikely. One explanation is that the cosmic ray must have come from a very nearby source, such that it has a higher chance of not yet having interacted with a relic photon. Unfortunately, there are no known nearby sources of such high energy rays. One explanation involves the C$\nu$B: of course, as we know, neutrinos only interact weakly, and thus would not interact with the CMB. So an extremely high energy neutrino, coming from a distant source, would be able to pass through the fog of relic photons without its energy being diminished. There is, of course, a chance that this cosmic ray neutrino interacts with a relic neutrino from the C$\nu$B. This interaction is a neutral current, and thus would create a Z-boson which could then decay into many possible combinations of particles. The energy threshold, calculating as before, would be 
	
	\begin{equation}
		E_i = \frac{m_Z^2}{2m_\nu} = 4.158 \times 10^{21}  \left(\frac{1\mathrm{eV}}{m_\nu}\right) \mathrm{eV},
	\end{equation}

	where the relic neutrino was taken to be at rest, and the mass of the Z-boson much larger than the neutrino. Theoretically, this cosmic ray neutrino could interact with a relic neutrino very close to earth, emitting particles with energies higher than the GZK cutoff and allowing them to arrive on earth without interacting with a CMB photon, such that it appears as if they have come from a nearby source. \\
	
	Another way in which the C$\nu$B might be seen is in the power spectrum of observed cosmic rays \cite{Wigmans:2002rb}. If a cosmic ray (say, for example, a proton) were to interact with a relic neutrino (continuing our example, via inverse beta decay), this should cause a change to the cosmic ray spectrum, depending on the mass of the neutrino. Performing a similar calculation to the GZK cutoff, we find that the threshold energy for a proton in this case would be around $E_i\approxeq 1.695\times 10^{15} \left(\frac{1\mathrm{eV}}{m_\nu}\right)$ eV. For example, a neutrino mass of $m_\nu = 5$ meV could explain the ``knee" seen in the spectrum of energies at around $10^{17.5}$ eV \cite{Yanagisawa} . Unfortunately, both of these methods are not only explicable by a relic neutrino population, and many other theories exist to explain both the kinks in the cosmic ray energy spectrum and the existence of rays with energies higher than the GZK cutoff \cite{Yanagisawa}. Thus, perhaps this is not the most fool-proof method for detecting the C$\nu$B.\\
	
	\underline{Capture by Beta-Decaying Nuclei}\\
	
	The most promising mode of detection seems to stem from the use of inverse beta decay, known as neutrino capture by beta-decaying nuclei (NCB), proposed by Weinberg \cite{Weinberg1962}. Recall that our relic neutrinos' energies are extremely small, with maximal values around the neutrinos' masses. Thus, an important feature of NCB is that the nuclei used must readily undergo beta decay, such that there is no energy threshold on the reaction. 
	
	The difference between regular beta decay, which the nuclei will also go through, and beta decay owing to neutrino capture is a difference in the possible energies of the emitted electron. As we shall see, the gap between the beta decay endpoint (the maximum possible energy for an emitted electron during a regular beta decay event) and the energy of an electron as the result of neutrino capture is $E_{\mathrm{C\nu B}}+m_{\mathrm{lightest}}$, with $m_{\mathrm{lightest}}$ being the mass of the lightest neutrino species.\\
	
	Since regular beta decay is the main source of background noise, it is preferable to choose an isotope for the experiment wherein the ratio between the C$\nu$B neutrino capture rate and the inverse of the half life of the isotope is maximised. Some candidates include tritium, $^3$H and Rhenium-187, $^{187}$Re \cite{Yanagisawa}. Rhenium-187 is an interesting candidate, as its half life is massive - over 40 billion years \cite{Galeazzi:2001ih}. This is useful as it will minimise background noise in the neutrino capture experiment. Tritium's half life, on the other hand, is much smaller at around 12.32 years \cite{LucasLL}. On the other side of the coin, the cross section of neutrino capture on tritium (as we will calculate explicitly later) is found to be of the order $10^{-45}$ cm$^2$, while for rhenium-187 it is much lower at around $10^{-52}$ cm$^2$ \cite{Yanagisawa}. So even though striking a balance is key, as we shall see in the next chapter, a cross section 7 orders smaller will make the capture rate - an already very small quantity - 7 orders smaller. For this reason, albeit creating a large amount of background noise for our experiment, it seems that NCB on tritium is our best bet for detecting the C$\nu$B, as such an interaction would not be explicable in any other way, and as we shall observe in the next chapter, the characteristic peaks in the emitted electron's spectrum will be a smoking gun indication. \\
	
	For astrophysical neutrinos - particularly from supernovae - the detectors of choice seem to utilise the Cherenkov effect, as they are able to offer not only energetic measurements, but also track the angle of incidence, to some degree. While scintillation detectors are adept at the former, owing to the isotropic nature of the emission of de-excitation photons, the incident direction of the neutrino is more difficult to detect. 
	
	For relic neutrinos, as we saw, capturing them on tritium seems to be the best option for the foreseeable future. While interactions with cosmic ray ultra-high energy neutrinos may explain the existence of ``Oh-My-God" particles, and distortions to the cosmic ray energy spectrum may prove the existence of the C$\nu$B, it is simply another indirect measurement - we are still not able to detect the energies, masses, and fermionic type of these elusive ancient particles. Elastic scattering - while indeed being a direct probe - is not a viable candidate owing to the miniscule reaction rates and the expected lack of lepton asymmetry. Should a lepton asymmetry be found in the universe, perhaps the Stodolsky effect will hold more promise. Other isotopes could indeed be used for capturing cosmic neutrinos, but factoring in half life and reaction rate among other variables, tritium seems to be the best bet.

	\newpage
	\section{The Capture Rate and Energy Spectra}
	
	As we have seen in the previous chapter, the best chance we have of detecting the extremely elusive cosmic neutrinos lies in inverse beta decay. The PTOLEMY experiment, projected to create a detector made up of 100 grams of tritium ($^3\mathrm{H}$) \cite{Betti:2019ouf}, is the leading hope for the C$\nu$B's imminent detection. The reasons for the choice of tritium are that firstly, and most importantly, the isotope needs to readily undergo inverse beta decay without any energy threshold, as the relic neutrinos' energies are extremely small. Secondly, the isotope must have a fairly large half-life, so that the experimental setup does not decay too quickly - tritium's half life is $t_{1/2}\approx 12.32$ years. Finally, a low Q-value (the energy released by the reaction) is preferred. This is because the energy difference between this inverse beta decay and regular beta decay is very small (as we will see kinematically), and so working at smaller energies will make it easier to observe. The reaction utilised is given by
	
	\begin{equation}
			\nu_i\ +\ ^3\mathrm{H} \ \rightarrow \ ^3\mathrm{He}\ +\ e^- \label{truereaction}
	\end{equation}

	and in calculating the scattering amplitude for this reaction, we will be able to obtain the expected capture rate - that is, the rate at which we should detect these relic neutrinos.
	
	\subsection{Cross Section of Neutrino-Tritium Interaction}
	
	The full theory needed to calculate the scattering amplitude of the reaction in \eqref{truereaction} is the electroweak theory described in chapter 3. However, since this reaction is occurring at energies much smaller than the weak boson masses, we can safely utilise 4-Fermi theory, the low-energy effective field theory at tree level. 
	
	To begin, following the lead of \cite{Long:2014zva}, we consider the process 
	
	\begin{equation}
		\nu_i + n \ \rightarrow \ p + e^- \label{reaction}.
	\end{equation}

	From our theory, we then get the matrix element
	
	\begin{equation}
		i\mathcal{M}_i = -i\frac{G_F}{\sqrt{2}}V_{ud}U^*_{ei}\left[\bar{u}_e\gamma^\alpha(1-\gamma^5)u_{\nu_i}\right]\left[\bar{u}_p\gamma_{\alpha}(f-g\gamma^5)u_n\right],
	\end{equation}

	where the factors of $f$ and $g$ have been added post-hoc to account for the fact that the neutron and proton are not elementary particles but have internal structure, consisting of quarks and gluons. These are known as the nucleonic form factors, with $f$ being for the proton and $g$ the neutron. These form factors are actually momentum dependent, but owing to the extremely small momenta of the relic neutrinos, we can take the values of $f$ and $g$ to be those in the limit as the transfer momentum goes to 0.
	
	Squaring this, we have
	
	\begin{equation}
		|\mathcal{M}_i|^2=\frac{G_F^2}{2}|V_{ud}|^2|U_{ei}|^2\mathcal{N}_1^{\alpha\beta}\mathcal{N}_{2\alpha\beta},
	\end{equation}

	with
	 \begin{equation}
	 	\mathcal{N}_1^{\alpha \beta} = {\rm tr}\left[\gamma^{\alpha}(1-\gamma^5)u_{\nu}\bar{u}_{\nu}\gamma^{\beta}(1-\gamma^5)u_e\bar{u}_e \right],
	 \end{equation}
 	 \begin{equation}
	 	\mathcal{N}_2^{\gamma \delta} = {\rm tr}\left[\gamma^{\gamma}(f-g\gamma^5)u_{n}\bar{u}_{n}\gamma^{\delta}(f-g\gamma^5)u_p\bar{u}_p \right].
	 \end{equation}
 
 	We need to sum over all the possible spins that the proton, neutron and electron may have (as well as averaging over the neutron spins), so we need to use the well-known completeness relations that are the outer products of the spinors:
 	
 	\begin{equation}
 		\sum_{s_j=\pm 1/2}u_j\bar{u}_j=(\slashed{p}_j+m_j),\label{outerprod}
 	\end{equation}
 
 	but we don't want to sum over the neutrino spins yet. Thus, we need to derive their outer product without summation using spinor formalism.\\
 	
 	Two-component spinors (or bi-spinors) have 2 indices each ranging from 1 to 2. Spinor indices are raised and lowered using the Levi-Civita tensor in 2D\footnote{We use the convention $\epsilon^{12}=1$ so $\epsilon_{12}=-1$.}. Next, we define the ``soldering forms", which can be used to change vectors into two-component spinors, using the Pauli matrices $\bar{\sigma}$:
 	
 	\begin{equation}
 		\sigma^\mu_{\alpha\dot{\beta}} = (1,\bar{\sigma}) \ \ \ \ ;  \ \ \ \ \sigma_{\mu\alpha\dot{\beta}} = (1,-\bar{\sigma}),
 	\end{equation}
 
 	\begin{equation}
 		\tilde{\sigma}^{\mu\dot{\alpha}\beta} = (1,-\bar{\sigma}) \ \ \ \ ;  \ \ \ \ \tilde{\sigma}_\mu^{\dot{\alpha}\beta} = (1,\bar{\sigma}).
 	\end{equation}
 
 	In the above, $\mu$ is our vector index ranging from 0 to 3, while $\alpha$ and $\beta$ are our spinor components, ranging from 1 to 2. Keep in mind that the ``1" is actually a 2D identity matrix.
 	
 	We also know that our neutrinos satisfy the Dirac equation, so writing their 4-component spinors (and their Dirac conjugates $\bar{u}_s =u_s^\dagger\gamma^0)$ as 
 	
 	\begin{equation}
 		u_s(\bar{p}) = \begin{pmatrix} x_{s,\alpha}(\bar{p}) \\ y_s^{\dagger\dot{\alpha}}(\bar{p}) \end{pmatrix} \ \ \ \ ;  \ \ \ \ \bar{u}_s^T(\bar{p}) = \begin{pmatrix} y_s^\alpha(\bar{p}) \\ x_{s,\dot{\alpha}}^\dagger(\bar{p}) \end{pmatrix},
 	\end{equation}
 
 	with $s$ being their spin, and selecting our basis (by setting a phase) to be 
 	
 	\begin{equation}
 		(\chi_{\frac{1}{2}})_\alpha = (\eta^\dagger_{\frac{1}{2}})^{\dot{\alpha}} = \begin{pmatrix}
 			1 \\ 0
 		\end{pmatrix} \ \ \ \ ;  \ \ \ \ (\chi_{-\frac{1}{2}})_\alpha = (\eta^\dagger_{-\frac{1}{2}})^{\dot{\alpha}} = \begin{pmatrix}
 		0 \\ 1 \label{basisspinors}
 	\end{pmatrix},
 	\end{equation}
 
 	we can solve the Dirac equation for our spinors $u_{s,\alpha}(\bar{p})$. We can also solve for the antiparticle's spinors $v_{s,\alpha}(\bar{p})$, but these are not necessary for our ends. Plugging $u_{s,\alpha}(\bar{p})$ into the Dirac equation and solving similarly to in \cite{Schwartzbook}, we find
 	
 	\begin{equation}
 		x_{s,\alpha}(\bar{p}) = \tensor{\left(\sqrt{p\cdot\sigma}\right)}{_\alpha^\beta}(\chi_s)_\beta \ \ \ \ ;  \ \ \ \ \ y_s^{\dagger\dot{\alpha}}(\bar{p}) = \tensor{\left(\sqrt{p\cdot\tilde{\sigma}}\right)}{^{\dot{\alpha}}_{\dot{\beta}}}(\eta^\dagger_s)^{\dot{\beta}},
 	\end{equation}
 	\begin{equation}
 		x^\dagger_{s,\dot{\alpha}}(\bar{p}) = (\chi_s^\dagger)_{\dot{\beta}}\tensor{(\sqrt{p\cdot\sigma})}{^{\dot{\beta}}_{\dot{\alpha}}} \ \ \ \ ;  \ \ \ \ y_s^\alpha(\bar{p}) = (\eta_s)^\beta \tensor{(\sqrt{p\cdot\tilde{\sigma}})}{_\beta^\alpha}.
 	\end{equation}
 
 	These give the most general solutions for our spinor and antispinor. Let us now consider what is known as the helicity basis - this is when we choose the spin's direction to align with the momentum's: $\hat{s} = \hat{p}$. This is named as such because the helicity, defined as $\hat{h} = \frac{\bar{\sigma} \cdot \bar{p}}{|\bar{p}|}$ (which is a conserved quantity) satisfies the eigenvalue equation \cite{Fidler:2017pkg}
 	
 	\begin{equation}
 		\hat{h}u_s = \pm2su_s
 	\end{equation}
 
 	for the case of a spin-$\frac{1}{2}$ particle. Thus, we see that $\bar{p}\cdot\bar{\sigma} = \pm2s|\bar{p}|$. For the basis spinors in \eqref{basisspinors}, we have that 
 	
 	\begin{equation}
 		\bar{p}\cdot\bar{\sigma} \chi_s = 2s|\bar{p}| \chi_s \ \ \ \ ;  \ \ \ \ \bar{p}\cdot\bar{\sigma} \eta_s = 2s|\bar{p}| \eta_s.
 	\end{equation}
 
 	Plugging this in, and using that $p\cdot\sigma = p^\mu \sigma_{\mu\alpha\dot{\beta}} = E\mathds{1} - \bar{p}\cdot\bar{\sigma}$ and similarly $p\cdot\tilde{\sigma} = E\mathds{1} + \bar{p}\cdot\bar{\sigma}$, we have 
 	
 	\begin{equation}
 		x_{s,\alpha}(\bar{p}) = \sqrt{E-2s|\bar{p}|}(\chi_s)_\alpha \ \ \ \ ;  \ \ \ \ y_s^{\dagger\dot{\alpha}}(\bar{p}) = \sqrt{E+2s|\bar{p}|}(\eta^\dagger_s)^{\dot{\alpha}},
 	\end{equation}
 \begin{equation}
 	x^{\dagger\dot{\alpha}}_s(\bar{p}) = \sqrt{E-2s|\bar{p}|}(\chi_s^\dagger)^{\dot{\alpha}}  \ \ \ \ ;  \ \ \ \ y_{s,\alpha}(\bar{p}) = \sqrt{E+2s|\bar{p}|}(\eta_s)_\alpha.
 \end{equation}

	Thus, we have our spinor and its conjugate: 
	
	\begin{equation}
		u_s = \begin{pmatrix}
			\sqrt{E-2s|\bar{p}|}(\chi_s)_\alpha \\ \sqrt{E+2s|\bar{p}|}(\eta^\dagger_s)^{\dot{\alpha}}
		\end{pmatrix} \ \ ;  \ \
		\bar{u}_s = \begin{pmatrix}
			\sqrt{E+2s|\bar{p}|}\epsilon^{\alpha\beta}(\eta_s)_\beta & \sqrt{E+2s|\bar{p}|}\epsilon_{\dot{\alpha}\dot{\beta}}(\chi^\dagger_s)^{\dot{\beta}}
		\end{pmatrix}. \label{refme}
	\end{equation}
 
 	The outer product of the spinors for each neutrino species is then
 	
 	\begin{equation}
 		u_{\nu_i}\bar{u}_{\nu_i} = \begin{pmatrix}
 			\sqrt{E^2-4s^2|\bar{p}|^2}(\chi_s)_\alpha\epsilon^{\beta\gamma}(\eta_s)_\gamma &
 			(E-2s|\bar{p}|)(\chi_s)_\alpha\epsilon_{\dot{\beta}\dot{\gamma}}(\chi^\dagger_s)^{\dot{\gamma}} \\
 			(E+2s|\bar{p}|)(\eta^\dagger_s)^{\dot{\alpha}}\epsilon^{\beta\gamma}(\eta_s)_\gamma &
 			\sqrt{E^2-4s^2|\bar{p}|^2}(\eta^\dagger_s)^{\dot{\alpha}}\epsilon_{\dot{\beta}\dot{\gamma}}(\chi^\dagger_s)^{\dot{\gamma}}
 		\end{pmatrix}.
 	\end{equation}
 
 	Luckily, this is equivalent to a far neater expression \cite{Fidler:2017pkg}: 
 	
 	\begin{equation}
 		u_{\nu_i}\bar{u}_{\nu_i}=\frac{1}{2}\bigl(\slashed{p}_{\nu_i}+m_{\nu_i}\bigl)\bigl(1+2s_{\nu}\gamma^5\slashed{S}_{\nu_i}\bigl), \label{neutouter}
 	\end{equation}
 
 	where we have specified $s=s_\nu$, and where the spin vector $S_{\nu_i}$ is given by 
 	
 	\begin{equation}
 		S_{\nu_i}^{\mu}=\left(\frac{|\bar{p}_{\nu}|}{m_{\nu_i}}, \frac{E_{\nu}}{m_{\nu_i}}\frac{\bar{p}_{\nu}}{|\bar{p}_\nu|}\right),
 	\end{equation}
 
 	again specifying now $E=E_\nu$ , $\bar{p} = \bar{p}_\nu$, etc. Note that summing \eqref{neutouter} over spins reduces it to \eqref{outerprod}, as must be.
 	
 	Plugging in \eqref{neutouter} and the other outer products, we have
 	
 	\begin{equation}
 		\frac{1}{2}\sum_{s_n,s_p,s_e=\pm 1/2}|\mathcal{M}_i|^2=\frac{G_F^2}{4}|V_{ud}|^2|U_{ei}|^2\tilde{\mathcal{N}}_1^{\alpha\beta}\tilde{\mathcal{N}}_{2\alpha\beta},
 	\end{equation}
 
with
 
 	\begin{equation}
 		\tilde{\mathcal{N}}_1^{\alpha\beta}=\frac{1}{2}\mathrm{tr}\left[\gamma^\alpha\bigl(1-\gamma^5\bigl)\bigl(\slashed{p}_{\nu_i}+m_{\nu_i}\bigl)\bigl(1+2s_\nu\gamma^5\slashed{S}_{\nu_i}\bigl)\gamma^\beta\bigl(1-\gamma^5\bigl)\bigl(\slashed{p}_e+m_e\bigl)\right],
 	\end{equation}
 \begin{equation}
 	\tilde{\mathcal{N}}_2^{\gamma\delta}=\mathrm{tr}\left[\gamma^\gamma\bigl(f-g\gamma^5\bigl)\bigl(\slashed{p}_n+m_n\bigl)\gamma^\delta\bigl(f-g\gamma^5\bigl)\bigl(\slashed{p}_p+m_p\bigl)\right].
 \end{equation}
 
 Multiplying these two expressions, and taking the traces, we have
 \begin{multline}
 	\tilde{\mathcal{N}}_1^{\alpha\beta}\tilde{\mathcal{N}}_{2\alpha\beta}=32\left\{\left(g+f\right)^2\left[\left(p_e\cdot p_p\right)\left(p_{\nu_i} \cdot p_n\right)\right]+\left(g-f\right)^2\left[\left(p_e\cdot p_n\right)\left(p_{\nu_i}\cdot p_p\right)\right]+\left(g^2-f^2\right)m_nm_p\left(p_e\cdot p_{\nu_i}\right)\right\} \\
 	-64s_\nu m_{\nu_i} \left\{\left(g+f\right)^2\left[\left(p_e\cdot p_p\right)\left(S_{\nu_i} \cdot p_n\right)\right]+\left(g-f\right)^2\left[\left(p_e\cdot p_n\right)\left(S_{\nu_i}\cdot p_p\right)\right]+\left(g^2-f^2\right)m_nm_p\left(p_e\cdot S_{\nu_i}\right)\right\}.
 \end{multline}

	Before continuing, we select a particular frame in which to work. We will work in the neutron's rest frame, such that
	
	\begin{equation}
		p_n^\mu = (m_n, 0),\ \ \ \ p_{\nu}^\mu=(E_{\nu}, \bar{p}_{\nu}),\ \ \ \ p_p^{\mu}=(E_p, \bar{p}_p),\ \ \ \ p_e^{\mu}=(E_e, \bar{p}_e).
	\end{equation}
 		
 	In this case we have 
 	
 	\begin{align}
 		\tilde{\mathcal{N}}_1^{\alpha\beta}\tilde{\mathcal{N}}_{2\alpha\beta}&=32m_nE_pE_eE_{\nu_i}\biggl\{\left(g+f\right)^2\left(1-\frac{\bar{p}_e\cdot\bar{p}_p}{E_eE_{p}}\right)+(g-f)^2\left(1-\frac{\bar{p}_{\nu}\cdot \bar{p}_p}{E_{\nu_i}E_p} \right) \nonumber \\
 		&\ \ \ \ +(g^2-f^2)\frac{m_p}{E_p}\left(1-\frac{\bar{p}_e\cdot\bar{p}_{\nu}}{E_eE_{\nu_i}}\right) \biggl\} -64s_{\nu}m_nE_pE_eE_{\nu_i}\biggl\{v_{\nu_i}(g+f)^2\left(1-\frac{\bar{p}_e\cdot\bar{p}_p}{E_eE_p}\right) \nonumber \\
 		&\ \ \ \ +(g-f)^2\left(v_{\nu_i}-\frac{\bar{p}_{\nu}\cdot\bar{p}_p}{|\bar{p}_{\nu}|E_p}\right) +(g^2-f^2)\frac{m_p}{E_p}\left(v_{\nu_i}-\frac{\bar{p}_{\nu}\cdot\bar{p}_e}{|\bar{p}_{\nu}|E_e} \right) \biggl\} \nonumber \\
 		&=32m_nE_pE_eE_{\nu_i}\biggl\{2\left(g^2+f^2 \right)\left(1-2s_{\nu}v_{\nu_i}\right)
 		+\left(g^2-f^2 \right)\frac{m_p}{E_p}\left(1-2s_{\nu}v_{\nu_i}\right) \nonumber \\
 		&\ \ \ \ +\left(f^2-g^2\right)\frac{m_p}{E_p}\left(v_{\nu_i}-2s_{\nu}\right)v_e\cos\theta_{e\nu} 
 		-\left(g+f\right)^2(1-2s_{\nu}v_{\nu_i})v_ev_p\cos\theta_{ep} \nonumber \\
 		&\ \ \ \ -\left(g-f\right)^2(v_{\nu_i}-2s_{\nu})v_p\cos\theta_{\nu p} \biggl\},
 		\label{NN}
 	\end{align}
 	where $v_j=|\bar{p}_j|/E_j$ is the velocity of each particle and $\cos\theta_{jk}=\bar{p}_j\cdot\bar{p}_k/(|\bar{p}_j||\bar{p}_k|)$ is the angle between the $j$th and $k$th particles. Next, we can use momentum conservation ($\bar{p}_\nu=\bar{p}_p+\bar{p}_e$) to rewrite all these angles in terms of one: the angle between the incoming neutrino and outgoing electron, $\cos\theta_{e\nu}$: 
 	
 	\begin{equation}
 		\cos\theta_{ep}=-\frac{|\bar{p}_e|}{|\bar{p}_p|}+\frac{|\bar{p}_{\nu}|}{|\bar{p}_p|}\cos\theta_{e\nu},\ \ \ \ 
 		\cos\theta_{\nu p}=\frac{|\bar{p}_{\nu}|}{|\bar{p}_p|}-\frac{|\bar{p}_e|}{|\bar{p}_p|}\cos\theta_{e\nu}.
 	\end{equation}
 	
 	Using the above equations then, our squared matrix element becomes
 	
 	\begin{align}
 		&\frac{1}{2}\sum_{s_n,s_p,s_e=\pm 1/2}|\mathcal{M}_i|^2= \nonumber \\
 		&\ \ \ \ 8G_F^2|V_{ud}|^2|U_{ei}|^2m_nE_pE_eE_{\nu_i} \biggl\{2\left(g^2+f^2 \right)\left(1-2s_{\nu}v_{\nu_i}\right)
 		+\left(g^2-f^2 \right)\frac{m_p}{E_p}\left(1-2s_{\nu}v_{\nu_i}\right) \nonumber \\
 		&\ \ \ \ +\left(f^2-g^2\right)\frac{m_p}{E_p}\left(v_{\nu_i}-2s_{\nu}\right)v_e\cos\theta_{e\nu} 
 		+\left(g+f\right)^2(1-2s_{\nu}v_{\nu_i})v_e\left(\frac{|\bar{p}_e|}{E_p}-\frac{|\bar{p}_{\nu}|}{E_p}\cos\theta_{e\nu} \right) \nonumber \\
 		&\ \ \ \  -\left(g-f\right)^2(v_{\nu_i}-2s_{\nu})\biggl(\frac{|\bar{p}_{\nu}|}{E_p}-\frac{|\bar{p}_e|}{E_p}\cos\theta_{e\nu} \biggl) \biggl\}.
 		\label{M2}
 	\end{align}
 
 	Finally we reach our cross section. Just before though, note that the Mandelstam variable $t$ is given by
 	\begin{equation}
 		t = (p_e-p_{\nu_i})^2=(E_e-E_{\nu_i})^2-|\bar{p_e}-\bar{p}_{\nu_i}|^2=(E_e-E_{\nu_i})^2-|\bar{p}_e|^2-|\bar{p}_{\nu_i}^2|+2|\bar{p}_e||\bar{p}_{\nu_i}|\cos\theta_{e\nu_i}
 	\end{equation}
 
 	and also that the centre of mass neutrino momentum $\bar{p}_{\nu_i}^{\mathrm{com}}$ is related to the neutrino momentum in the neutron's rest frame $\bar{p}_{\nu_i}$ by 
 	
 	\begin{equation}
 		m_n^2|\bar{p}_{\nu_i}|^2 = s|\bar{p}_{\nu_i}^{\mathrm{com}}|^2,
 	\end{equation}
 
 	where $s$ is the Mandelstam variable $s=(p_n+p_{\nu_i})^2$.
 	
 	Thus, using that 
 	
 	\begin{equation}
 		\frac{d\sigma_i}{dt}=\frac{1}{64 \pi s}\frac{1}{|\bar{p}_{\nu}^{\rm com}|^2}\ \frac{1}{2}\sum_{s_n,s_e,s_p=\pm 1/2}|\mathcal{M}_i|^2,
 	\end{equation}
 
 	we have
 	
 	\begin{equation}
 		\frac{d\sigma_i}{d\cos\theta_{e\nu_i}} = \frac{d\sigma_i}{dt} \frac{dt}{d\cos\theta_{e\nu_i}} = \frac{1}{32\pi}\frac{1}{m_n^2}\frac{|\bar{p}_e|}{|\bar{p}_{\nu_i}|}\  \frac{1}{2}\sum_{s_n,s_e,s_p=\pm 1/2}|\mathcal{M}_i|^2.
 	\end{equation}
 
 	So, plugging in our matrix element, we have
 	
 	\begin{align}
 		\frac{d \sigma_i}{d \cos\theta_{e\nu_i}}&=\frac{G_F^2}{4\pi}|V_{ud}|^2|U_{ei}|^2\frac{E_pE_e|\bar{p}_e|}{m_nv_{\nu_i}} \biggl\{2\left(g^2+f^2 \right)\left(1-2s_{\nu}v_{\nu_i}\right)
 		+\left(g^2-f^2 \right)\frac{m_p}{E_p}\left(1-2s_{\nu}v_{\nu_i}\right) \nonumber \\
 		&\ \ \ \ +\left(f^2-g^2\right)\frac{m_p}{E_p}\left(v_{\nu_i}-2s_{\nu}\right)v_e\cos\theta_{e\nu} 
 		+\left(g+f\right)^2(1-2s_{\nu}v_{\nu_i})v_e\left(\frac{|\bar{p}_e|}{E_p}-\frac{|\bar{p}_{\nu}|}{E_p}\cos\theta_{e\nu} \right) \nonumber \\
 		&\ \ \ \  -\left(g-f\right)^2(v_{\nu_i}-2s_{\nu})\biggl(\frac{|\bar{p}_{\nu}|}{E_p}-\frac{|\bar{p}_e|}{E_p}\cos\theta_{e\nu} \biggl) \biggl\}.
 	\end{align}
 
 	and finally integrating over $\cos\theta_{e\nu_i}$ and taking $v_{\nu_i}$ over, we have
 	
 	\begin{align}
 		\sigma_iv_{\nu_i}&=\frac{G_F^2}{2\pi}|V_{ud}|^2|U_{ei}|^2\frac{E_pE_e|\bar{p}_e|}{m_n}\biggl\{2\left(g^2+f^2 \right)\left(1-2s_{\nu}v_{\nu_i}\right)
 		+\left(g^2-f^2 \right)\frac{m_p}{E_p}\left(1-2s_{\nu}v_{\nu_i}\right) \nonumber \\
 		&\ \ \ \  +\left(g+f\right)^2(1-2s_{\nu}v_{\nu_i})v_e\frac{|\bar{p}_e|}{E_p}  -\left(g-f\right)^2(v_{\nu_i}-2s_{\nu})\frac{|\bar{p}_{\nu}|}{E_p} \biggl\}.
 	\end{align}
	
	Note that actually, when integrating over $\cos\theta_{e\nu_i}$ an approximation was made, since 
	\begin{align}
		E_p &\simeq m_p\left(1+\frac{|\bar{p}_p|}{2m_p^2}\right)\simeq m_p\left(1+ \frac{|\bar{p}_e|^2}{2m_p^2}-\frac{|\bar{p}_e||\bar{p}_{\nu}|}{m_p^2}\cos\theta_{e\nu_i} \right).\label{protonE}
	\end{align}
	
	When integrating, this last term actually falls away. However, this change is only around $10^{-18}$ of the proton's energy, and so we assume that $E_p$ is left unchanged. In fact, to a very high accuracy we can take $E_p$ to be $m_p$, as the other terms in  \eqref{protonE} are at least $10^{5}$ times smaller.
	
	For consistency, in this regard we are able to discard the last 2 terms, proportional to $\frac{|\bar{p}_e|}{E_p}$ and $\frac{|\bar{p}_\nu|}{E_p}$, as they are around $10^{-5}$ and $10^{-13}$ times the leading order terms. Thus, we are left with
	
	\begin{align}
		\sigma_iv_{\nu_i}&=\frac{G_F^2}{2\pi}|V_{ud}|^2|U_{ei}|^2\frac{m_pE_e|\bar{p}_e|}{m_n}\left(3g^2+f^2 \right)\left(1-2s_{\nu}v_{\nu_i}\right).
	\end{align}

	To obtain the full expression for $\sigma_iv_{\nu_i}$ of the reaction in \eqref{truereaction}, we need to now shoehorn some added factors in. Firstly, the most obvious change is that the neutron and proton masses become the $^3$H and $^3$He masses respectively, and the nucleonic form factors are replaced by transition probabilities, with $f^2$ becoming the Fermi transition probability $\braket{f_F}^2$ and $3g^2$ becoming $\frac{g_A^2}{g_V^2}\braket{g_{GT}}^2$ \cite{Baroni:2016xll}, with $\braket{g_{GT}}$ the Gamow-Teller transition probability and $g_V$ and $g_A$ the vector and axial coupling constants, respectively.
	
	Next, the Fermi function $F(Z,E_e)$ is needed to account for the Coulombic attraction between the outgoing electron and helium nucleus, and is given by 
	
	\begin{equation}
		F(Z,E_e)=\frac{2\pi\alpha ZE_e/|\bar{p}_e|}{1-e^{-2\pi\alpha ZE_e/|\bar{p}_e|}},
	\end{equation}

	where $Z$ is the atomic number of the nucleus - in this case $Z=2$ - and $\alpha$ is the fine structure constant. Putting this all in, we have
	
	\begin{align}
		\sigma_iv_{\nu_i}&=\frac{G_F^2}{2\pi}|V_{ud}|^2|U_{ei}|^2\frac{m_\mathrm{^3He}E_e|\bar{p}_e|}{m_\mathrm{^3H}}F(2,E_e)\left(\braket{f_F}^2 + \frac{g_A^2}{g_V^2}\braket{g_{GT}}^2 \right)\left(1-2s_{\nu}v_{\nu_i}\right).\label{crosssection}
	\end{align}
	
	The term including $v_{\nu_i}$ is important, and is the next-to leading order term. Even for the most massive possible neutrinos, this is still of the order $10^{-2}$, and so for the accuracy to which we are working, it is not negligible. For massless neutrinos, it is even a leading order term. The minute terms can be neglected as other miniscule effects have been as well, such as the differences incurred by using the full electroweak theory, as well as those brought about by the fact that atomic tritium is used - not free nuclei.
	
	\subsection{The Capture Rate of Relic Neutrinos on Tritium}
	
	Considering some lone tritium atom in a sea of neutrinos, the usual rate for this atom to capture a neutrino would be $\Gamma=\braket{n\sigma v}$, with $n$ the number density of the surrounding neutrinos, $\sigma$ the interaction cross section and $v$ the neutrino's velocity. In actuality, this is just an approximation, with the real capture rate being
	
	\begin{equation}
		\Gamma_i = \sum_{s_\nu=\pm 1/2} \int \frac{d^3p_\nu}{(2\pi)^3}f_{\nu_i}(\bar{p}_\nu)\sigma_{\nu_i}(\bar{p}_\nu,s_\nu)v_{\nu_i}.
	\end{equation}

	Considering an experimental setup that involves a mass $M_T$ of atomic tritium with atomic mass $m^{\mathrm{atom}}_{^3\mathrm{H}}$, there are now $N_T=\frac{M_T}{m^{\mathrm{atom}}_{^3\mathrm{H}}}$ interaction points. Atop this, there are 3 species in the mass basis which need to be summed over. Thus, the full capture rate of cosmic neutrinos on this setup of tritium is
	
	\begin{equation}
		\Gamma_{\mathrm{C\nu B}} = N_T\sum_{i=1}^{3}\sum_{s_\nu=\pm 1/2} \int \frac{d^3p_\nu}{(2\pi)^3}f_{\nu_i}(\bar{p}_\nu)\sigma_{\nu_i}(\bar{p}_\nu,s_\nu)v_{\nu_i}.
	\end{equation}

	The factor of $\sigma_{\nu_i}(\bar{p}_\nu,s_\nu)v_{\nu_i}$ was just calculated in \eqref{crosssection} above. As discussed in chapters 4 and 5, besides for non-standard physics, the distribution function can be written as
	
	\begin{equation}
		f_{\nu_i} = f_0(1+\delta f_{\nu_i}^d +\delta f_{\nu_i}^c)
	\end{equation}

	for both left- and right-handed neutrinos if they are Majorana fermions, and for just left-handed neutrinos if they are Dirac particles. As a reminder, $f_0$ is the massless Fermi-Dirac distribution function, with $\delta f_{\nu_i}$ being the change to this distribution function owing to non-instantaneous decoupling and gravitational clustering.\\
	
	Before continuing, we need to consider the kinematics of the situation to see how the ejected electron's energy and momentum are related to the incoming neutrino's momentum. To begin, we start by analysing normal beta decay by a tritium nucleus : $^3\mathrm{H} \ \rightarrow \ ^3\mathrm{He} + e^- + \bar{\nu}_i$. In this case, the kinetic energy of the emitted electron, $K_e=E_e-m_e$, is given by 
	
	\begin{equation}
		K_e=\frac{(m_{\mathrm{^3H}}-m_e)^2-m_{\nu_i}^2-m_{\mathrm{^3He}}^2-2E_{\nu_i}E_{\mathrm{^3He}}+2|\bar{p}_{\nu}||\bar{p}_{\mathrm{^3He}}|\cos \theta_{\mathrm{^3He}\nu }}{2m_{\mathrm{^3H}}}.
	\end{equation}

	Recall that during beta decay, the emitted electron can take on many possible values, and has a characteristic spectrum - this was one of the first signs that a neutrino was needed in the standard model. The electron's possible energy does have a cutoff, however: the ``endpoint", the maximal possible kinetic energy that the electron can have, is achieved when the electron is emitted anti-parallel to the helium and anti-neutrino, and when the anti-neutrino emitted is in the lightest mass eigenstate. In this case then, $\cos \theta_{\mathrm{^3He}\nu } = 1$, $m_{\nu_i}=m_{\mathrm{lightest}}$, and the maximisation of $E_e$ with respect to $|\bar{p}_\nu|$ leads to a minimisation of $E_\nu+E_{^3\mathrm{He}}$ with respect to $|\bar{p}_\nu|$, which gives
	
	\begin{equation}
		\frac{|\bar{p}_\nu|}{|\bar{p}_{^3\mathrm{He}}|} = \frac{m_{\nu_i}}{m_{^3\mathrm{He}}}.
	\end{equation}

	Plugging these in, we have 
	
	\begin{equation}
		K_{\mathrm{end}} = \frac{(m_{\mathrm{^3H}}-m_e)^2-(m_{\rm lightest}+m_{\mathrm{^3He}})^2}{2m_{\mathrm{^3H}}},\label{Kend}
	\end{equation}
	
	independent of everything except masses. We can define $K_{\mathrm{end}}^0$, the maximum possible kinetic energy for the emitted electron if the lightest species of neutrinos is massless:
	
	\begin{equation}
		K_{\mathrm{end}}^0 = \frac{(m_{\mathrm{^3H}}-m_e)^2-m_{\mathrm{^3He}}^2}{2m_{\mathrm{^3H}}}. \label{Kend0}
	\end{equation}

	Next, let us move on to the reaction at hand: neutrino capture. In this case, the kinetic energy of the emitted electron is slightly altered, and is given by
	
	\begin{align}
		K_e^{\rm{C\nu B}}&=\frac{(E_{\nu_i}+m_{\mathrm{^3H}}-m_e)^2-|\bar{p}_{\nu}|^2+2|\bar{p}_{\nu}||\bar{p}_e|\cos \theta_{e\nu}-m_{\mathrm{^3He}}^2}{2(E_{\nu_i}+m_{\mathrm{^3H}})} \nonumber \\
		&\simeq \frac{(E_{\nu_i}+m_{\mathrm{^3H}}-m_e)^2-m_{\mathrm{^3He}}^2}{2(E_{\nu_i}+m_{\mathrm{^3H}})},
	\end{align}

	where in the second line we have neglected terms proportional to $|\bar{p}_{\nu}|^2$ and $2|\bar{p}_{\nu}||\bar{p}_e|$, as these terms are much smaller than the mass scale of tritium and helium. By similar logic, keeping only the leading order term in $E_{\nu_i}$, we have that 
	
	\begin{equation}
		K_e^{\rm{C\nu B}} \approxeq K_{\mathrm{end}}^0 + E_{\nu_i}.
	\end{equation}

	Thus, for the $E_e$ appearing in our capture rate, we have $E_e \approxeq m_e + K_{\mathrm{end}}^0 + E_{\nu_i}$, with $|\bar{p}_e| = \sqrt{E_e^2-m_e^2}$.
	
	In order to simplify our calculation, we can make the approximation 
	
	\begin{equation}
		E_e \approxeq m_e + K_{\mathrm{end}}^0 + m_{\nu_i},
	\end{equation}

	so that we can remove both $E_e$ and $|\bar{p}_e|$ from the integral over $p_{\nu}$. This approximation is certainly justified: for massive neutrinos, their masses are much larger than the characteristic momentum seen in \eqref{lowenergy}, and for massless neutrinos, this selfsame momentum in \eqref{lowenergy} is obviously much smaller than the energy scale of $m_e$.\\
	
	Going back to our capture rate then, we now have 
	
	\begin{multline}
			\Gamma_{\mathrm{C\nu B}} = N_T\sum_{i=1}^{3}\sum_{s_\nu=\pm 1/2}\frac{G_F^2}{2\pi}|V_{ud}|^2|U_{ei}|^2\frac{m_\mathrm{^3He}E_e|\bar{p}_e|}{m_\mathrm{^3H}}F(2,E_e)\left(\braket{f_F}^2 + \frac{g_A^2}{g_V^2}\braket{g_{GT}}^2 \right) \times \\
			 \int \frac{d^3p_\nu}{(2\pi)^3}f_0(1+\delta f_{\nu_i}^d +\delta f_{\nu_i}^c)\left(1-2s_{\nu}v_{\nu_i}\right),
	\end{multline}

	or rather
	
	\begin{multline}
		\Gamma_{\mathrm{C\nu B}} = N_T\sum_{i=1}^{3}\sum_{s_\nu=\pm 1/2}\frac{G_F^2}{2\pi}|V_{ud}|^2|U_{ei}|^2\frac{m_\mathrm{^3He}E_e|\bar{p}_e|}{m_\mathrm{^3H}}F(2,E_e)\left(\braket{f_F}^2 + \frac{g_A^2}{g_V^2}\braket{g_{GT}}^2 \right)\times \\
		(n_0+\delta n_{\nu_i} -2s_\nu n_0 \braket{v_{\nu_i}} - 2s_\nu\braket{\delta v_{\nu_i}}),\label{capturerate}
	\end{multline}

	with $n_0$ given by \eqref{n0}, $\delta n_{\nu_i} = \delta n_{\nu_i}^d + \delta n_{\nu_i}^c$ given by \eqref{numdens} and whose numerical values were seen in chapter 5, $\braket{v_{\nu_i}}$ the normalised expectation value of the neutrino's momentum using the massless Fermi-Dirac distribution, and 
	
	\begin{equation}
		\braket{\delta v_{\nu_i}} = \int\frac{d^3p_\nu}{(2\pi)^3} f_0(\bar{p}_\nu)(\delta f_{\nu_i}^d + \delta f_{\nu_i}^c)v_{\nu_i}
	\end{equation}
	
	the unnormalised expectation of the change to the velocity owing to the clustering and spectral distortive effects. 
	
	Some interesting quantities to seperate are 
	
\begin{align}
	\delta \Gamma_i^d&=N_T\frac{G_F^2}{2\pi}|V_{ud}|^2|U_{ei}|^2\frac{m_\mathrm{^3He}}{m_\mathrm{^3H}}\left(\langle f_F \rangle^2+\frac{g_A^2}{g_V^2}\langle g_{GT} \rangle^2 \right)F(2,E_e)E_e|\bar{p}_e| 
	\sum_{s_{\nu}=\pm\frac{1}{2}}( \delta n_{\nu_i}^d  -2s_\nu \langle \delta v_{\nu_i}^d \rangle),  \nonumber \\
	\delta \Gamma_i^c&=N_T\frac{G_F^2}{2\pi}|V_{ud}|^2|U_{ei}|^2\frac{m_\mathrm{^3He}}{m_\mathrm{^3H}} \left(\langle f_F \rangle^2+\frac{g_A^2}{g_V^2}\langle g_{GT} \rangle^2 \right)F(2,E_e)E_e|\bar{p}_e|
	\sum_{s_{\nu}=\pm\frac{1}{2}}( \delta n_{\nu_i}^c  -2s_\nu \langle \delta v_{\nu_i}^c \rangle),
\end{align}

	the changes to the capture rate for each species $i$ owing to the effects of spectral distortion due to non-instantaneous decoupling and gravitational clustering.
	
	\subsection{Numerical Results}
	
	As we have seen, particularly in this chapter and the discussions in chapter 5, there are many factors that play a role in the expected capture rate of relic neutrinos. The two main factors, making the largest difference to the situation are the choice of hierarchy (normal or inverted) and the type of fermion (Dirac or Majorana). Thus, this is how our results will be divided, with other smaller effects being considered within each category. In this section, we consider only standard physics, with the added effects of new physics being covered in a later subsection.\\
	
	The numerical values calculated were done so using both Python and Matlab, and the values arrived at were in agreement to the accuracy stated below. Below are the values used for the numerical calculations that were not already stated.
	
	\begin{center}
		\begin{tabular}{||c|c||}
			$m_{^3\mathrm{H}}$ & $2.808921\times 10^9$ eV \cite{Meng2017}\\
			$m^{\mathrm{atom}}_{^3\mathrm{H}}$ & $2.809432\times 10^9$ eV \cite{Meng2017}\\
			$m_{^3\mathrm{He}}$ & $2.808391\times 10^9$ eV \cite{Meng2017}\\
			$m_{e}$ & $5.1099891\times 10^5$ eV \cite{Beringer:1900zz}\\
			$\braket{f_F}$ & 0.9998 \cite{Baroni:2016xll}\\
			$\braket{g_{GT}}$ & $\sqrt{3}\times (0.9511\pm0.0013)$ \cite{Baroni:2016xll}\\
			$g_A$ & 1.2723 \cite{Baroni:2016xll}\\
			$g_V$ & 1 \cite{Baroni:2016xll}\\
			$V_{ud}$ & 0.974 \cite{Zyla:2020zbs}\\
			$|U_{ej}|^2$ & [0.681 , 0.297 , 0.0222]  \cite{Esteban:2020cvm}\\
		\end{tabular}
	\end{center}

	Using these values, we can calculate a characteristic cross section for this reaction following the lead in \cite{Long:2014zva}:
	
	\begin{equation}
		\bar{\sigma} = \frac{G_F^2}{2\pi}|V_{ud}|^2\frac{m_\mathrm{^3He}E_e|\bar{p}_e|}{m_\mathrm{^3H}}F(2,E_e)\left(\braket{f_F}^2 + \frac{g_A^2}{g_V^2}\braket{g_{GT}}^2 \right) \approxeq 3.835 \times 10^{-45} \mathrm{cm}^2. \label{charcrosssec}
	\end{equation}

	For our numerical results we considered a tritium target weighing 100 grams, in line with the planned project PTOLEMY. From the mass differences seen in \eqref{massdiff} and the constraint in \eqref{massupperlim}, the possible masses for neutrinos are limited but not fixed. For the sake of this calculation, for both the normal and inverted hierarchies, the minimal allowed masses were used: that is, the lightest species was taken to be massless.
	
	This was done for two reasons. Firstly, it allows us to analyse relic neutrinos that are still relativistic today, and secondly, it is the only way that the effects of non-instantaneous decoupling could be observed behind the shroud of gravitational clustering, as we saw in chapter 5.
	
	Finally, before stating the results, for the sake of the calculation the $\braket{\delta v_{\nu_i}}$ terms were neglected. Working to next-to leading order accuracy, both the velocity and effects altering the distribution function (non-instantaneous decoupling and gravitational clustering) were included, but their product would be an even smaller correction, akin to annual modulation, helicity flipping and anistropy, already neglected above.\\
	
	For the normal ordering case, with masses given by \eqref{normalorder}, for the case where neutrinos are Majorana particles, we have the total capture rate for each massive species $\Gamma_i^M$, its deviation originating from the spectral distortion from non-instantaneous decoupling $\delta \Gamma_i^{Md}$, and that from the gravitational clustering effects $\delta \Gamma_i^{Mc}$, given by (considering $100$ g of tritium)
	\begin{align}
		\Gamma_1^M\simeq 5.48\ {\rm yr^{-1}},  \ \ \ \  \Gamma_2^M\simeq 2.40\ {\rm yr^{-1}},  \ \ \ \   \Gamma_3^M\simeq 0.200 \ {\rm yr^{-1}},
		\label{ValueM}
	\end{align}
	\begin{align}
		\delta \Gamma_1^{Md}\simeq 0.061\ {\rm yr^{-1}},  \ \ \ \  \delta \Gamma_2^{Md}\simeq 0.024\ {\rm yr^{-1}},  \ \ \ \  \delta \Gamma_3^{Md}\simeq 1.6 \times 10^{-3} \ {\rm yr^{-1}},
		\label{DGMd}
	\end{align}
	\begin{align}
		\delta \Gamma_1^{Mc}\simeq 0\ {\rm yr^{-1}},  \ \ \ \  \delta \Gamma_2^{Mc}\simeq 0.013\ {\rm yr^{-1}},  \ \ \ \  \delta \Gamma_3^{Mc}\simeq 0.021 \ {\rm yr^{-1}},
		\label{DGMc}
	\end{align}

	and therefore $\Gamma_{C\nu B}^M\approxeq 8.08{\rm yr^{-1}}$. 
	
	Similarly for the Dirac case, we have 
	
	\begin{align}
		\Gamma_1^D\simeq 5.48\ {\rm yr^{-1}},  \ \ \ \ \Gamma_2^D\simeq 1.27\ {\rm yr^{-1}},  \ \ \ \  \Gamma_3^D\simeq 0.101 \ {\rm yr^{-1}}, 
	\end{align}
	\begin{align}
		\delta \Gamma_1^{Dd}\simeq 0.061\ {\rm yr^{-1}},  \ \ \ \  \delta \Gamma_2^{Dd}\simeq 0.012\ {\rm yr^{-1}},  \ \ \ \   \delta \Gamma_3^{Dd}\simeq 8.0 \times 10^{-4} \ {\rm yr^{-1}},
	\end{align}
	\begin{align}
		\delta \Gamma_1^{Dc}\simeq 0\ {\rm yr^{-1}},  \ \ \ \  \delta \Gamma_2^{Dc}\simeq 6.3 \times 10^{-3}\ {\rm yr^{-1}},  \ \ \ \   \delta \Gamma_3^{Dc}\simeq 0.011\ {\rm yr^{-1}},
	\end{align}

	and so $\Gamma_{C\nu B}^D\approxeq 6.851{\rm yr^{-1}}$.

	It is interesting to note that the ratios between the capture rates for Dirac and Majorana neutrinos are
	\begin{align}
		\Gamma_1^M/\Gamma_1^D=1, \ \ \ \ \Gamma_2^M/\Gamma_2^D\simeq 1.89, \ \ \ \ \Gamma_3^M/\Gamma_3^D\simeq 1.98,
	\end{align}
	
	as expected. For relativistic neutrinos, owing to the factor of $1-2s_\nu v_\nu$, even in the Majorana case where right-helical states contain some left-chiral component, and therefore can interact, right handed neutrinos would not contribute as $1-2s_\nu v_\nu=0$ when the spin is $\frac{1}{2}$ and the velocity is 1.
	
	For non-relativistic (massive) neutrinos, this factor of approximately 2 (altered slightly by clustering and velocity effects) is of course owing to the fact that for Dirac neutrinos only left-helical neutrinos can interact with the tritium, whereas for Majorana neutrinos both left- and right-helical neutrinos can be captured.
	
	It is interesting to note that in the Majorana case, the capture rate's mass dependence only comes in in the clustering effect - the only other term that the neutrino mass affects is the velocity term, which cancels out in the Majorana case as the left- and right-helical components are summed over.\\
	
	For the inverted hierarchy case, with masses given by \eqref{invertedorder}, once again we have the total capture rate for each massive species $\Gamma_i^M$, its deviation originating from the spectral distortion from non-instantaneous decoupling $\delta \Gamma_i^{Md}$, and that from the gravitational clustering effects $\delta \Gamma_i^{Mc}$, given by (considering $100$ g of tritium)
	
	\begin{align}
		\Gamma_1^M\simeq 6.13\ {\rm yr^{-1}},  \ \ \ \  \Gamma_2^M\simeq 2.67\ {\rm yr^{-1}},  \ \ \ \   \Gamma_3^M\simeq 0.178 \ {\rm yr^{-1}},
	\end{align}
	\begin{align}
		\delta \Gamma_1^{Md}\simeq 0.061\ {\rm yr^{-1}},  \ \ \ \  \delta \Gamma_2^{Md}\simeq 0.024\ {\rm yr^{-1}},  \ \ \ \  \delta \Gamma_3^{Md}\simeq 1.6 \times 10^{-3} \ {\rm yr^{-1}},
	\end{align}
	\begin{align}
		\delta \Gamma_1^{Mc}\simeq 0.65\ {\rm yr^{-1}},  \ \ \ \  \delta \Gamma_2^{Mc}\simeq 0.28\ {\rm yr^{-1}},  \ \ \ \  \delta \Gamma_3^{Mc}\simeq 0 \ {\rm yr^{-1}},
	\end{align}
	
	and so $\Gamma_{C\nu B}^M\approxeq 8.978{\rm yr^{-1}}$. Note that we take the same values of $\delta n_1^c$ for $m_1=49.3\ {\rm meV}$ and $\delta n_2^c$ for $m_2=50\ {\rm meV}$ \cite{Akita:2020jbo}.
	
	Next, for the case of Dirac neutrinos, we have
	\begin{align}
		\Gamma_1^D\simeq 3.10\ {\rm yr^{-1}},  \ \ \ \ \Gamma_2^D\simeq 1.35\ {\rm yr^{-1}},  \ \ \ \  \Gamma_3^D\simeq 0.178 \ {\rm yr^{-1}},
	\end{align}
	\begin{align}
		\delta \Gamma_1^{Dd}\simeq 0.031\ {\rm yr^{-1}},  \ \ \ \  \delta \Gamma_2^{Dd}\simeq 0.012\ {\rm yr^{-1}},  \ \ \ \   \delta \Gamma_3^{Dd}\simeq 1.6 \times 10^{-3}\ {\rm yr^{-1}},
	\end{align}
	\begin{align}
		\delta \Gamma_1^{Dc}\simeq 0.33\ {\rm yr^{-1}},  \ \ \ \  \delta \Gamma_2^{Dc}\simeq 0.14\ {\rm yr^{-1}},  \ \ \ \   \delta \Gamma_3^{Dc}\simeq 0\ {\rm yr^{-1}}.
	\end{align}
	Once again, we see the ratios between the capture rates of each species for the Dirac and Majorana cases are
	\begin{align}
		\Gamma_1^M/\Gamma_1^D \simeq 1.98, \ \ \ \ \Gamma_2^M/\Gamma_2^D\simeq 1.98, \ \ \ \ \Gamma_3^M/\Gamma_3^D = 1,
	\end{align}
	
	again displaying unity for relativistic species and approximately 2 for the massive species.
	
	As stated in chapter 5, we could also consider helicity flipping. In the case of complete helicity flipping, which in the Dirac case entails half of the left-helical population to become right-helical, at leading order this does not make a difference as helicities are summed over. However, the next-to leading order term which involves the velocity now cancels out. In the relativistic case, this makes no difference as the $\sum_{s_\nu=\pm 1/2}(1-2s_\nu v_{\nu_i})$ factor is equal to 2 either way. In the non-relativistic case however, the small correction of $v_{\nu_i}$ is lost. In a more realistic case, where there is only partial helicity flipping, and only clustered neutrinos' helicities being flipped, only a fraction of this will be lost, changing the capture rate at the order $\approx10^{-4} - 10^{-3} \mathrm{yr}^{-1}$.

\begin{figure}[htbp]
	\begin{minipage}{0.5\hsize}
		\begin{center}
			\includegraphics[width=85mm]{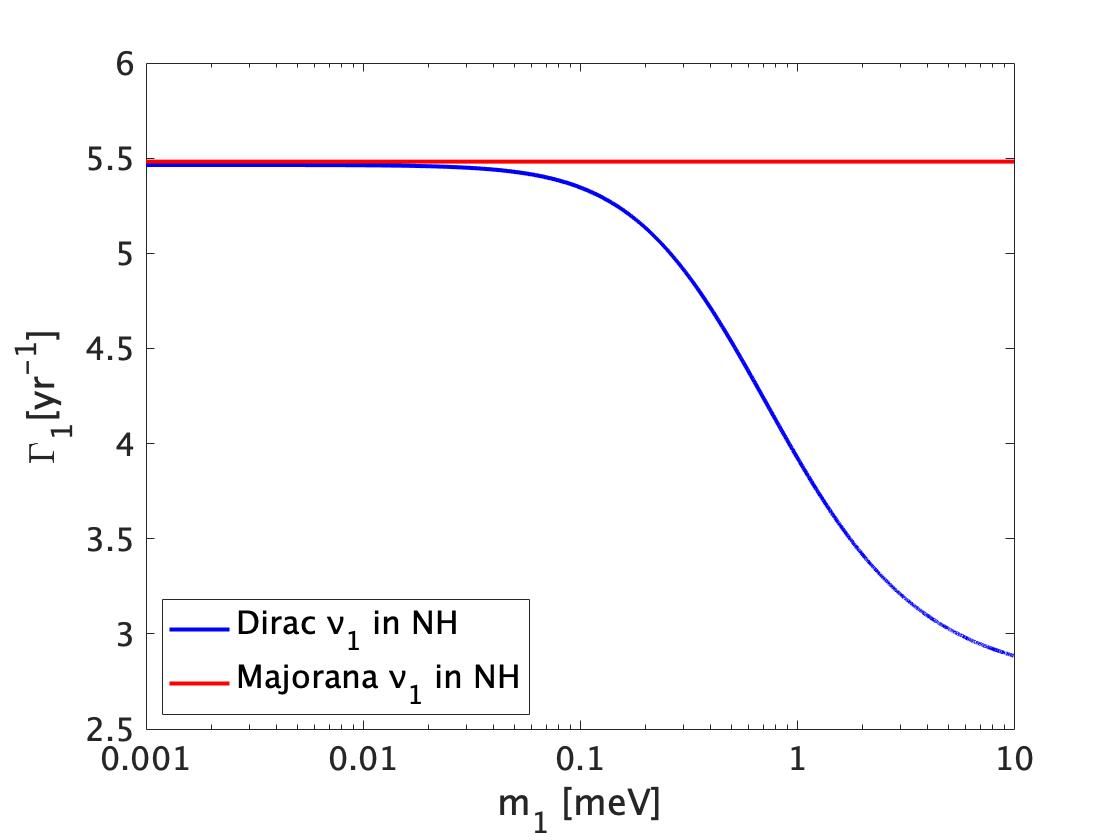}
		\end{center}
	\end{minipage}
	\begin{minipage}{0.5\hsize}
		\begin{center}
			\includegraphics[width=85mm]{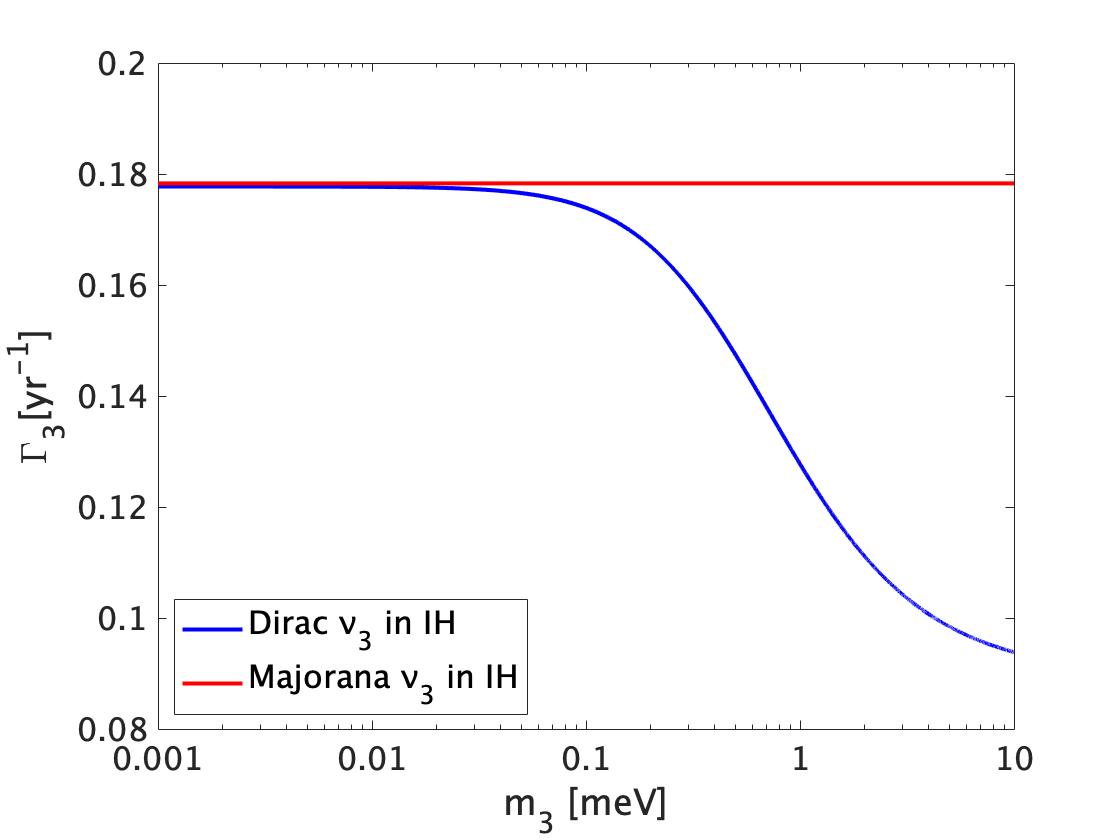}
		\end{center}
	\end{minipage}
	\vspace{-4mm}
	\caption{The relationship between the capture rate and the mass of the lightest neutrino, with Dirac cases shown in blue and Majorana in red. The left panel depicts the normal hierarchy case (that is, the lightest neutrino is $m_1$) and the right panel is for the inverted hierarchy case (where $m_3$ is the smallest mass). Taken from \cite{Akita:2020jbo}.}
	\label{fig:Gammanu}
\end{figure}

	As stated already, the mass dependence of this capture rate is tied to its velocity dependence, which cancels out in the Majorana case. We can see in figure \ref{fig:Gammanu} how the capture rate is affected in the Dirac case for the lightest species, with the Majorana capture rate as a benchmark.

	\subsection{Beta Decay Rate's Energy Dependence}
	
	The main source of background noise in our chosen experiment is the regular beta decay of the tritium nuclei. In order to see how the spectrum of this noise will look, we consider the three body decay process of tritium into helium-3, an electron and an anti-neutrino:
	\begin{equation}
		^3\mathrm{H}(M,0)\rightarrow\ ^3\mathrm{He}(E',\bar{p}') + e^-(E_e,\bar{p}_e) + \bar{\nu}_e(E_\nu,\bar{p}_\nu).
	\end{equation}
	We wish to analyse the dependence of the decay rate $\Gamma$ on the electrons energy: $\frac{d\Gamma}{dE_e}$. To do so, we begin with the differential decay rate \cite{Schwartzbook}:
	\begin{equation}
		d\Gamma = \frac{1}{2E_i}|\mathcal{M}|^2d\Pi_{LIPS},
	\end{equation}
	where $E_i=M$, $|\mathcal{M}|^2$ is the squared matrix element, and $d\Pi_{LIPS}$ is the Lorentz invariant phase space. Expanding this yields
	
	\begin{equation}
		d\Gamma = \frac{1}{2^9\pi^5M}\int\frac{d^3p'd^3p_ed^3p_\nu}{E'E_eE_\nu}|\mathcal{M}|^2\delta^4(p_H-p_{He}-p_e-p_\nu).
	\end{equation}
	
	We can then use the 3 dimensional delta function to kill the $d^3p'$ integral. This will leave $E'=\sqrt{(\bar{p}_e+\bar{p}_\nu)^2+M'^2}=\sqrt{\bar{p}_e^2+\bar{p}_\nu^2+2|\bar{p}_e||\bar{p}_\nu|\cos\theta+M'^2}$, where $M'$ is the mass of the helium-3.
	
	The remaining delta function is used to remove one angular integral. We have that
	\begin{equation}
		\delta(M-E'-E_e-E_\nu)=\delta(M-E_e-E_\nu-(\bar{p}_e^2+\bar{p}_\nu^2+2|\bar{p}_e||\bar{p}_\nu|\cos\theta+M'^2)^{1/2}).
	\end{equation}
	We then use the property of delta functions that $\delta(f(x))=\frac{\delta(x-x*)}{f'(x*)}$, for every root $x*$ of $f(x)$ to obtain
	\begin{equation}
		\delta(M-E'-E_e-E_\nu)=\delta(\cos\theta-\frac{(M-E_e-E_\nu)^2-\bar{p}_e^2-\bar{p}_\nu^2-M'^2}{2|\bar{p}_e||\bar{p}_\nu|})\frac{M-E_e-E_\nu}{|\bar{p}_e||\bar{p}_\nu|}.
	\end{equation}
	
	From here we will use the notation that $p_i=\bar{p}_i$, unless otherwise specified. We have from the above energy delta function that $E'=M-E_e-E_\nu$. Plugging this in and doing the remaining 3 trivial angular integrals yields
	\begin{equation}
		d\Gamma = \frac{2\cdot2\pi\cdot4\pi}{2^9\pi^5M}\int\frac{dp_edp_\nu}{E_eE_\nu(M-E_e-E_\nu)}\frac{p_e^2p_\nu^2(M-E_e-E_\nu)}{p_ep_\nu}|\mathcal{M}|^2.
	\end{equation}
	We now change our variable of integration from momentum to energy, and since $E_i^2=p_i^2+m_i^2$, $p_idp_i=E_idE_i$ and we have that
	\begin{equation}
		d\Gamma = \frac{1}{(2\pi)^38M}\int dE_edE_\nu|\mathcal{M}|^2.
	\end{equation}
	\\
	Next we consider the form of $|\mathcal{M}|^2$. Using symmetry, the general form, up to momenta squared, is given by
	\begin{equation}
		|\mathcal{M}|^2=A+Bp_ep_\nu+Cp_Hp_{He}=A+B(E_eE_\nu-\bar{p}_e\cdot\bar{p}_\nu)+CM(M-E_e-E_\nu),
	\end{equation}
	
	where in the above we used 4-momenta and used bars to clearly denote 3-momenta. We also used the fact that $E'=M-E_e-E_\nu$.
	
	We can use 4-momentum conservation, $p_H-p_{He}=p_e+p_\nu$ to rewrite $\bar{p}_e\cdot\bar{p}_\nu$. We have that
	\begin{equation}
		(p_H-p_{He})^2=(p_e+p_\nu)^2 \rightarrow M^2+M'^2-2M(M-E_e-E_\nu) = m_e^2+m_\nu^2+2E_eE_\nu-2\bar{p}_e\cdot\bar{p}_\nu,
	\end{equation}
	and so
	\begin{equation}
		\bar{p}_e\cdot\bar{p}_\nu = \frac{1}{2}(M^2-M'^2+m_e^2+m_\nu^2-2ME_e+2E_\nu E_e-2ME_\nu),
	\end{equation}
	and thus we will insert into the $d\Gamma$ integral that
	\begin{equation}
		|\mathcal{M}|^2 = A+\frac{B}{2}(M'^2-M^2-m_e^2-m_\nu^2+2ME_e+2ME_\nu)+CM(M-E_e-E_\nu).
	\end{equation}
	
	Now recall that we want an expression for $\frac{d\Gamma}{dE_e}$. So we have now that
	\begin{equation}
		\frac{d\Gamma}{dE_e}= \frac{1}{(2\pi)^38M}\int_{E_\nu^{min}}^{E_\nu^{max}}dE_\nu\left[A+\frac{B}{2}(M'^2-M^2-m_e^2-m_\nu^2+2ME_e+2ME_\nu)+CM(M-E_e-E_\nu)\right].
	\end{equation}
	
	This integration is trivial and we obtain 
	
	\begin{multline}
		\frac{d\Gamma}{dE_e} = \frac{1}{(2\pi)^38M}[(A+\frac{B}{2}(M'^2-M^2-m_e^2-m_\nu^2+2ME_e)+CM(M-E_e))(E_\nu^{max}-E_\nu^{min})\\
		+M(B-C)(E_\nu^{max}-E_\nu^{min})(E_\nu^{max}+E_\nu^{min})]
	\end{multline}
	
	where we have used that $(E_\nu^{max})^2-(E_\nu^{min})^2=(E_\nu^{max}-E_\nu^{min})(E_\nu^{max}+E_\nu^{min})$.
	
	Now, following the notation used in \cite{Eidelman:2004wy} and \cite{Masood:2007rc}, we define
	\begin{equation}
		m_{ij}^2=(p_i+p_j)^2,
	\end{equation}
	where $p_i$ is the 4-momentum of a decay product. Naming $E_1=E_\nu$, $E_2=E'$ and $E_3=E_e$, specifically, we have 
	\begin{equation}
		m_{12}^2=(p'^2+p_\nu^2)^2=(p_H-p_e^2)^2=M^2+m_e^2-2ME_e.
	\end{equation}
	
	Similarly, we have $m_{23}^2=M^2+m_\nu^2-2ME_\nu$, and so the maximisation/minimisation of $m_{23}^2$ corresponds with that of $E_\nu$, as the masses are constant.
	Expanding $m_{23}^2=(p'+p_e)^2$, we have that it is maximised/minimised when $\bar{p}'$ is parallel/anti-parallel to $\bar{p}_e$.
	
	This leads to the maximisation/minimisation conditions:
	\begin{equation}
		m_{23}^{2max/min}=(E'^*+E_e^*)^2-(\sqrt{E'^{*2}-M'^2}\mp\sqrt{E_e^{*2}-m_e^2})^2,
	\end{equation}
	
	where $E^*$ is the energy of each particle in the rest frame of $m_{12}$. For example,
	\begin{equation}
		(m_{12}-p_2)^2=m_{12}^2-2m_{12}E^{*}_2+m_2^2=m_1^2\ \ \ \rightarrow \ \ \ E^*_2=E'^*=\frac{m_{12}^2-m_\nu^2+M'^2}{2m_{12}}.
	\end{equation}
	
	Similarly $E_e^*=\frac{M^2-m_{12}^2-m_e^2}{2m_{12}}$.
	
	Adding and subtracting $m_{23}^{2max}$ and $m_{23}^{2min}$ gives 2 factorisations:
	
	\begin{equation}
		E_\nu^{max}-E_\nu^{min}=\frac{2}{M}\sqrt{(E'^{*2}-M'^2)(E_e^{*2}-m_e^2)},
	\end{equation}
	
	\begin{equation}
		E_\nu^{max}+E_\nu^{min}=\frac{1}{M}(M^2+m_\nu^2-(E'^*+E_e^*)^2+E'^{*2}-M'^2+E_e^{*2}-m_e^2).
	\end{equation}
	
	After plugging in the expressions for each $E^*$, and defining $E_e^{max}=\frac{1}{2M}(M^2+m_e^2-(m_\nu+M')^2)$ as the maximum electron energy, we are left with
	
	\begin{equation}
		E_\nu^{max}-E_\nu^{min}=\frac{2Mp_e}{m_{12}^2}\sqrt{(E_e^{max}-E_e)(E_e^{max}-E_e+\frac{2m_\nu M'}{M})},
	\end{equation} 
	
	\begin{equation}
		E_\nu^{max}+E_\nu^{min}=\frac{2M}{m_{12}^2}(M-E_e)(E_e^{max}-E_e+\frac{m_\nu}{M}(M'+m_\nu)).
	\end{equation}

	In the 4-Fermi approximation, we saw previously that $|\mathcal{M}|^2$ is proportional to $E_eE_\nu$ with a constant of order $G_F^2$. Thus, we can set $A=C=0$ and we get
	\begin{multline}
		\frac{d\Gamma}{dE_e} = \frac{B}{(2\pi)^3}\frac{p_e}{4m_{12}^2}\sqrt{y(y+\frac{2m_\nu M'}{M})}(\frac{M^2}{m_{12}^2}(M-E_e)(y+\frac{m_\nu}{M}(M'+m_\nu))\\
		-\frac{1}{2}(M^2-M'^2+m_e^2+m_\nu^2-2ME_e))
	\end{multline}
	
	wherein $y=E_e^{max}-E_e$. We then perform some algebra on the last bracket above:
	\begin{equation}
		M^2-M'^2+m_e^2+m_\nu^2-2ME_e=-M^2-m_e^2+(m_\nu+M')^2-2m_\nu^2-2M'm_\nu+2ME_e = 2M(-E_e^{max}+E_e-\frac{m_\nu^2+M'm_\nu}{M}),
	\end{equation}
	and, plugging this back in and using the definition of $y$, we add the 2 terms and we obtain
	\begin{equation}
		\frac{d\Gamma}{dE_e} = \frac{B}{(2\pi)^3}\frac{Mp_e}{4m_{12}^4}\sqrt{y(y+\frac{2m_\nu M'}{M})}(ME_e-m_e^2)(y+\frac{m_\nu}{M}(M'+m_\nu)).
	\end{equation}
	
	Taking out some factors and cancelling, we arrive at the form
	\begin{equation}
		\frac{d\Gamma}{dE_e} = B'E_ep_e\frac{1-\frac{m_e^2}{E_eM}}{(1-\frac{2E_e}{M}+\frac{m_e^2}{M^2})^2}\sqrt{y(y+\frac{2m_\nu M'}{M})}(y+\frac{m_\nu}{M}(M'+m_\nu))
	\end{equation}
	
	and, in defining $H(E_e,m_\nu)$, we have $\frac{d\Gamma}{dE_e} = B'E_ep_eH(E_e,m_\nu)$.
	
	Now, we know that $B'$ must be of order $G_F^2$ as it is a Fermi interaction, we know it must have a factor $|V_{ud}|^2$ and $|U_{ej}|^2$ for each species in the mass basis and it must be multiplied by $N_T$ for the number of beta-decaying sites. Adding the small correction factors of Fermi and Gamow-Teller transition probabilities owing to the tritium and helium-3, a factor of $\frac{M'}{M}$ from the kinematics, as well as a Fermi function to account for the Coulombic attraction of the outgoing electron to the nucleus, we are left with
	
	\begin{equation}
		\frac{d\Gamma}{dE_e} = \sum_{j=1}^{3}|U_{ej}|^2\frac{\bar{\sigma}}{\pi^2}N_TH(E_e,m_\nu),\label{betadec}
	\end{equation}
	
	where $\bar{\sigma}$ is defined as in \eqref{charcrosssec} as 
	\begin{equation}
		\bar{\sigma} = \frac{G_F^2}{2\pi}|V_{ud}|^2F(Z,E_e)\frac{M'}{M}E_ep_e (\braket{f_F}^2+(\frac{g_A}{g_V})^2\braket{g_{GT}}^2),
	\end{equation} 
	which is obtained when computing the cross section of the 4-Fermi interaction. Of course, we must remember that this expression is not perfectly relativistic, as even though $H(E_e,m_\nu)$ is, this $\bar{\sigma}$ comes from an effective Lagrangian.
	
	\subsection{The Expected Energy Spectra}
	
	From the previous subsections, we can now model both neutrino capture and beta decay and see graphically how they may look in an experiment like PTOLEMY. In this experiment, recall that it is the emitted electron that is observed, and its energy is what signals a C$\nu$B interaction. As we will see below, one of the main challenges for observing the C$\nu$B is the distinction of the signal from the background noise that is the constant supply of electrons from usual beta decay.\\
	
	Recall from the kinematical discussions above that the maximum possible energy that an electron can have as a result of beta decay - the ``beta decay endpoint" - is 
	
	\begin{equation}
		E_{\mathrm{end}}\approxeq K_{\mathrm{end}}^0+m_e-m_{\mathrm{lightest}}.\label{Eend}
	\end{equation}

	A more accurate value is given by $m_e+K_{\mathrm{end}}$, taken from \eqref{Kend}, but \eqref{Eend} is correct to linear order and since $m_{\mathrm{lightest}}\ll m_e$, this approximation is valid. \\
	
	On the other hand, when an electron is emitted owing to the tritium capturing a relic neutrino, we have
	
	\begin{equation}
		E^{\mathrm{C\nu B}}\approxeq K_{\mathrm{end}}^0+m_e +E_{\nu_i},
	\end{equation}
	
	which is clearly always larger that \eqref{Eend}. Again, this was approximated to linear order as above. What is important is that the difference in energy between the two is
	
	\begin{equation}
		\Delta E_e = E^{\mathrm{C\nu B}} - E_{\mathrm{end}} \approxeq E_{\nu_i} + m_{\mathrm{lightest}}.
	\end{equation}

	Thus, the ``smoking gun" characteristic signal of detections of the C$\nu$B would be an emitted electron energy spectrum with peaks a distance $\Delta E_e$ away from the beta decay endpoint.\\
	
	From \eqref{betadec} and its derivation, we know the spectrum of usual beta decay, and from \eqref{capturerate} we know the spectrum of electrons emitted owing to C$\nu$B interactions. Unfortunately, in reality, the energy resolution of the detector is not perfect, and can only measure the electron's energy to a certain accuracy. In order to take this experimental fact into account, instead of plotting the theoretical, exact spectra, we must rather plot Gaussian-smeared versions of them, with the FWHM (full width at half max - the spread of the Gaussian at the half-way point of its height) $\Delta$, where $\Delta$ is the energy resolution of the detector. These Gaussian-smeared spectra are given by \cite{Long:2014zva}
	
	\begin{align}
		\frac{d\tilde{\Gamma}_i}{dE_e} &=\frac{1}{\sqrt{2\pi}\sigma}\int^{\infty}_{-\infty}dE_e'\  \Gamma_i(E_e')\ \delta[E_e'-(E_{\rm end}+E_{\nu_i}+m_{\rm lightest})]\exp\left[-\frac{(E_e'-E_e)^2}{2\sigma^2} \right], 
		\label{tildeGi}
		\\
		\frac{d\tilde{\Gamma}_{\beta}}{dE_e}&=\frac{1}{\sqrt{2\pi}\sigma}\int^{\infty}_{-\infty}dE_e' \ \frac{d \Gamma_{\beta}}{dE_e}(E_e')\ \exp\left[-\frac{(E_e'-E_e)^2}{2\sigma^2} \right]\label{tildebeta},
	\end{align}

	with $\Gamma_i(E_e')$ from \eqref{capturerate} and $\frac{d \Gamma_{\beta}}{dE_e}(E_e')$ from \eqref{betadec}. Here $\sigma$ is the standard deviation of the Gaussian (not a cross section), and can be related to $\Delta$.
	
	To see how, consider a normalised continuous Gaussian distribution with standard deviation $\sigma$, given by $y=\frac{1}{\sqrt{2\pi}\sigma}e^{-x^2/2\sigma^2}$, with $y$ its height and $x$ half its width. Inverting this relation yields $x=\sigma\sqrt{-\ln(2\pi\sigma^2y^2)}$. At the maximum (when $x=0$), $y=\frac{1}{\sqrt{2\pi}\sigma}$ so at half this maximum $y=\frac{1}{2\sqrt{2\pi}\sigma}$. At this height, the full width $\Delta$ is given by twice $x$, so 
	
	\begin{equation}
		\Delta = 2x|_{\mathrm{HM}} = 2\sigma\sqrt{-\ln(\frac{2\pi\sigma^2}{8\pi\sigma^2})} = \sigma\sqrt{8\ln(2)}.
	\end{equation}
	
\begin{figure}[htbp]
		\begin{center}
			\includegraphics[width=0.7\linewidth]{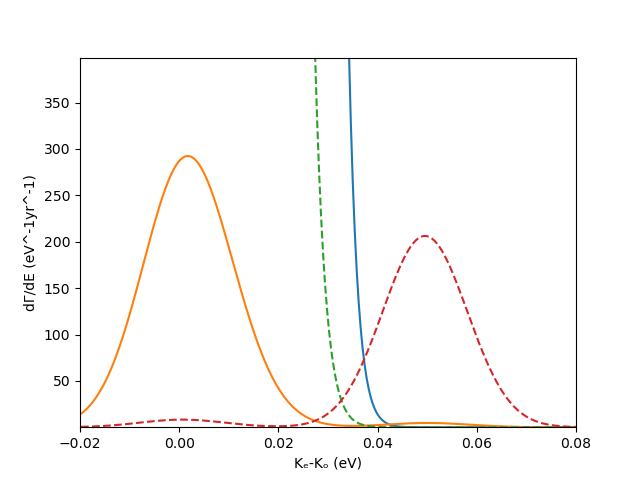}
		\end{center}
		\caption{The expected spectra for both the normal and inverted hierarchies for the case wherein neutrinos are Dirac particles, and the energy resolution of the detector is 20 meV - a realistic but perhaps optimistic figure. The solid lines show the normal ordering case, with orange being the C$\nu$B spectrum and blue being the background noise from the tritium beta decay. Similarly the dashed lines are the inverted hierarchy case, with red the C$\nu$B spectrum and green the background noise from the tritium beta decay. The vertical axis describes the capture rate for each energy while the horizontal axis denotes the emitted electron's kinetic energy $K_e$, less the kinetic energy of the endpoint of beta decay for the case of the lightest neutrino being massless $K_0$, which is denoted $K_{\mathrm{end}}^0$ in the body of this work - see \eqref{Kend0}. The lightest mass neutrino was chosen in both cases to be massless, and the values shown are for the case of a 100g tritium experiment. Plotted using Python.}
		\label{fig:Spectrum20}
	\end{figure}

\begin{figure}
		\begin{center}
			\includegraphics[width=0.7\linewidth]{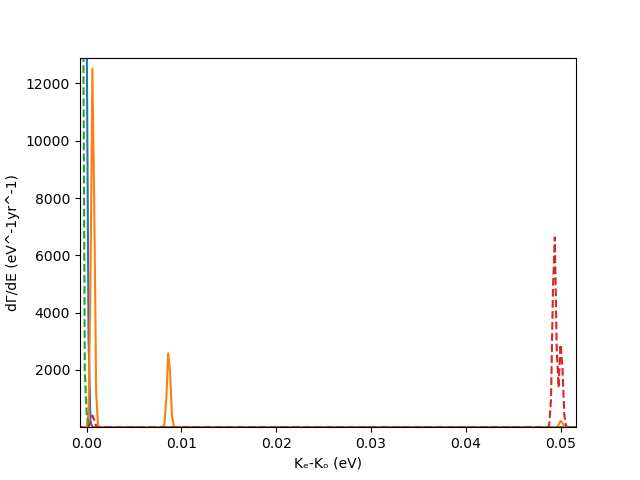}
		\end{center}
		\caption{The expected spectra for both the normal and inverted hierarchies for the case wherein neutrinos are Dirac particles, and the energy resolution of the detector is 0.4 meV - the maximum necessary to fully resolve all C$\nu$B signals. The solid lines show the normal ordering case, with orange being the C$\nu$B spectrum and blue being the background noise from the tritium beta decay. Similarly the dashed lines are the inverted hierarchy case, with red the C$\nu$B spectrum and green the background noise from the tritium beta decay. The vertical axis describes the capture rate for each energy while the horizontal axis denotes the emitted electron's kinetic energy $K_e$, less the kinetic energy of the endpoint of beta decay for the case of the lightest neutrino being massless $K_0$, which is denoted $K_{\mathrm{end}}^0$ in the body of this work - see \eqref{Kend0}. The lightest mass neutrino was chosen in both cases to be massless, and the values shown are for the case of a 100g tritium experiment. Plotted using Python.}
		\label{fig:Spectrum04}
	\end{figure}

	The results of these smeared spectra \eqref{tildeGi} and \eqref{tildebeta} are seen in figures \ref{fig:Spectrum20} and \ref{fig:Spectrum04} - this is what we would actually expect to observe.
	
	These figures are plotted for two specific values of the energy resolution - 20 meV, a realistic  but perhaps optimistic figure, and 0.4 meV - the threshold resolution necessary to fully resolve all C$\nu$B signals, as we shall see.  The beta decay background near the end point is larger in the normal hierarchy case than that in the inverted hierarchy case since the lightest neutrinos contribute to the endpoint, proportionally to $|U_{ei}|^2$, and $|U_{ei}|^2$ is obviously larger in the normal ordering case, for when $i=1$.
	
	In order to see how small a resolution is needed, we need to consider the capture rate within the energy range $\Delta$ around a C$\nu$B signal's $E^{\mathrm{C\nu B}}\approxeq K_{\mathrm{end}}^0+m_e +E_{\nu_i}$ for both the signal and the background noise:
	
	\begin{align}
		\tilde{\Gamma}_i(\Delta)=\int^{E^{{\rm C\nu B},i}_e+\Delta/2}_{E^{{\rm C\nu B},i}_e-\Delta/2}dE_e\frac{d\tilde{\Gamma}_i}{dE_e}(E_e), 
		\label{OGi} \\
		\tilde{\Gamma}_{\beta, i}(\Delta)=\int^{\langle E^{{\rm C\nu B},i}_e \rangle+\Delta/2}_{\langle E^{{\rm C\nu B},i}_e \rangle-\Delta/2}dE_e\frac{d\tilde{\Gamma}_\beta}{dE_e}(E_e).
		\label{OGbeta}
	\end{align}

	This $\langle E^{{\rm C\nu B},i}_e \rangle$ is defined as the average energy for a C$\nu$B signal electron, so $E_{\nu_i}=\sqrt{\braket{p_0}^2+m_{\nu_i}^2}$ with $\braket{p_0}$ taken from \eqref{lowenergy}.
	
	From this, we can define a signal-to-noise ratio:
	
	\begin{equation}
		r^i_{\mathrm{SN}}(\Delta) = \frac{\tilde{\Gamma}_i(\Delta)}{\tilde{\Gamma}_{\beta, i}(\Delta)}.
	\end{equation}

	This ratio is how to tell quantitatively if a C$\nu$B signal will be drowned out by the normal beta decay. An $r^i_{\mathrm{SN}}$ value of 1 would mean that there are as many signals in that energy range from beta decay as there are from C$\nu$B interactions - this is the tipping point of observation. An $r^i_{\mathrm{SN}}$ that is $\gg1$ would mean extreme certainty of observation, and $\ll1$ would mean any C$\nu$B signal would be completely masked behind the veil of tritium beta decay.
	
	In order to confirm the effects of non-instantaneous decoupling and gravitational clustering, we need to measure the C$\nu$B signals to within 1\% precision. In order for this to ever be possible, we would need an $r^i_{\mathrm{SN}}$ of at least 100. As it turns out, the energy resolution needed to achieve $r^i_{\mathrm{SN}}=1$ or $r^i_{\mathrm{SN}}=100$ is almost identical, as $r^i_{\mathrm{SN}}$ is an exponentially rising function of $\Delta$.
	
	In Tables \ref{tb:DeltaNH} and \ref{tb:DeltaIH}, we show the energy resolution required to distinguish between the C$\nu$B signals and the beta decay background in both the Dirac and Majorana cases for $r^i_{\mathrm{SN}}=1$ and $r^i_{\mathrm{SN}}=100$, considering both the normal and inverted mass orderings.
	In both the Dirac and Majorana cases, we find almost the same required $\Delta$ - this makes sense, as the position of the C$\nu$B peaks are dependent on the neutrino masses, not their fermionic type. In the normal ordering case, the required $\Delta$ values for $\nu_1$ with $m_{\nu_1}=0\ {\rm meV}$, $\nu_2$ with $m_{\nu_2}=8.6\ {\rm meV}$ and $\nu_3$ with $m_{\nu_1}=50\ {\rm meV}$ are $0.46\ {\rm meV}, 4.2\ {\rm meV}$ and $19\ {\rm meV}$ respectively.
	
	In the inverted hierarchy case, owing to the mass and mass difference scales, it is difficult to distinguish between the signals for $\nu_1$ and $\nu_2$ due to their nearly degenerate masses. Because of this, we estimate the required $\Delta$ to distinguish the combined signal for $\nu_1$ and $\nu_2$ from the $\beta$-decay background, as seen in table \ref{tb:DeltaIH}. The required $\Delta$ values for $\nu_1$ and $\nu_2$ with $m_{\nu_1}=49.3\ {\rm meV}$ and $m_{\nu_2}=50\ {\rm meV}$, and $\nu_3$ with $m_{\nu_3}=0\ {\rm meV}$ are $23\ {\rm meV}$ and $0.46\ {\rm meV}$ respectively.
	 
	In both cases, if the lightest neutrino is taken to be massive rather, the required energy resolution to detect the lightest neutrino becomes larger and the difficulty in doing so lessened. Thus, the required energy resolution calculated herein is the absolute minimum needed.
	
	\begin{table}[h]
		\begin{center}
			\begin{tabular}{|l|l|l|l|l|l|l|}
				\hline
				NH case& $r^M_{\mathrm{SN}}=1$ & $r^M_{\mathrm{SN}}=100$ & $r^D_{\mathrm{SN}}=1$ & $r^D_{\mathrm{SN}}=100$ \\
				\hline
				$\nu_1\ (m_{\nu_1}=0\ {\rm meV})$ & $\Delta =0.83\ {\rm meV}$ &  $\Delta =0.46\ {\rm meV}$ & $\Delta =0.83\ {\rm meV}$ & $\Delta =0.46\ {\rm meV}$   \\
				$\nu_2\ (m_{\nu_2}=8.6\ {\rm meV})$ & $\Delta =5.3\ {\rm meV}$ &  $\Delta =4.2\ {\rm meV}$ & $\Delta =5.1\ {\rm meV}$ & $\Delta =4.1\ {\rm meV}$   \\
				$\nu_3\ (m_{\nu_3}=50\ {\rm meV})$ & $\Delta =21\ {\rm meV}$ &  $\Delta =19\ {\rm meV}$ & $\Delta =21\ {\rm meV}$ & $\Delta =18\ {\rm meV}$   \\
				\hline
			\end{tabular}
			\caption{Table showing the energy resolutions needed to achieve certain signal-to-noise ratios $r_{\mathrm{SN}}$ for both the cases wherein neutrinos are Dirac and Majorana particles, for each massive species in the normal hierarchy. Note that, as expected, the massless species requires the smallest resolution of under 0.5 meV, while the most massive only requires around 20 meV. Taken from \cite{Akita:2020jbo}.}
			\label{tb:DeltaNH}
		\end{center}
	\end{table}
	
	\begin{table}[h]
		\begin{center}
			\begin{tabular}{|l|l|l|l|l|l|l|}
				\hline
				IH case& $r^M_{\mathrm{SN}}=1$ & $r^M_{\mathrm{SN}}=100$ & $r^D_{\mathrm{SN}}=1$ & $r^D_{\mathrm{SN}}=100$ \\
				\hline
				$\nu_1+\nu_2 $ & $\Delta =29\ {\rm meV}$ &  $\Delta =23\ {\rm meV}$ & $\Delta =28\ {\rm meV}$ & $\Delta =22\ {\rm meV}$   \\
				\hspace{-0.1cm}$(m_{\nu_1, \nu_2}=49.3, 50\ {\rm meV})$ \hspace{-0.3cm} & &   &  &    \\
				$\nu_3\ (m_{\nu_3}=0\ {\rm meV})$ & $\Delta =0.83\ {\rm meV}$ &  $\Delta =0.46\ {\rm meV}$ & $\Delta =0.83\ {\rm meV}$ & $\Delta =0.46\ {\rm meV}$   \\
				\hline
			\end{tabular}
			\caption{Table showing the energy resolutions needed to achieve certain signal-to-noise ratios $r_{\mathrm{SN}}$ for both the cases wherein neutrinos are Dirac and Majorana particles, for each massive species in the inverted hierarchy. Note that, as expected, the massless species requires the smallest resolution of under 0.5 meV, while the most massive only requires around 20 meV. Taken from \cite{Akita:2020jbo}.}
			\label{tb:DeltaIH}
		\end{center}
	\end{table}
	
	Finally we can consider the energy resolution needed to distinguish between the signals from $\nu_1$ and $\nu_2$ neutrinos in the inverted hierarchy case. For this purpose, we can define
	
	\begin{align}
		r^{12}_{\mathrm{SN}}(\Delta)&=\frac{\int^{E^{{\rm C\nu B},1}_e+\Delta/2}_{E^{{\rm C\nu B},1}_e-\Delta/2}dE_e\frac{d\tilde{\Gamma}_1}{dE_e}(E_e)}{\int^{E^{{\rm C\nu B},1}_e+\Delta/2}_{E^{{\rm C\nu B},1}_e-\Delta/2}dE_e\frac{d\tilde{\Gamma}_2}{dE_e}(E_e)}, \label{r12}
	\end{align}

	and the inverse would be $r^{21}_{\mathrm{SN}}(\Delta)$. We find numerically that for $r^{12}_{\mathrm{SN}}(\Delta)=100$ we would need $\Delta\approxeq 0.50$ meV and for $r^{21}_{\mathrm{SN}}(\Delta)=100$ we would need $\Delta\approxeq 0.42$ meV \cite{Akita:2020jbo}. Thus, to both distinguish between $\nu_1$ and $\nu_2$ in the inverted hierarchy, as well as distinguishing between $\nu_3$ and the beta decay background, we would need an energy resolution of about $\Delta\approx 0.4$ meV.
	
	\subsection{Beyond the Standard Model}
	
	We have now established what standard physics tells us about the expected capture rate and energy spectra in a capture-by-tritium experiment. Any deviations from this outside of the usual error margins would be a clear sign of new physics.
	
	As discussed in chapter 5, if a large deviation was detected, the leading theories to explain it include sterile neutrinos, lepton asymmetry, neutrino decay and a change to the C$\nu$B's temperature.\\
	
	The number density for sterile neutrinos would be given by \eqref{sterilen}, accounting for the remainder of the observed $N_{\mathrm{eff}}$. From the latest calculation in \cite{Akita:2020szl}, the theoretical value for $N_{\mathrm{eff}}$ is 3.044 when accounting for only 3 flavours, while the observed value from \cite{Aghanim:2018eyx} is maximally 3.33, to within a 95\% confidence level. It is important to note that this maximal experimental value depends on the fixing of other parameters in the data, and as such can be increased. The reason why the observed effective number of species might not be closer to 4 can be seen in \eqref{Neff}: if we consider a fourth neutrino species as well, it would be altered to 
	
		\begin{equation}
		N_{\mathrm{eff}}=\left(\frac{1.40102}{1.39797}\right)^4 \left(4+\frac{\delta\rho_{\nu_e}}{\rho_{\nu_0}}+\frac{\delta\rho_{\nu_\mu}}{\rho_{\nu_0}}+\frac{\delta\rho_{\nu_\tau}}{\rho_{\nu_0}}+\frac{\delta\rho_{\nu_\chi}}{\rho_{\nu_0}}\right),
	\end{equation}

	where $\chi$ is some random symbol for the fourth flavour of neutrino. Thus, if $\frac{\delta\rho_{\nu_\chi}}{\rho_{\nu_0}}$ is both negative and fairly large relative to 1, it is possible to achieve an $N_{\mathrm{eff}}$ of 3.33. 
	
	Defining $\Delta N_{\mathrm{eff}} = N_{\mathrm{eff}}^{\mathrm{exp}}-N_{\mathrm{eff}}^{\mathrm{theo}} \approx 0.284$, we can begin to see what the capture rate for these sterile neutrinos might be. Recall that, in order to explain certain excesses in laboratory experiments, the parameters chosen for these sterile neutrinos are a mass between 0.3 eV and 3 eV  (based on $\Delta m_{41}^2 \approx (0.1 - 10) \mathrm{eV}^2$) and $|U_{e4}|^2\approx|U_{\mu4}|^2\approx 0.03$. For the sake of calculation, we can consider a 1 eV neutrino species. 
	
	According to \cite{Mertsch:2019qjv}, the clustering effect $\delta n^c_{\nu_s}$ would actually be the new leading order term, being approximately equal to 10. Putting this all together in \eqref{sterilen} yields 
	
	\begin{equation}
		n_4 \approxeq \Delta N_{\mathrm{eff}}(1+\delta n^c_{\nu_4})n_0 \approxeq 3.124n_0.
	\end{equation} 

	For a mass this large, the velocity and the effect of non-instantaneous decoupling would be negligible, and to leading order, the capture rate would be approximately
	
	\begin{equation}
		\Gamma_4^M = N_T\frac{G_F^2}{\pi}|V_{ud}|^2|U_{e4}|^2\frac{m_\mathrm{^3He}E_e|\bar{p}_e|}{m_\mathrm{^3H}}F(2,E_e)\left(\braket{f_F}^2 + \frac{g_A^2}{g_V^2}\braket{g_{GT}}^2 \right)n_4 \approxeq  3.124\frac{|U_{e4}|^2}{|U_{e1}|^2}\Gamma_1^M \approxeq 0.754 \mathrm{yr}^{-1},\label{sterilerate}
	\end{equation} 

	with the rate in the Dirac case being roughly half of this. While this capture rate appears small, it is larger than those of some active neutrinos seen in the section of numerical results. On top of this, however, is the fact that sterile neutrinos (should they exist) would be far easier to detect, as their large mass would cause their peak in the electron energy spectrum to be much further away from both the beta decay endpoint and those of other active species, circumventing the need for extremely good energy resolution. Atop this, including a fourth massive species in one's model allows for parameters to be varied so as to fit the data. In the model with only one added species, the data from Planck can be reinterpreted and $N_{\mathrm{eff}}$ can be close to 4 \cite{Zhang:2014dxk}, making \eqref{sterilerate} even larger, by a factor of around 3.
	
	For the case of a keV-scale sterile neutrino to account for dark matter, the detection prospects are far less likely. Though the characteristic peak would be unmistakable, the capture rate would simply be far too small. As stated in chapter 5 already, in this case, $|U_{e4}|^2$ would be of the order $10^{-11}$, constrained to be maximally $<10^{-9}$ \cite{Long:2014zva}. Even though, owing to their much larger mass, they would cluster extremely effectively, this would only increase their number density at earth by around a factor of 800, $n_4\approxeq 800n_0$ and so in the best case scenario, the capture rate would be 
	
	\begin{equation}
		\Gamma_4^M \approxeq 800 \frac{|U_{e4}|^2}{|U_{e1}|^2}\Gamma_1^M \approxeq 6.4\times 10^{-6} \mathrm{yr}^{-1},
	\end{equation}

	corresponding to roughly, on average, one detection every 156 thousand years. Clearly this experiment would not be the best way to detect these keV neutrinos, should they exist.\\
	
	Next we can turn to considering the effect a lepton asymmetry would have on the capture rate. If there was some primordial asymmetry between neutrinos and anti-neutrinos, a chemical potential would arise as an attempt to balance it out, entering into the distribution function in the usual way. As we showed in \eqref{asym}, we saw that including this effect leads to additional terms in the number density: 
	
	\begin{equation}
		n_{\nu} = T_\nu^3\left(\frac{3\zeta(3)}{4\pi^2} +\frac{\xi}{12} +\frac{\ln(2)}{2\pi^2}\xi^2\right),
	\end{equation}

	with $\xi$ related to the chemical potential by $\xi=\frac{\mu}{T_\nu}$. Thus, the enhancement factor owing to the chemical potential is given by
	
	\begin{equation}
		\frac{n_\nu}{n_0} \approxeq 1 + \frac{\pi^2\xi}{9\zeta(3)} + \frac{2\ln(2)\xi^2}{3\zeta(3)}.
	\end{equation}
	
	This factor is actually always there, but in the standard model, $\xi$ would be negligibly small. If we consider other scenarios however, we can see that the chemical potential can indeed affect the capture rate.
	
	For the Dirac case, where only left-helical neutrinos are detected, this value of $\xi$ could be either positive or negative: that is, there could be more neutrinos or anti-neutrinos in the universe today. Thus, for certain values of $\xi$, this factor might actually diminish the capture rate of our relic neutrinos. For example, for some arbitrary value of $\xi=\pm0.1$, we have 
	
	\begin{equation}
		n_\nu = n_0\left(1\pm\frac{\pi^2}{90\zeta(3)} + \frac{2\ln(2)}{300\zeta(3)}\right) \approxeq n_0(1\pm 0.0912 + 0.0038),
	\end{equation}

	changing the capture rate by almost 10\% in either direction. For Majorana neutrinos however, both the left-helical neutrinos and the right-helical neutrinos are detected, and these two components - taking the place of ``neutrinos" and ``anti-neutrinos" - will have opposite signs for their chemical potentials. Thus, when calculating the capture rate and summing over spins, the linear term will cancel out and we are left with only the quadratic contribution. Thus, in this case, the capture rate is always enhanced by a lepton asymmetry. Considering again some arbitrary value of $\xi=\pm0.1$, we would get $n_\nu=n_0(1+0.0038)$, increasing the capture rate around 0.4\%.
	
	Observations have set limits on what $\xi$ can actually take on for its values. The current bound is $-0.091<\xi<0.051$ \cite{Schwarz:2012yw}, meaning that this can change the capture rate maximally in the Dirac case by either diminishing it by 7.98\% or increasing it by 4.75\%, and in the Majorana case increasing it by approximately 0.32\%.\\
	
	Next up we may consider the case wherein all neutrinos decay into the lightest species. Of course, it is possible to consider some scenario wherein only some of the more massive species have decayed, but for the sake of the calculation, let us consider the case where in the past almost 14 billion years, all relic neutrinos have decayed into the lightest species. In this case then, the lightest eigenstate of neutrino has 3 times as many neutrinos in it, with the others being unpopulated. At leading order, ignoring all other higher-order terms and effects, this will affect the capture rate purely via the elements of the PMNS matrix. Instead of summing over all 3 species, we simply have one term with the PMNS element corresponding to the lightest eigenstate, but multiplied by 3. That is, $\sum_{i=1}^3|U_{ei}|^2 \ \rightarrow \ 3|U_{e\ell}|^2$, with $\ell$ the lightest mass eigenstate. Owing to unitarity, $\sum_{i=1}^3|U_{ei}|^2=1$ and so the change to the capture rate would be a factor of $3|U_{e\ell}|^2$. As we shall see, in any scenario this leads to a smaller chance of detection.
	
	Conversely, the presence of multiple signals with different masses will put a new upper constraint on the lifetime of neutrinos - the present age of the universe.\\
	
	Let us consider first the normal ordering case. In this scenario, at first glance, the capture rate seems to be significantly improved :
	
	\begin{equation}
		\Gamma^{M,D}_{C\nu B} = \Gamma^{M,D}_{C\nu B,0}\times 3|U_{e1}|^2 \approxeq 2.043 \  \Gamma^{M,D}_{C\nu B,0},
	\end{equation}

	essentially doubling the total capture rate. However, recall from the previous section that the lightest neutrino species is also the most difficult to observe owing to the poor energy resolution of cosmic neutrino detectors and the experimental noise of regular beta decay. Thus, even though the capture rate would be doubled, unless the energy resolution is improved significantly or unless the lightest species' mass is actually around 20 meV, none of these captured neutrinos would actually be observed.\\
	
	In the inverted ordering case, where the lightest mass is $m_3$, the diminishment of the capture rate is far more apparent. In this case, we have 
	
	\begin{equation}
		\Gamma^{M,D}_{C\nu B} = \Gamma^{M,D}_{C\nu B,0}\times 3|U_{e3}|^2 \approxeq 0.0666 \  \Gamma^{M,D}_{C\nu B,0},
	\end{equation}
	
	a diminishment of the capture rate by over 93\%. Atop this, the added effect of the lightest species' being difficult to detect owing to the noise of beta decay and poor energy resolution leaves us with an essentially invisible signal.\\
	
	Finally, we may consider a change to the C$\nu$B's temperature and how this might affect the capture rate. As we have seen many times now, the number density is related to the effective temperature by $n\propto T_\nu^3$. Any changes to this temperature would then affect the capture rate. In fact, we have already seen this in the section on the distortive effects of non-instantaneous decoupling: as the temperature was changed from the leading order calculation of $T_\nu= 1.9454$ K to the more accurate $T_\nu= 1.9496$ K, the number density (and thus, the capture rate) was increased by around 0.65\%, seen in \eqref{newnumdens}.
	
	A more striking change may come about if there was some ``dark radiation" (an exotic, yet undetected species of matter that was relativistic at the epoch considered) which annihilated between the times of neutrino and photon decoupling. In this case, the presence of the dark radiation would change the effective degrees of freedom and, as we saw at the end of chapter 5, would change the ratio between the neutrino effective temperature and the CMB's temperature today: 
	
	\begin{equation}
		T_{\nu,0}\propto T_{\gamma,0}\left(\frac{g_*(T_0)}{g_*(T_{dec})}\right)^{1/3}=T_{\gamma,0}\left(\frac{2}{11/2+\Delta g_*}\right)^{1/3}.
	\end{equation}

	Thus, the presence of any more species would decrease the temperature, and in turn decrease the number density and capture rate by a factor of 
	
	\begin{equation}
		n_\nu = n_0\left(\frac{T^{\mathrm{new}}_\nu}{T_\nu^{\mathrm{std}}}\right)^3 = \frac{22}{22+\Delta g_*}n_0.
	\end{equation}
	
	For some numerical examples, the inclusion of another fermionic spin-$\frac{1}{2}$ particle (with its antiparticle) would reduce the capture rate by about 13.7\% \footnote{This hypothetical particle does not interact strongly, and so does not have colour degrees of freedom.}, while the inclusion of 3 of these would reduce the capture rate by approximately 32.3\%.\\
	
	One final point must be made here. The calculations leading up to this section are what we \itshape expect \normalfont to be detected. In this section, then, we provided some explanations, should the capture rate deviate from our predictions. Unfortunately though, should some deviation be found, it seems extremely difficult, for the most part, to differentiate between these models. As we saw in our numerical calculations, the capture rate varies considerably based on factors like whether neutrinos are Dirac or Majorana particles, as well as their mass. Other standard factors like gravitational clustering are only approximate, and rely on numerical simulations, whose error margins can certainly be argued to be large. On top of this, the elements of the PMNS matrix have current error margins of around 10\% at the 3$\sigma$ level. The question is, how then, given this miasma of factors and parameters, are we expected to eke out information and conclusions from data collected?
	
	While the detection of the C$\nu$B on tritium may not instantly answer all of our questions about the Dirac versus Majorana debacle, the questions of sterile neutrinos and lepton asymmetry and other such mysteries, it will certainly provide many more answers than we currently have available. For example, the observed peaks in the energy spectrum of the emitted electrons will certainly provide tight bounds on the neutrino masses. The capture rates - applying Occam's razor - should provide a sign as to whether they are Majorana or Dirac neutrinos, owing to the factor of 2 difference for non-relativistic species. The presence of sterile neutrinos would have the smoking gun of a characteristic peak in the electrons' spectra, and neutrino decay would be seen by the presence of only a single peak. In conclusion, while competing theories of non-standard physics may arise to assuage the tension if the capture rates are not as expected, varying parameters to fix their models to the data, this first step into observing the C$\nu$B will already provide some interesting, if not concrete, information. 
	
	\newpage
	\section{Lensing as a Multipurpose Tool}
	
	Gravitational lensing is a natural corollary of general relativity, and has been experimentally confirmed since 1919 \cite{Dyson1920}. Fortunately, it is more than just a consequence - it is a useful tool we can use to our advantage in our quest to learn about neutrinos and our universe. We begin by covering the basics of lensing, then showing how we can use it to massively increase the neutrino influx from a supernova event. After that we show how the C$\nu$B can actually be used to observe the evolution of the ``lens", and in the next chapter we will show how lensed supernova neutrinos might help us uncover their masses.
	
	\subsection{A Brief Introduction to Lensing}
	
	To begin our study of gravitational lensing, let us consider a spacetime with a single object of mass $M$ surrounded by vacuum. We can describe this spacetime using the Schwarzschild metric or a Minkowski metric perturbed by a Newtonian potential:
	
	\begin{equation}
		ds^2=-(1+2\phi)dt^2+(1-2\phi)d\bar{x}^2,
	\end{equation}

	wherein $\phi=-\frac{GM}{r}$. Recall that this is also the weak field limit of the Schwarzschild metric. Now, since light travels on null geodesics, we can obtain the ``effective" speed of light for when $ds^2=0$. In that case we have
	
	\begin{equation}
		(1-2\phi)d\bar{x}^2=(1+2\phi)dt^2 \ \ \ \rightarrow \ \ \ \left|\frac{d\bar{x}}{dt}\right|=\sqrt{\frac{1+2\phi}{1-2\phi}}.
	\end{equation}

	To linear order then, $\left|\frac{d\bar{x}}{dt}\right|\approxeq1+2\phi$. Then, since the refractive index $n$ is defined as the ratio of the speed of light in vacuum to the actual speed,
	
	\begin{equation}
		n\approxeq1-2\phi.
	\end{equation}

	Since the potential is always $<0$, we can write $n=1+2|\phi|$.\\
	The same effects seen in usual geometric optics are observed owing to the presence of gravity. Similarly to how light is slowed and its path bent when entering a glass prism, analogously the presence of our mass bends and slows light in its gravitational field. The first of these connections is the deflection angle, $\alpha$, which is related to the gradient of the refractive index \cite{Narayan:1996ba}:
	
	\begin{equation}
		\bar{\alpha}=-\int\bar{\nabla}_{\perp}nd\ell \ = \ 2\int\bar{\nabla}_{\perp}\phi d\ell.
	\end{equation}

\begin{figure}
	\centering
	\includegraphics[width=\linewidth]{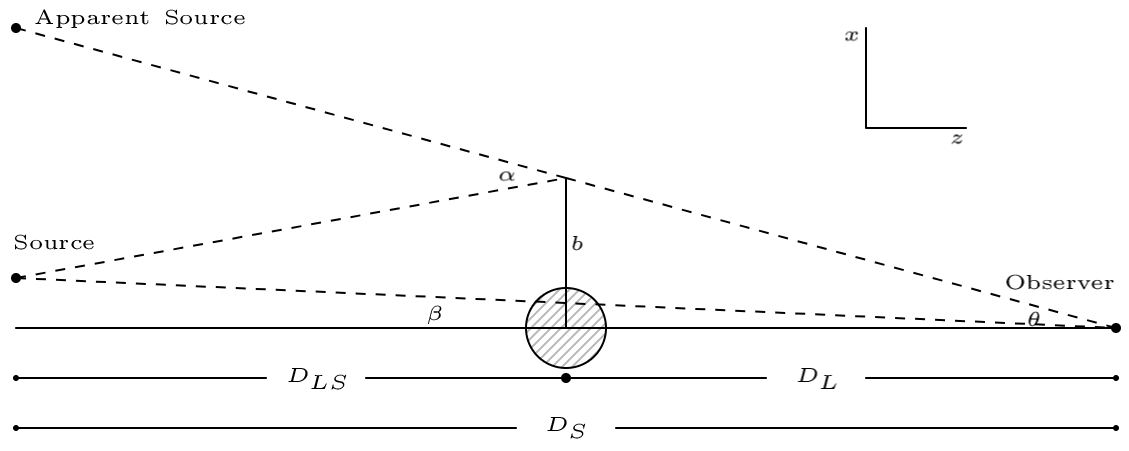}
	\caption{Lensing by a point mass, drawn using \cite{Drawingtool}.}
	\label{fig:lensing}
\end{figure}

	So, consider our point mass $M$ placed at the origin, and our light will move in the $x-z$ plane, as seen in figure \ref{fig:lensing}. Then the $r$ in our potential is $\sqrt{x^2+z^2}$, and we have
	
	\begin{equation}
		\bar{\alpha}=2\int\bar{\nabla}_{\perp}\phi d\ell = 2\int\bar{\nabla}_x\frac{-GM}{\sqrt{x^2+z^2}} dz.
	\end{equation}

	Most of the bending will occur near the point of closest approach to the mass, and so this is the most important region of the path. At the point of closest approach, the direction perpindicular to the path is $x$. So, performing the differentiation and taking the source to be infinitely far away and us (the observer) to be as well in the other direction, we have
	
	\begin{equation}
		\bar{\alpha}=2\int_{-\infty}^{\infty}\frac{GM\bar{x}}{(x^2+z^2)^{3/2}}dz = \frac{2GM\bar{x}}{x^2}\frac{z}{\sqrt{x^2+z^2}}\bigg|^\infty_{-\infty}.
	\end{equation}

	So the magnitude of the angle of deflection will be
	
	\begin{equation}
		\alpha=\frac{4GM}{x}\lim_{z\rightarrow\infty}\frac{z}{\sqrt{x^2+z^2}} = \frac{4GM}{x}.\label{alpha}
	\end{equation}

	This $x$ is measured at the point of closest approach, and so is known as the impact parameter $b$. Thus, the deflection angle comes out to be $\alpha=\frac{4GM}{b}=\frac{2r_s}{b}$, where $r_s$ is the Schwarzschild radius of the lensing object.\\

	In the above, we assumed that the source, lens and observer were collinear, but this may not always be the case. Depicted in figure \ref{fig:lensing}, we can relate the angle of the \itshape apparent \normalfont source $\theta$ to that of the actual source $\beta$ by simple geometry:
	
	\begin{equation}
		\theta D_S = \beta D_S+\alpha D_{LS},\label{lensinit}
	\end{equation}

	where $D_S$ is the distance to the source, $D_{LS}$ the distance from the lens to the source, and $\alpha$ our deflection angle as before. In the above, all angles are taken to be small enough such that the $\sin x\approxeq x$ approximation can be used. Thus, for the case of a single point mass, we have $\alpha$ from \eqref{alpha} and 
	
	\begin{equation}
		\theta-\beta=\frac{D_{LS}}{D_S}\frac{4GM}{b}.
	\end{equation}

	$b$, the impact parameter, is the distance of closest approach and geometrically (again using the small angle approximation) $b\approxeq D_L\theta$, with $D_L$ the distance to the lens. An important distance is the Einstein radius: the distance from the point mass that the apparent source appears, when the source, lens and observer form a syzygy. In this case, $\beta=0$ and we have 
	
	\begin{equation}
		\theta=\frac{D_{LS}}{D_S}\frac{4GM}{D_L\theta} \ \ \rightarrow \ \ \theta=\sqrt{\frac{4GMD_{LS}}{D_SD_L}}.
	\end{equation}
	
	The Einstein ``radius", though actually an angle, is commonly symbolised by $\theta_E$ and is given by (under the small angle approximation)
	
	\begin{equation}
		\theta_E=\theta-0=\sqrt{\frac{4GMD_{LS}}{D_SD_L}}=\sqrt{\frac{4GMy}{D_L(1+y)}},\label{einsteinring}
	\end{equation}

	where $y=\frac{D_{LS}}{D_L}$. This angle, multiplied by $D_L$, gives the radius of the ``ring" formed by the source when lensed. This is because in the case of a syzygy, the light from the source is lensed towards the observer symmetrically around the lens, forming a circular image. Lensed light from sources that do not lie exactly on the observer-lens line of sight do not form rings, and instead form arcs or arclets \cite{Narayan:1996ba}. The Einstein ring is the most extreme case of a lensing event.\\
	
	One extremely useful consequence of lensing is that the source is ``magnified" - that is, more light (or any particles) reaches us, as even though the surface brightness is conserved \cite{Lieu:2004ns}, the apparent solid angle is increased. The magnification $\mu$, when the lens is circularly symmetric (which is the case for a point mass), is defined as 
	
	\begin{equation}
		\mu = \frac{\theta}{\beta}\frac{d\theta}{d\beta}\label{mag}.
	\end{equation}

	To this end, for our point mass lens, we can use \eqref{alpha}, \eqref{lensinit}, \eqref{einsteinring} and $b\approxeq D_L\theta$ to see that 
	
	\begin{equation}
		\beta = \theta-\frac{\theta_E^2}{\theta}.\label{lens}
	\end{equation}

	Solving this yields $\theta=\frac{1}{2}\left(\beta\pm\sqrt{\beta^2+4\theta_E^2}\right)$. For a point mass, any source is always lensed twice, creating two images or arc(lets)\footnote{In the case of an Einstein ring, the two arcs are semicircles and thus appear as one ring.}(for more complex lenses, this is not the case). These two images occur at what are commonly denoted $\theta_\pm$, and unless $\theta_\pm=\theta_E$, one image is always outside the ring and one inside. Thus, using \eqref{mag} and \eqref{lens}, the full magnification is given by the sum of the magnification of the two images:
	
	\begin{equation}
		\mu = \mu_++\mu_- = \left(1-\left(\frac{\theta_E}{\theta_+}\right)^4\right)^{-1} + \left(1-\left(\frac{\theta_E}{\theta_-}\right)^4\right)^{-1} = \frac{\frac{\beta^2}{\theta_E^2}+2}{\frac{\beta}{\theta_E}\sqrt{\frac{\beta^2}{\theta_E^2}+4}}.\label{mag2}
	\end{equation}

	For a point mass this takes on a maximum value as $\beta\rightarrow0$, in which case $\mu \rightarrow \frac{\theta_E}{\beta}$. Physically, $\beta$ cannot be smaller than the radius of the source over the distance to it, so $\mu_{max}=\frac{\theta_ED_S}{R_S}$.\\
	
	Owing to the presence of the lens, the light from the source must take a different, longer path to reach us, resulting in a change to the travel time known as the Shapiro time delay. Since the light from the source can travel along two different paths (or more, for more complex mass distributions) there is another time delay between the two detections, $\Delta t = t_--t_+$. The light from the innermost apparent source arrives last, as it has the longest path to travel, while that coming from the outside of the ring arrives first. Thus, $\Delta t$ is always positive. These time delays will be calculated for supernova neutrinos below.\\
	
	Before continuing, for lensing objects such as galaxies, the point-mass model is perhaps a little too far from reality. For more complex lensing objects that have some structure, there are other models available. We will discuss one here: the singular isothermal sphere model.
	
	In this model, galaxies are modelled with their constituents behaving like particles in an ideal gas - an approximation that is quite fair, as most astrophysical objects do not regularly interact. Thus, the equation of state of these constituents is of course the usual $P=\frac{NT}{V}$, where the Boltzmann constant $k_B=1$ and we can rewrite this as 
	
	\begin{equation}
		P=\frac{\rho T}{m},\label{idealgas}
	\end{equation}

	with $\rho$ and $m$ the mass density and mass of the constituents, respectively\footnote{Of course in reality not every object in a galaxy has the same mass, and this mass is usually taken to be an average stellar mass.}. In thermal equilibirium, the temperature is related to the velocity dispersion by equating the thermal and kinetic energies:
	
	\begin{equation}
		\frac{3}{2}T = \frac{1}{2}m\sigma_v^2.\label{energies}
	\end{equation}
	
	The word ``isothermal" refers to the fact that the temperature is taken to be constant throughout the galaxy, and so in turn is the velocity dispersion. Of course, a more general model could be used for a radially-dependent temperature, but since most of the galaxy consists of interstellar gas which is approximately isothermal, it is not a terrible approximation.
	
	Next, from hydrostatic equlibirium (wherein the force of thermal pressure balances that of gravity) we have that 
	
	\begin{equation}
		\frac{1}{\rho}\frac{dP}{dr} = -\frac{GM}{r^2}.
	\end{equation}

	So, since $M=\frac{4\pi}{3}\rho r^3$, we have that $\frac{dP}{dr} =-\frac{4\pi G\rho^2r}{3}$ and so $P=-\frac{2\pi G\rho^2r^2}{3}$. And so, using this along with \eqref{idealgas} and \eqref{energies}, we have
	
	\begin{equation}
		\rho = \frac{\sigma_v^2}{2\pi Gr^2}.
	\end{equation}

	In more general lenses that are not point masses, the surface mass density is important, as this essentially describes the ``lens" nature of the object, and the surface mass density distribution is all that is needed to derive the characteristics of a lens. It is given by
	
	\begin{equation}
		\Sigma(r) = \int \rho(r,z)dz,
	\end{equation}

	where $r$ is the radial distance from the centre of the lens and $z$ the distance along the observer-lens axis. So, in the case of a singular isothermal sphere (SIS), we have 
	
	\begin{equation}
		\Sigma(r) = \int \frac{\sigma_v^2}{2\pi Gr^2} dz = \int_0^\pi \frac{\sigma_v^2}{2\pi Gr} d\theta = \frac{\sigma_v^2}{2Gr},
	\end{equation}

	where we used that $dz\approxeq rd\theta$. To obtain the deflection angle, we need to sum over all deflections from all the masses in the plane. This gives a two dimensional integral:
	
	\begin{equation}
		\alpha(r) = 4G \int d^2r' \frac{(\bar{r}-\bar{r}')\Sigma(r')}{|\bar{r}-\bar{r}'|^2}.
	\end{equation}
	
	To solve this, we put $r'$ into radial coordinates, such that $d^2r' = r'dr'd\theta'$ and then use circular symmetry to integrate over the angular coordinate. Next, we use a change of variables from $r-r'$ to $r'$, similar to integrals in electrostatics. Then, we are left with 
	
	\begin{equation}
		\alpha(r) = \frac{8\pi G}{r} \int_0^r \Sigma(r')r'dr'.
	\end{equation}

	This can be written as $\alpha(r) = \frac{4GM(r)}{r}$, which is remniscient of \eqref{alpha}, except instead of $M$ being a known value, it is now given by $M(r) = 2\pi \int_0^r \Sigma(r')r'dr'.$ Thus, for our SIS, we have
	
	\begin{equation}
		\alpha(r) = \frac{4GM(r)}{r} = \frac{4G}{r}\cdot2\pi\int_0^r \frac{\sigma_v^2}{2Gr'}r'dr' = 4\pi \sigma_v^2.
	\end{equation}

	This deflection angle can then be inserted into \eqref{lensinit} to obtain the Einstein radius, and that in turn can be used with the lens equation to calculate the magnification and time delays.
	
	As already stated, there are many models used to describe lenses. As long as the surface distribution $\Sigma(r)$ is known, all other quantities can be derived from there.
	
	\subsection{Lensing of Massive Particles} 
	
	While the previous subsection served as our springboard, for the case of massive particles we require the full Schwarzschild metric to describe our spacetime. 
	
	The Schwarzschild metric is given by
	
	\begin{equation}
		ds^2=-d\tau^2=-(1-\frac{2M}{r})dt^2+(1-\frac{2M}{r})^{-1}dr^2+r^2d\Omega^2, \label{SC}
	\end{equation}
	
	where $d\Omega^2=r^2d\theta^2+r^2\sin^2\theta d\phi^2$, and for this calculation we have taken both $c$ and $G$ to be unity.\\
	Let us consider some particle moving in this spacetime around our central mass. We can always align our coordinate system such that the particle's orbit is at $\theta=\frac{\pi}{2}$, so that $d\theta=0$ and $\sin^2\theta=1$.
	
	Dividing \eqref{SC} by $d\tau^2$ gives us 
	
	\begin{equation}
		g_{\mu\nu}\frac{dx^\mu}{d\tau}\frac{dx^\nu}{d\tau} = -(1-\frac{2M}{r})\left(\frac{dt}{d\tau}\right)^2+ (1-\frac{2M}{r})^{-1}\left(\frac{dr}{d\tau}\right)^2-r^2\left(\frac{d\phi}{d\tau}\right)^2.\label{r}
	\end{equation}

	Now, requiring that the line element is invariant, we can use calculus of variations:
	
	\begin{equation}
		\delta s = 0 \ \ \rightarrow \ \ \delta\int ds = 0 \ \ \rightarrow \ \ \delta\int\sqrt{ds^2} = 0 \ \ \rightarrow \delta\int\sqrt{g_{\mu\nu}\frac{dx^\mu}{d\tau}\frac{dx^\nu}{d\tau}}d\tau = 0.
	\end{equation}

	Thus, in order for this to be invariant, the integrand must obey the Euler-Lagrange equations. However, if $\mathcal{L}$ is conserved, so is $\mathcal{L}^{1/2}$. Thus, instead of considering the above, our integrand can simply be taken to be
	
	\begin{equation}
		\mathcal{L} = -(1-\frac{2M}{r})\left(\frac{dt}{d\tau}\right)^2+ (1-\frac{2M}{r})^{-1}\left(\frac{dr}{d\tau}\right)^2-r^2\left(\frac{d\phi}{d\tau}\right)^2.
	\end{equation}

	Applying the E-L equations, $\frac{d}{d\tau}\frac{\partial\mathcal{L}}{\partial\dot{x}}=\frac{\partial\mathcal{L}}{\partial x}$ for the cases $x=t$ and $\phi$ yield 
	
	\begin{equation}
		\frac{d}{d\tau}\left[\left(1-\frac{2M}{r}\right)\frac{dt}{d\tau}\right]=0 \ \ \rightarrow \ \ \left(1-\frac{2M}{r}\right)\frac{dt}{d\tau}=E,
	\end{equation}
\begin{equation}
	\frac{d}{d\tau}\left[r^2\frac{d\phi}{d\tau}\right] = 0 \ \ \rightarrow \ \ r^2\frac{d\phi}{d\tau}=J,
\end{equation}

	where $E$ and $J$ are constants of integration and represent conserved quantities (the former being energy and the latter angular momentum per unit mass). Rearranging \eqref{r}, we have that 
	
	\begin{equation}
		\left(\frac{dr}{d\tau}\right)^2 =\left(1-\frac{2M}{r}\right)^2\left(\frac{dt}{d\tau}\right)^2 -\left(1-\frac{2M}{r}\right)-r^2\left(1-\frac{2M}{r}\right)\left(\frac{d\phi}{d\tau}\right)^2,
	\end{equation}

\begin{equation}
	\rightarrow \ \ \left(\frac{dr}{d\tau}\right)^2 + \left(1-\frac{2M}{r}\right)\left(1+\frac{J^2}{r^2}\right)=E^2.
\end{equation}
	
	Our next act of manipulation involves $u=\frac{1}{r}$ and $\lambda=\frac{E^2-1}{J^2}$. Since $\frac{dr}{d\tau}=-\frac{1}{u^2}\frac{du}{d\tau}$ and $\frac{du}{d\tau}=\frac{du}{d\phi}\frac{d\phi}{d\tau}=\frac{du}{d\phi}Ju^2$, we can rewrite our previous result as 
	
	\begin{equation}
		\left(\frac{du}{d\phi}\right)^2=\lambda+\frac{2Mu}{J^2}-u^2+2Mu^3.
	\end{equation}

	Now, we can write this as 
	
	\begin{equation}
		\left(\frac{du}{d\phi}\right)^2=2M(u-u_0)(u-u_1)(u-u_2),\label{uphi}
	\end{equation}

	where $u_i$ are the roots of $\lambda+\frac{2Mu}{J^2}-u^2+2Mu^3$. Now, though this is the fully general relativistic equation, in the Newtonian limit, the $u^3$ term would disappear, and so we can obtain the Newtonian roots of $\lambda+\frac{2Mu}{J^2}-u^2$ as 
	
	\begin{equation}
		u_{1,2}=\frac{M}{b^2v_0^2}\left(1\pm\sqrt{1+\frac{\lambda b^4v_0^4}{M^2}}\right),
	\end{equation}
	
	where we have used that the angular momentum per unit mass $J$ at the point of closest approach to the lensing object is equal to the impact parameter $b$ times the velocity of the particle at infinity, $v_0$.
	
	Now these $u_1$ and $u_2$ can be used in our relativistic case approximately, and we can rewrite $u_0$ in terms of them as well. We have that
	
	\begin{equation}
		2Mu^3-u^2+\frac{2M}{J^2}u+\lambda = 2M(u-u_0)(u-u_1)(u-u_2),
	\end{equation}

	and expanding the right hand side and equating coefficients of powers of $u$ by uniqueness of power series we get from the coefficient of $u^2$ that
	
	\begin{equation}
		2Mu_0=1-2M(u_1+u_2).
	\end{equation}
	
	Going back to \eqref{uphi}, we have that
	
	\begin{equation}
		\phi-\phi_0=2\int_{0}^{u_1}\frac{du}{\sqrt{2M(u-u_0)(u-u_1)(u-u_2)}},
	\end{equation}
	
	where the factor of 2 in front of the integral comes from the fact that we have the path from the source to the point of closest approach, and that from the point of closest approach to the observer. Using our relationship between the $u_i$, and taking the initial angle to be $-\pi$ (that is, the source is approximately infinitely far away) we have
	
	\begin{equation}
		\phi=2\int_{0}^{u_1}\frac{du}{\sqrt{(u_1-u)(u-u_2)(1-2M(u+u_1+u_2))}}-\pi.
	\end{equation}
	
	We now make the change of variables $u=u_2+(u_1-u_2)\sin^2x$, such that at $u=0$ we have $x=x_*=\arcsin\sqrt{\frac{u_2}{u_2-u_1}}$ and at $u=u_1$ we have $x=\frac{\pi}{2}$. Using that $du=2(u_1-u_2)\sin x\cos xdx$, we have
	
	\begin{equation}
		\phi + \pi = 4\int_{x_*}^{\frac{\pi}{2}}\frac{dx}{\sqrt{1-2M(u_1+u_2+u_1\sin^2x+u_2\cos^2x)}}.
	\end{equation}

	Now, $2Mu_i$ goes as $\frac{M^2}{b^2v_0^2}$, which in all cases of interest is a small quantity. Thus, we will take $\frac{M}{bv_0}$ to be our small quantity and perform perturbation theory. \\
	For example, in the solar system case, considering a solar-mass star and a particle with relativistic speed at an impact parameter of characteristic distance around 1 AU, this quantity is of the order $10^{-8}$. For a super massive black hole at the centre of our galaxy, using a galactic characteristic distance of around 8 kpc, this quantity is as small as $10^{-11}$.
	
	So, using the binomial approximation this integral becomes
	
	\begin{equation}
		\phi+ \pi = 4\int_{x_*}^{\frac{\pi}{2}}dx(1+M(u_1+u_2+u_1\sin^2x+u_2\cos^2x) +\frac{3M^2}{2}(u_1+u_2+u_1\sin^2x+u_2\cos^2x)^2+...).
	\end{equation}

	Performing this integration, and keeping terms only linear in $\frac{M}{bv_0}$, we have
	
	\begin{equation}
		\phi+ \pi \approxeq 4\left[\frac{\pi}{2}-x_*+M(u_1+u_2)\left(\frac{\pi}{2}-x_*\right) +\frac{M}{2}u_1\left(\frac{\pi}{2}-x_*+\sin x_*\cos x_*\right) +\frac{M}{2}u_2\left(\frac{\pi}{2}-x_*-\sin x_*\cos x_*\right)\right].
	\end{equation}

	Now, we know that $\sin x_*=\sqrt{\frac{u_2}{u_2-u_1}}$. Plugging in our expressions for $u_i$, and once again using the binomial expansion owing to our small parameter, we have
	
	\begin{equation}
		\sin x_*=\sqrt{\frac{\sqrt{1+\frac{\lambda b^4v_0^4}{M^2}}-1}{2\sqrt{1+\frac{\lambda b^4v_0^4}{M^2}}}} \approxeq \frac{1}{\sqrt{2}}\left(1-\frac{M}{2\sqrt{\lambda}b^2v_0^2}\right).
	\end{equation}

	Next we use some trigonometric identities to find $x_*$. First, we see that
	
	\begin{equation}
		\sin\left(\frac{x_*}{2}-\frac{\pi}{8}\right)\cos\left(\frac{x_*}{2}+\frac{\pi}{8}\right) = \frac{1}{2}\sin x_* -\frac{1}{2}\sin\frac{\pi}{4} = -\frac{M}{4\sqrt{2\lambda}b^2v_0^2}
	\end{equation}

	and so if we take $x_*$ close to $\frac{\pi}{4}$, using the small angle approximation for the sine function we obtain $x_*\approxeq\frac{\pi}{4}-\frac{M}{2\sqrt{\lambda}b^2v_0^2}$. Using this, as well as $\cos x_*\approxeq \frac{1}{\sqrt{2}}\left(1+\frac{M}{2\sqrt{\lambda}b^2v_0^2}\right)$, $\lambda = \frac{1}{b^2}$ (since $J=bv_0$ and $E^2-1=v_0^2$), and that $u_1+u_2=\frac{2M}{b^2v_0^2}$, we have to linear order in $M$
	
	\begin{equation}
		\phi + \pi \approxeq 2\pi - \pi + \frac{2M}{bv_0^2}+\frac{2M}{b} + ...,
	\end{equation}

\begin{equation}
	\rightarrow \ \ \phi = \frac{2M}{bv_0^2}(1+v_0^2) +\mathcal{O}\left(\left(\frac{M}{bv_0}\right)^2\right). \label{massdeflec}
\end{equation}

	This deflection angle for massive particles reduces exactly to that of massless particles when $v_0^2=1$, equation \eqref{alpha}. Similarly, in the Newtonian limit when $v_0^2\ll1$, we obtain the Newtonian deflection angle: $\phi=\frac{2M}{bv_0^2}$.
	
	 Plugging in \eqref{massdeflec} as our $\alpha$, we can get the Einstein radius for massive particles as 
	
	\begin{equation}
		\theta_E = \sqrt{\frac{2M(1+v_0^2)D_{LS}}{v_0^2D_SD_L}} = \sqrt{\frac{2My(1+v_0^2)}{v_0^2D_L(1+y)}},
	\end{equation}

	which can be used to calculate the magnification in \eqref{mag}.
	
	Though this derivation was done using a point mass, the result for the SIS model has a similar effect from the mass \cite{Lin:2019lko}, with a factor of $\frac{1+v_0^2}{v_0^2}$ entering the deflection angle.
	
	\subsection{Lensing of Supernova Neutrinos}
	
	Though we have the cosmic neutrino background, there is no constant source of high energy neutrinos that we may study. On top of this, owing to their very high velocities, they will not cluster gravitationally to assist us with detection. One intense source of these high energy neutrinos seems to be supernovae, but supernovae that are close enough to be of use are rare, and as we shall see, those that are lensed are even less frequent. Adding to our troubles are that these intense bursts have a pulse time of only around 10 seconds \cite{Mena:2006ym}.\\
	
	Despite these hinderences, with some luck, a supernova event lensed on its way to earth can provide some interesting information, and has distinct characteristics. Deflection by the sun would cause too small an angular deviation for detection, as current neutrino detectors' angular resolution is not very small. For example, a neutrino skimming our sun would have a deflection angle of $\alpha=\frac{4GM}{R}\approxeq 8.48\ \mu\mathrm{rad}$, while detectors like Super Kamiokande have angular resolution of the order of degrees \cite{Mena:2006ym}.
	
	Similarly, lensing by very far away (extragalactic) sources (such as other galaxies) would cause the flux to be diminished too greatly to be compensated by the amplification owing to magnification. The most suitable lens seems to be the black hole at the centre of the Milky Way, as it is extremely massive and relatively close.\\
	
	Though any lensing would provide a wealth of information, for our purposes here we will focus on the case of a supernova occurring on the opposite side of our galaxy. The two main characteristic effects of lensing are the amplification of the flux and a time dispersion of their arrival. The usual characteristic behaviour of a supernova can be described as a time-dependent luminosity, modelled as \cite{Fogli:2003dw} 
	
	\begin{equation}
		L(t)=\frac{E_B}{6t_P}e^{-t/t_P},
	\end{equation}

	where $E_B$ is the energy released by the supernova, done so over the time period $t_P$. The factor of 6 arises to account for the 3 flavours of neutrinos and their anti-particles. The exponential decay following the initial pulse has an e-fold time equal to that of the initial pulse, and using $t_P\approx3s$, this model is consistent with observations. Thus, the effects of magnification and time dispersion will be clearly seen if the detected luminosity pattern differs significantly from our model.\\
	
	As shown above in \eqref{mag}, the magnification goes as 
	
	\begin{equation}
		\mu = \frac{\frac{\beta^2}{\theta_E^2}+2}{\frac{\beta}{\theta_E}\sqrt{\frac{\beta^2}{\theta_E^2}+4}},
	\end{equation}
	
	going as $\frac{\theta_E}{\beta}$ at very small $\beta$ values, and with its maximal value $\mu_{max}=\frac{\theta_ED_S}{R_S}$. From \eqref{einsteinring}, this gives 
	
	\begin{equation}
		\mu_{max}=\sqrt{\frac{4GMyD_S^2}{R_S^2D_L(1+y)}} =\frac{1}{R_S}\sqrt{4GMD_Ly(1+y)}.
	\end{equation}
	
	As a numerical estimate, for a usual source around 20 km in diameter, and using that the supermassive black hole at the centre of our galaxy is around $3.61\times10^6$ solar masses and a distance approximately 8 kpc from us \cite{Mena:2006ym}, we get $\mu_{max}=2.3\times10^{11}\sqrt{y(1+y)}$. Thus, varying $y$ from very small values (around 0.01) to 2 (approximately the edge of the Milky Way) only changes $\mu_{max}$ by an order of magnitude. From this, we can see that a perfect supernova-black hole-earth syzygy would result in an absolutely massive amplification of the number of neutrinos reaching earth.\\
	
	Next, we need to consider the time delays of these neutrinos. For this we once again use the Schwarzschild metric. Of course, this spacetime implies that the only object in the spacetime is the black hole, but for the case of neutrinos this is a valid approximation, as the Milky Way is extremely ``transparent" to neutrinos: the mean free path of neutrinos is much larger than the size of the galaxy.\\
	If the mean free path is given by 
	\begin{equation}
		\ell_{mfp}=\frac{1}{\sigma n},
	\end{equation}
	
	with $\sigma$ of the order $\bar{\sigma}\approx10^{-45}\mathrm{cm}^{2}$ as seen in \eqref{charcrosssec} and $n$ for the interstellar medium at its densest in molecular clouds is around $10^6\mathrm{cm}^{-3}$ \cite{Vidarsson}, we have an $\ell_{mfp}$ of around $10^{20}$pc, while the Milky Way's diameter is only around 30kpc.\\
	
	So, using our Schwarzschild metric, we can consider the geodesic equations for $t,r$ and $\phi$. Recall that $g_{00}=-\left(1-\frac{2GM}{r}\right)$ and $g_{11}=\left(1-\frac{2GM}{r}\right)^{-1}$. Writing them suggestively, we have:
	
	\begin{equation}
		\frac{d^2t}{d\tau^2}+\frac{\partial_rg_{00}}{g_{00}}\frac{dr}{d\tau}\frac{dt}{d\tau}=0,
	\end{equation}
\begin{equation}
	\frac{d^2\phi}{d\tau^2}+\frac{2}{r}\frac{d\phi}{d\tau}\frac{dr}{d\tau}=0,
\end{equation}
\begin{equation}
	\frac{d^2r}{d\tau^2}+\frac{\partial_rg_{11}}{2g_{11}}\left(\frac{dr}{d\tau}\right)^2 -\frac{r}{g_{11}}\left(\frac{d\phi}{d\tau}\right)^2 -\frac{\partial_rg_{00}}{2g_{11}}\left(\frac{dt}{d\tau}\right)^2 =0.\label{rgeo}
\end{equation}

These first two equations can be written as complete derivatives, leading to conserved quantities. Dividing the $t$ equation by $\frac{dt}{d\tau}$, and using chain rule that $\frac{d}{dr}\frac{dr}{d\tau}=\frac{d}{d\tau}$, we have:

\begin{equation}
	\frac{d}{d\tau}\left[\ln\left(\frac{dt}{d\tau}g_{00}\right)\right]=0 \ \ \rightarrow \ \ \frac{dt}{d\tau} = \frac{C}{g_{00}}.\label{tgeo}
\end{equation}

We can use $C$ to define our parameter $\tau$, and we can choose $C=1$. Performing similar manipulation, for the $\phi$ geodesic equation we have

\begin{equation}
	\frac{d}{d\tau}\left[\ln\left(\frac{d\phi}{d\tau}r^2\right)\right]=0 \ \ \rightarrow \ \ \frac{d\phi}{d\tau}=\frac{J}{r^2},
\end{equation}

	where $J$ is another conserved quantity, and is once again the angular momentum per unit mass. Plugging these in to \eqref{rgeo}, we have
	
	\begin{equation}
		\frac{d^2r}{d\tau^2}+\frac{\partial_rg_{11}}{2g_{11}}\left(\frac{dr}{d\tau}\right)^2 -\frac{J^2}{g_{11}r^3}-\frac{\partial_rg_{00}}{2g_{11}g_{00}^2}=0.
	\end{equation}

Multiplying by $2g_{11}\frac{dr}{d\tau}$ and again using chain rule, we have that

\begin{equation}
	\frac{d}{d\tau}\left[g_{11}\left(\frac{dr}{d\tau}\right)^2+\frac{J^2}{r^2}+\frac{1}{g_{00}}\right] = 0 \ \ \rightarrow \ \ \frac{dr}{d\tau}=\sqrt{-\frac{E}{g_{11}}-\frac{J^2}{r^2g_{11}}-\frac{1}{g_{00}g_{11}}},\label{Eeqn}
\end{equation}

where $E$ is another constant of integration, and is the conserved quantity related to the energy of the particle. To see this, we consider the line element of our metric (recalling that $\theta=\frac{\pi}{2}$ and that $d\theta=0$):
\begin{equation}
	ds^2=g_{00}dt^2+g_{11}dr^2+r^2d\phi^2 \ \ \rightarrow \ \ ds^2=d\tau^2\left(\frac{1}{g_{00}}-E-\frac{J^2}{r^2}-\frac{1}{g_{00}}+\frac{J^2}{r^2}\right)=-Ed\tau^2.
\end{equation}

Thus, for massless particles with $ds^2=0$, $E=0$ and for massive particles we have $E>0$ as they are timelike.

Using \eqref{tgeo} and \eqref{Eeqn}, we have

\begin{equation}
	\frac{dt}{dr}=\sqrt{\frac{g_{11}}{g_{00}^2\left(-E-\frac{J^2}{r^2}-\frac{1}{g_{00}}\right)}}.\label{tr}
\end{equation}

 If we consider \eqref{tr} at infinity, $g_{00}$ and $g_{11}$ go to -1 and 1 respectively and we have 

\begin{equation}
	\left(\frac{dr}{dt}\right)^2\big|_\infty=-E+1,
\end{equation}

and so $E$ is related to the particle's velocity at infinity $v_0$ by $E=1-v_0^2$.\\
Now, to calculate the time delay we need to integrate \eqref{tr}. For our case, we are considering extremely relativistic neutrinos, and so $E\approxeq0$. Similarly, for our case of interest, $\frac{GM}{r}$ is a small quantity. Numerically, for our central black hole we have $GM\approx1.73\times10^{-7}$pc \cite{Mena:2006ym}, and $r$ on the order of kpc, so this is a very small quantity. Thus, it is reasonable to consider the time delay to linear order in $\frac{GM}{r}$. In this case, $g_{00}=-1+\frac{2GM}{r}$ and $g_{11}\approxeq1+\frac{2GM}{r}$. Then we have

\begin{equation}
	\frac{dt}{dr}\approxeq-\left(1+\frac{3GM}{r}\right)\left(1+\frac{2GM}{r}-\frac{J^2}{r^2}\right)^{-1/2}.
\end{equation}
	 
	 Next, if the particle's closest approach is at $r_0$, at $r_0$, $\frac{dr}{dt}=0$. So, at $r_0$ we have that $J^2=r_0^2(1+\frac{2GM}{r_0})$. But $J$ is constant, so this is true for all $r$. Plugging this in, we have
	 
	 \begin{equation}
	 	\frac{dt}{dr}\approxeq -\left(1+\frac{3GM}{r}\right)\left(1-\frac{r_0^2}{r^2} +2GM(\frac{1}{r}-\frac{r_0}{r^2})\right)^{-1/2}
	 \end{equation} 
 
 \begin{equation}
 	=-\frac{1}{\sqrt{1-\frac{r_0^2}{r^2}}}\left(1+\frac{3GM}{r}\right)\left(1+\frac{2GM}{r+r_0}\right)^{-1/2} = -\frac{1}{\sqrt{1-\frac{r_0^2}{r^2}}}\left(1+\frac{GM(2r+3r_0)}{r(r+r_0)}\right),
 \end{equation}

and so, integrating from the source $r_i$ to the point of closest approach $r_0$, and then from $r_0$ to the observer $r_f$, we have

\begin{equation}
	t=\int_{r_0}^{r_i}\frac{dr}{\sqrt{1-\frac{r_0^2}{r^2}}}\left(1+\frac{GM(2r+3r_0)}{r(r+r_0)}\right) + \int_{r_0}^{r_f}\frac{dr}{\sqrt{1-\frac{r_0^2}{r^2}}}\left(1+\frac{GM(2r+3r_0)}{r(r+r_0)}\right).
\end{equation}

Luckily, this integral can be performed analytically. The first term in each integral gives what is known as the ``geometric" contribution: i.e, the time taken for the particle to trace out its path. Though called geometric, it of course still comes to be owing to gravity, as without the lensing object, the path would simply be a straight line. Integrating this term gives

\begin{equation}
	t_{geo}=\sqrt{r_i^2-r_0^2}+\sqrt{r_f^2-r_0^2},
\end{equation}

the standard Pythagorean distance. The second term gives the ``gravitational" contribution (despite both terms owing to gravity) and is the gravitational time dilation owing to the lensing mass. Integrating gives 

\begin{equation}
t_{grav}=GM\left[\sqrt{\frac{r_i-r_0}{r_i+r_0}}+\sqrt{\frac{r_f-r_0}{r_f+r_0}} + 2\ln\left(\frac{r_i+\sqrt{r_i^2-r_0^2}}{r_0}\right) + 2\ln\left(\frac{r_f+\sqrt{r_f^2-r_0^2}}{r_0}\right)\right].	
\end{equation}

Defining $\eta=\beta D_S$, we have $r_i=\sqrt{D_{LS}^2+\eta^2}=D_L\sqrt{y^2+\beta^2(1+y)^2}$ and $r_f=D_L$, where as before $y=\frac{D_{LS}}{D_L}$. $r_0$ takes on two values, for the two images, $r_0=D_L|\theta_\pm|$. Thus, $\Delta t$, the time dispersion of the signals, is given by $t_--t_+$, which is $t=t_{geo}+t_{grav}$ with $\theta_\pm$ substituted in from solving \eqref{lens}:

\begin{equation}
	\Delta t \approxeq 2GM\left[2\frac{\beta}{\theta_E}\sqrt{1+\frac{\beta^2}{4\theta^2_E}} +\ln\left(\frac{1+\frac{\beta^2}{2\theta^2_E}-\frac{\beta}{\theta_E}\sqrt{1+\frac{\beta^2}{4\theta^2_E}}}{1+\frac{\beta^2}{2\theta^2_E}+\frac{\beta}{\theta_E}\sqrt{1+\frac{\beta^2}{4\theta^2_E}}}\right)\right].
\end{equation}

\begin{figure}
	\centering
	\includegraphics[width=0.7\linewidth]{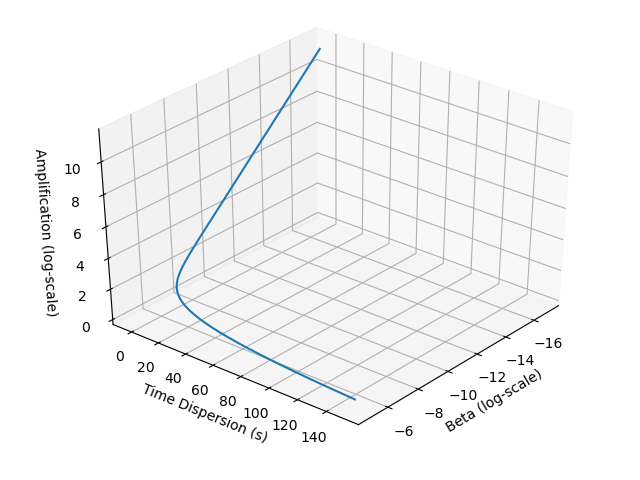}
	\caption{Relationship between the time dispersion, signal amplification and angle $\beta$ between the source and the observer-lens axis. The values of $\beta$ considered range from the minimum possible, $\beta=\frac{R_S}{D_S}$, up to the Einstein radius. Plotted using Python.}
	\label{Amp-TD-Beta}
\end{figure}

The above expression applies for a source on the exact opposite side of the black hole (that is, $y=1$). To achieve this expression, both $\beta$ and $\theta_E$ were taken to be much less than 1\footnote{But $\frac{\beta}{\theta_E}$ was not taken to be much less than 1.}, such that we could expand terms using binomial approximations. For very small values of $\beta$, as the situation approaches syzygy, we have 

\begin{equation}
	\Delta t \approxeq 2GM\left[2\frac{\beta}{\theta_E} +\ln\frac{1+\frac{\beta}{\theta_E}}{1-\frac{\beta}{\theta_E}}\right]\approxeq \frac{8GM\beta}{\theta_E},
\end{equation}

where we used that $\ln(1+x)\approx x$ for $x\ll1$. As we can see, for very small $\beta$, while the amplification gets larger, the time dispersion gets much smaller. Plotted in figure \ref{Amp-TD-Beta}, we see both the amplification's and the time dispersion's relations to the source's angle $\beta$. While a source close to the observer-lens axis will have a very large amplification of incoming neutrinos, a source closer to the Einstein radius will have a significant time dispersion. Thus, a lensed supernova event would either be characterised by an incredible increase to the neutrino flux, or by a significant increase to the signal's pulse length.\\

As we now see, there is a relationship in a lensing event between the magnification and the time dispersion. Thus, measuring the period of time that neutrinos are detected and comparing it with the observed amplification, more accurate models can be calculated for the pulse length and decay time of supernovae. Alternatively, (or, in conjunction), using the same relationship but in reverse, one can determine how much of the neutrino flux was owing to amplification, and thus how much of the energy of the supernova was released via neutrinos. In cases where the supernova was lensed by the Milky Way's supermassive black hole, neutrinos will be one of the main sources of information, as visible light will not be transparent to the galactic disk.\\

It is now clear that, using lensing, neutrinos can provide an interesting new lens through which to view supernovae. Unfortunately, as we shall see, the probability of such an event in our lifetime is vanishingly small. To see this, let us begin by approximating our galaxy as a disk, with supernovae being equally distributed according to the disk's mass density, modelled by

\begin{equation}
	\sigma(r)=\sigma_0e^{-r/r_0} \ \ ; \ \ r<r_G,
\end{equation}

	with $r_G$ the galactic radius (around 15 kpc) and $r_0$ a parameter of the model, with 3.5 kpc modelling the Milky Way fairly well \cite{Mena:2006ym}. Then, in this case, the fraction of supernovae that have some part at some radius $r$ is given by
	
	\begin{equation}
		f(r)=\frac{2\pi\int_{r+R_*}^{r-R_*}dr'r'\sigma(r')}{2\pi\int_{0}^{r_G}dr'r'\sigma(r')}\approxeq\frac{2\sigma(r)rR_*}{\int_{0}^{r_G}dr'r'\sigma(r')},\label{frac}
	\end{equation}

	where $R_*$ is the typical radius of a supernovae, taken to be $\approxeq10$ km, and in the final form we have used the approximation that between $r+R_*$ and $r-R_*$, $\sigma(r)$ is essentially constant. Integrating the denominator, halving the sky (as only supernovae on the other side of the supermassive black hole would be lensed towards us) and substituting in values as well as dimensionless parameters $y=\frac{r}{D_L}$ and $y_0=\frac{r_0}{D_L}$, we have
	
	\begin{equation}
		f(y)\approxeq 4.6\times 10^{-16}ye^{-y/y_0}.
	\end{equation}

	Next, we need to consider the probability that the supernova at this (dimensionless) radius $y$ is within the angular range to be lensed towards us. This gives an additional fraction of $\frac{2\theta_E}{2\pi}$, and so
	
	\begin{equation}
		P(y)\approxeq f(y)\frac{\theta_E}{\pi} \approxeq 1.35 \times 10^{-21} ye^{-y/y_0}\sqrt{\frac{y}{1+y}},
	\end{equation}

	where in the last step we used our expression for the Einstein radius in \eqref{einsteinring}. Integrating now over all radii, and then dividing by a factor of $\frac{2R_*}{D_L}$ as we integrated over radii twice (once now and before in \eqref{frac}), we have that the total fraction of supernovae lensed toward earth is $\approx 1.8\times 10^{-6}$, with different models for the galactic disk giving consistent results \cite{Mena:2006ym}. With the current rate of supernovae in our galaxy estimated to be around one every $47\pm12$ years, this would mean a lensing event towards earth would take place roughly once every 26 million years: the chance of this happening during our lifetime is miniscule.

	\subsection{Lensing of the C$\nu$B}

	Until recently, our only probe into the universe was electromagnetic radiation. These particles are massless and as a result, regardless of their energy or momentum, they always travel with the same speed. Relic neutrinos, however, have a momentum distribution characterised by the massless Fermi-Dirac distribution function, and along with the fact that there are 3 different massive species, there is an entire spectrum of velocities that the cosmic neutrinos can have. Owing to this, while the C$\nu$B was created long before the CMB - recall that the C$\nu$B is from when the universe was roughly 1 second old while the CMB came about almost 400 000 years later - the last scattering surface of it is actually much closer to us than that of the CMB \cite{Dodelson:2009ze}. The only time this is not the case is when there is a massless neutrino species.\\
	The total probability distribution over momenta $p_0$ is given by 
	
	\begin{equation}
		P(p_0)=\frac{1}{n_0}\int\frac{d^3p_0}{(2\pi)^3}\frac{1}{e^{p_0/T_0}+1}=1,
	\end{equation}
	
	where the factor of $\frac{1}{n_0}$ is the normalisation. Then, integrating over angular coordinates and using $n_0$ from \eqref{n0}, we have the differential probability as 
	
	\begin{equation}
		\frac{dP}{dp_0}=\frac{2}{3\zeta(3)T_0^3}\frac{p_0^2}{e^{p_0/T_0}+1}.\label{momdist}
	\end{equation}
	
	We can clearly see that there is a range of momenta, distributed according to the above equation, and for each momentum and mass there will be a different distance to the last scattering surface (LSS). For massive particles, the comoving distance $\chi$ travelled since the LSS is given by the integral \cite{Dodelson:2009ze}
	
	\begin{equation}
		\chi = \int_{t_i}^{t_0}\frac{dt}{a(t)}\frac{p}{E} = \int_{t_i}^{t_0}\frac{dt}{a(t)}\frac{p_0/a}{\sqrt{(p_0/a)^2+m_\nu^2}}, \label{comovingdist}
	\end{equation}
	 
	 which for the massless case reduces to the usual $\chi = \int_{t_i}^{t_0}\frac{dt}{a(t)}$. Since these possible values of $p_0$ have a range, this distance to the last scattering surface does not give a single value, as is the case with the CMB, and instead the LSS of the C$\nu$B is actually broad and spread out over many distances. The probability that a neutrino last scattered at a distance $\chi$ away from us is 
	 
	 \begin{equation}
	 	\frac{dP}{d\chi}=\frac{dP}{dp_0}\frac{dp_0}{d\chi} = \frac{2}{3\zeta(3)T_0^3}\frac{p_0^2}{e^{p_0/T_0}+1} \left(\int_{t_i}^{t_0}\frac{dt}{a^2(t)}\frac{m_\nu^2}{((p_0/a)^2+m_\nu^2)^{3/2}}\right)^{-1}.
	 \end{equation}
 
	 Note that there are 2 factors that broaden the LSS of relic radiation: the non-instantaneous nature of decoupling (as discussed in chapter 5) and the range of momenta. The former exists in both the CMB and C$\nu$B, and has a much smaller broadening effect on the LSS than the range of momenta.\\
	 
	 These broad ranges of last scattering surfaces - of which there are three, one for each massive species - can be extremely useful when lensed. While light signals arrive at us from what we conventionally call our past light ``cone" - that is, from events which lie on the \itshape surface \normalfont of a cone in a spacetime diagram - signals from the various last scattering surfaces of the C$\nu$B come from the entire volume of this light cone, and can therefore provide information on our entire causal volume theoretically. It has been said that the C$\nu$B could be used to watch the evolution of gravitational potentials in the observable universe ``as a movie, limited only by causality and the poor prospects for imminent detection" \cite{Lin:2019lko}.\\
	 
	 One of the most interesting uses of this feature is that we could observe the entire evolution of some lensing object - be it a black hole, galaxy, star, etc. This could provide the most concrete probe into galactic evolution, and would provide massive amounts of data which could allow us to better understand the process and construct more accurate models.\\
	 
	 Recall that massive particles, when lensed, are deflected according to their velocities, as was seen in \eqref{massdeflec}. Also recall from \eqref{moma} that the momentum today $p_0$ is related to an earlier time's momentum by $p_0=p(a)a(t)$, where we have taken $a_0=1$ today. Thus, using that $p=\gamma m_\nu v$, we have
	 
	 \begin{equation}
	 	p=\frac{p_0}{a} \ \ \rightarrow \ \ \gamma m_\nu v = \frac{\gamma_0v_0m_\nu}{a} \ \ \rightarrow \ \ \frac{v}{\sqrt{1-v^2}}=\frac{v_0}{a\sqrt{1-v_0^2}}
	 \end{equation}
	\begin{equation}
		\rightarrow v(a) = \frac{v_0}{\sqrt{a^2+v_0^2(1-a^2)}}.\label{velocitya}
	\end{equation}
	
	We can rewrite this in terms of the redshift $z=\frac{1}{a}-1$, where again we have taken the present scale factor to be unity. Then the velocity at a given redshift is given by 
	
	\begin{equation}
		v(z) = \frac{v_0}{\sqrt{\frac{1}{(1+z)^2}+v_0^2(1-\frac{1}{(1+z)^2})}}.
	\end{equation}
	
	If we consider the velocity at ``infinity" - that is, when $z\rightarrow\infty$, we retreive that all particles were fully relativistic at that time: $v(z)\rightarrow 1$. As $z$ decreases (that is, as we consider the velocity of the particle as it traverses spacetime and comes closer to us today), the velocity also decreases, as is expected.\\
	
	Let us begin our return to lensing. The velocity as a particle passes the point of closest approach is labelled $v_{lens}$. As we saw in \eqref{massdeflec}, the deflection angle is related to the velocity of the particle at infinity\footnote{Note here a slight confusion in notation: previously $v_0$ was the velocity at infinity, while in this section it is the velocity today.}. Let us consider then, if we were to replace this velocity at infinity with $v_{lens}$ how much our deflection angle would be altered.\\
	
	For the most alteration, we need to consider the smallest redshift. Let us consider lensing by a galaxy at a redshift $z=1$, for example's sake. Then our $v_{lens}$ is given by
	
	\begin{equation}
		v_{lens} = \frac{v_0}{\sqrt{\frac{1}{4}+v_0^2(1-\frac{1}{4})}} = \frac{2v_0}{\sqrt{1+3v_0^2}}.
	\end{equation}
	
	For particles still relativistic today with $v_0\approx1$, we have $v_{lens}\approx 1 = v_{inf}$, and so there is a negligible change to the formula. However, for particles that are ultra non-relativistic today, we have $v_{lens}\approx2v_0$, a very large change to the velocity at infinity\footnote{If we take the limit $z \rightarrow 0$, this only changes $v_{lens}$ by a factor of 2.}. For example, for a neutrino mass around 10 meV, from \eqref{lowenergy} we get on average $v_{lens}\approx 0.1$, an order lower than $v_{inf}$.
	
	If we consider this factor of 10 as the maximal deviation, then there is a maximum change to the deflection angle (truly proportional to $\frac{1+v_{inf}^2}{v_{inf}^2}$) of a factor of $\approx\frac{101}{2}$ when using $\frac{1+v_{lens}^2}{v_{lens}^2}$ instead. Thus, from \eqref{massdeflec} we can write the deflection angle as 
	
	\begin{equation}
		\alpha = \rho(z,v_0)\frac{2GM}{bv_{lens}^2}(1+v_{lens}^2) + \mathcal{O}\left(\left(\frac{GM}{bv_{lens}}\right)^2\right),
	\end{equation}

	where we have reinstated $G$ explicitly and $\rho(z,v_0)$ is a factor between 1 and $\approx\frac{1}{50}$ \footnote{The maximum when considering the $z\rightarrow0$ limit and a larger mass of about 50 meV is $\approx\frac{1}{1000}$.}, depending as we have seen on the redshift of the lens and the velocity of the particle at detection. It also obviously follows that the Einstein radius is 
	
	\begin{equation}
		\theta_E = \sqrt{\frac{2\rho(z,v_0)GM(1+v_{lens}^2)D_{LS}}{v_{lens}^2D_SD_L}},
	\end{equation}

	where the distances to the lens and source are given by
	
	\begin{equation}
		D_i = \eta \cdot v(a) = \int \frac{dt}{a(t)}v(a) = \int_{a_i}^{1} \frac{da}{H(a)a^2}v(a),
	\end{equation}

	where $i=L,S$ and $v(a)$ comes from \eqref{velocitya}. It follows that $D_{LS}=D_S-D_L$.
	
	It is interesting to note that there is a physical cutoff: since the LSS can be closer than the lensing object, obviously in that case the relic neutrinos will not be lensed. So, the cutoff is when $D_S=D_L$ or equivalently when $D_{LS}=0$.
	
	Let us consider then what would actually be detected. Recall that neutrinos are detected in flavour eigenstates (most probably the electron neutrino state), which are comprised of the 3 massive eigenstates. This flavour neutrino will be detected with some momentum $p_0$ coming from the direction of the lensing object, and based on which mass eigenstate it is in, will have some velocity $v_0$. Thus, for a given momentum, many observations will provide 3 superimposed results, whose amplitudes are related to the elements of the PMNS matrix. These 3 results, having different velocities, will thus correspond to 3 different LSSs (3 different $D_S$ values) and 3 different ``lookback times" ($D_L$ values) to the lensing object. Applying this to every possible momentum $p_0$ (whose distribution we saw in \eqref{momdist}), we get a continuous stream of information about the lensing object's gravitational potential: a ``movie" of its evolution.
	
	Of course, we also need to consider the source's position and its angle with the observer-lens axis, $\beta$. The most dramatic case is obviously that of $\beta=0$, wherein we have our Einstein rings. Since the C$\nu$B is expected to come from every direction, there is no reason why there would not be a source at $\beta=0$, and so we should be able to observe Einstein rings. Of course, arcs from weaker lensing would also be visible, but would require statistical analysis to fully understand, similarly to the weak lensing of the CMB. \\
	
	 As can be seen in \eqref{comovingdist}, neutrinos with larger masses have closer last scattering surfaces, and thus enter the lens' potential at a later time. If the lensing object is growing in time (as we expect galaxy clusters to do in the standard model of cosmological evolution), those entering its potential later will be deflected more, as its gravitational potential would be deeper. In this way, the lens would also act as a mass spectrometer, splitting the mass eigenstates in the angular plane for each momentum bin. Any deviation from this expected result (i.e. more massive neutrinos being deflected less) would be an indication that our current model of structure formation is incorrect - a revolutionary thought.
	 
	 Finally, there are some considerations that have not been taken into account for this discussion, which change results quantitatively but the gist of the concepts remain. These are the integrated Sachs-Wolfe (ISW) effect and the peculiar velocity of the lensing object. The integrated Sachs-Wolfe effect is the loss of energy of particles as they enter into gravitational potentials and must thereafter exit them, and this will then affect the $p_0$ of neutrinos today. Of course, as the object evolves and becomes more massive, the ISW effect would be more substantial and so $p_0$ would be altered based on itself. 
	 
	 The peculiar velocity of the lensing object, be it a galaxy, cluster or something else, will have a subtle effect if the velocity is towards or away from earth (as this will simply affect $D_L(a)$), but will complicate the situation immensely if it is perpendicular to our line of sight, as then neutrinos from different lookback times will come from different directions. 
	 
	 Though this seems exciting, we also need to consider the fact that neutrino detectors are miles from achieving this. While some high-energy neutrino detectors, like Cherenkov detectors discussed in chapter 6, have the ability to see from which direction the incident neutrino arrived, the C$\nu$B detectors in the near future (such as PTOLEMY) do not. For use of this method, an entirely new experiment would need to be devised wherein extremely low energy neutrinos can be detected along with their direction of incidence.\\
	 
	 This chapter took a massive turn from previous ones, and is for the most part self-contained. As we saw, the possibilities that gravitational lensing of neutrinos offers us are endless, and the field is still young. The effects of lensing discussed in this chapter were used primarily as a probe to study the cosmos - using supernova neutrinos to investigate supernovae, and relic neutrinos to study the evolution of objects ranging from stars to galaxy clusters. This is neutrino astronomy in its purest sense. In the next chapter, however, we will do the opposite: using the known properties of a lensing object, we will be able to learn about neutrinos.
	
	\newpage
	\section{The Effect of a Neutrino's Spin}
	\subsection{The Classical Mathisson–Papapetrou Equations}
	
	To begin this section, we start by modelling our neutrinos (or any particle with both mass and spin) as actual solid spinning spheres. As we will see in the next subsection, though this view of spin is outdated, the same result is obtained when taking the leading order term in a quantum calculation.\\
	
	For the calculations only in this and the next subsections, we will use the $(+,-,-,-)$ signature, which is the opposite of what we have done up to this point, such that our calculation follows that given originally in \cite{Papapetrou:1951pa}. This has very minimal effect, and when important will be mentioned.
	
	 Consider a body with radius $R$ such that this radius is tiny when compared to the length scales of the spacetime (such as a Schwarzschild radius), and which has an energy momentum tensor density $\tilde{T}^{\mu\nu}=\sqrt{-g}T^{\mu\nu}$ describing it.
	
	For our spinning particle whose centre follows a path labelled $X(t)$, we can consider a multipole expansion of the internal structure of the particle, with each point within the particle being at $x(t)$ and $\delta x^i = x^i - X^i$ being the distance coordinates from the centre of mass.

\begin{figure}[h]
	\centering
	\includegraphics[width=0.4\linewidth]{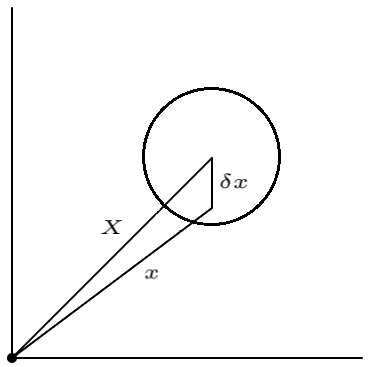}
	\caption{Diagram depicting a classical particle with internal structure and the relation between $X$,$x$, and $\delta x$. Drawn using \cite{Drawingtool}.}
\end{figure}

	The monopole terms will be of the form $\int T^{\mu\nu}\sqrt{-g} d^3x$, and the dipole terms $\int \delta x^i T^{\mu\nu}\sqrt{-g} d^3x$, with higher multipoles following the same pattern, and integration being done at a constant time $t$. As we shall see, when we consider only the particle's monopole moment, we retrieve the geodesic equation for a point particle. The reason higher-order terms are neglected is that each multipole is smaller by a factor of $\frac{R}{r_s}$, where $r_s$ is the characteristic length scale of the spacetime.  \\ 
	
	We begin with the usual conservation of energy-momentum equation:
	
	\begin{equation}
		\nabla_\mu(\sqrt{-g}T^{\mu\nu}) = 0 \ \rightarrow \ \partial_\mu (\sqrt{-g}T^{\mu\nu}) + \tensor{\Gamma}{^\nu_{\alpha\beta}}(\sqrt{-g}T^{\alpha\beta}) = 0,  \label{dynameq}
	\end{equation}

	where we used that $\nabla_\mu \sqrt{-g}=\partial_\mu\sqrt{-g}-\tensor{\Gamma}{^\nu_{\mu\nu}}\sqrt{-g}=0$.  That is,
	
	\begin{equation}
		\partial_\mu \tilde{T}^{\mu\nu} = -\tensor{\Gamma}{^\nu_{\alpha\beta}}\tilde{T}^{\alpha\beta}. \label{EMrelation}
	\end{equation}
	
	From here onwards we shall drop the tilde over $\tilde{T}^{\mu\nu}$, and shall simply use $T^{\mu\nu}$ as the label of our tensor density. Next, by the simple product rule and \eqref{EMrelation}, we have that 
	
	\begin{equation}
		\partial_\gamma(x^\alpha T^{\beta\gamma}) = \delta^\alpha_\gamma T^{\beta\gamma} + x^\alpha\partial_\gamma T^{\beta\gamma} = T^{\alpha\beta} - x^\alpha\tensor{\Gamma}{^\beta_{\mu\nu}}T^{\mu\nu},
	\end{equation}

	which, integrated over, gives
	
	\begin{equation}
		\frac{d}{dt}\int dV x^\alpha T^{\beta0} = \int dV T^{\alpha\beta} - \int dV x^\alpha\tensor{\Gamma}{^\beta_{\mu\nu}}T^{\mu\nu},
	\end{equation}

	where to obtain this expression, we used the fact that the metric is time-independent, and also used the divergence theorem to transform $\int dV \partial_i (x^\alpha T^{\beta i})$ into $\int dS x^\alpha T^{\beta i} \hat{n}_i=0$. The reason this integral is zero is that $T^{\mu\nu}$ is the stress-energy-momentum (density) of the particle, and so it cannot flow in or out of the particle's boundaries. Using the same tricks, we have from \eqref{EMrelation} that
	
	\begin{equation}
		\frac{d}{dt}\int dV T^{\alpha0} = -\int dV \tensor{\Gamma}{^\alpha_{\mu\nu}}T^{\mu\nu}. \label{SEeqn2}
	\end{equation}

	So using these equations we can write 
	
	\begin{equation}
		\int dV T^{\alpha\beta} = \int dV x^\alpha \frac{d}{dt}T^{\beta0} + \int dV \frac{d}{dt}x^\alpha T^{\beta0} + \int dV x^\alpha\tensor{\Gamma}{^\beta_{\mu\nu}}T^{\mu\nu}. \label{SEeqn1}
	\end{equation}

	To continue, we need to decide to which order of multipole we wish to work. For pedagogy, we begin by considering a monopole particle. In this case, we use that $x^\mu = X^\mu + \delta x^\mu$, and integrals containing $\delta x^\mu$ are taken to be negligible. Similarly, we must expand the Christoffel symbols in a Taylor series:
	
	\begin{equation}
		\tensor{\Gamma}{^\alpha_{\mu\nu}} = \ _X\tensor{\Gamma}{^\alpha_{\mu\nu}} + \partial_\sigma (_X\tensor{\Gamma}{^\alpha_{\mu\nu}}) \delta x^\sigma + \dots,
	\end{equation}

	where the subscript $X$ denotes the value of the Christoffel symbol at the particle's centre, and of course higher order terms will include higher powers of $\delta x^\sigma$. For our monopole toy particle then we have $\tensor{\Gamma}{^\alpha_{\mu\nu}} = \ _X\tensor{\Gamma}{^\alpha_{\mu\nu}}$. So using this and \eqref{SEeqn2}, \eqref{SEeqn1} becomes
	
	\begin{equation}
		\int dV T^{\alpha\beta} = -X^\alpha\int dV \tensor{\Gamma}{^\beta_{\mu\nu}}T^{\mu\nu} + \frac{dX^\alpha}{dt}\int dV T^{\beta0} + X^\alpha\int dV \tensor{\Gamma}{^\beta_{\mu\nu}}T^{\mu\nu} = \frac{dX^\alpha}{dt}\int dV T^{\beta0}. \label{SEeqn3}
	\end{equation}

	To continue, we define the quantities for a monopole particle
	
	\begin{equation}
		M^{\alpha\beta} = u^0\int dV T^{\alpha\beta},\label{Morigdef}
	\end{equation}

	where $u^\mu$ is the 4-velocity of the centre of mass of the particle with respect to the proper time, $u^\mu=\frac{dX^\mu}{d\tau}$ so that $u^0 = \frac{dt}{d\tau}$. Then, from \eqref{SEeqn2} and \eqref{SEeqn3}, we have 
	
	\begin{equation}
		\frac{d}{d\tau}\left(\frac{M^{\alpha0}}{u^0}\right) + \tensor{\Gamma}{^\alpha_{\mu\nu}}M^{\mu\nu} = 0 \ \ \ \ ; \ \ \ \ M^{\alpha\beta} = \frac{u^\alpha}{u^0}M^{\beta0}. \label{Meqn1}
	\end{equation}

    Taking the $\beta=0$ equation, we have $M^{\alpha0} = \frac{u^\alpha}{u^0}M^{00}$, and plugging this back in \eqref{Meqn1} we have
    
    \begin{equation}
    	 M^{\alpha\beta} = mu^\alpha u^\beta,
    \end{equation}

	where $m$, the rest mass of the particle, is given by $m = \frac{M^{00}}{(u^0)^2}$. Plugging this in to the former equation in \eqref{Meqn1}, we have that
	
	\begin{equation}
		\frac{d}{d\tau}\left(mu^\alpha\right) + \tensor{\Gamma}{^\alpha_{\mu\nu}}mu^\mu u^\nu = 0. 
	\end{equation}

	Finally, by multiplying this equation by $u_\alpha$ and using that $u_\alpha\frac{du^\alpha}{d\tau} + \tensor{\Gamma}{^\alpha_{\mu\nu}}u_\alpha u^\mu u^\nu = 0$ (since $u_\mu u^\mu=1$ in this signature), we obtain 2 equations:
	
	\begin{equation}
		\frac{dm}{d\tau}=0,
	\end{equation}
\begin{equation}
	\frac{du^\alpha}{d\tau} + \tensor{\Gamma}{^\alpha_{\mu\nu}}u^\mu u^\nu = 0.
\end{equation}

	The first of these shows that the particle's rest mass is conserved along its journey, while the second is the well-known geodesic equation that we expected to obtain. Next we repeat this calculation, but this time considering the dipole moments as well. \\
	
	To start, we consider
	
	\begin{equation}
		\partial_\delta(x^\alpha x^\beta T^{\gamma\delta}) = x^\beta T^{\gamma\alpha} + x^\alpha T^{\gamma\beta} - x^\alpha x^\beta \tensor{\Gamma}{^\gamma_{\mu\nu}}T^{\mu\nu},\label{SEeqn4}
	\end{equation}

	where the last term was obtained using \eqref{EMrelation}. For our dipole considerations, we will now use $\tensor{\Gamma}{^\alpha_{\mu\nu}} = \ _X\tensor{\Gamma}{^\alpha_{\mu\nu}} + \partial_\sigma (_X\tensor{\Gamma}{^\alpha_{\mu\nu}}) \delta x^\sigma$ but drop the $X$ subscript from here on. This firstly affects \eqref{SEeqn2} to give
	
	\begin{equation}
		\frac{d}{dt}\int dV T^{\alpha0} + \tensor{\Gamma}{^\alpha_{\mu\nu}}\int dV T^{\mu\nu} + \partial_\sigma\tensor{\Gamma}{^\alpha_{\mu\nu}} \int dV \delta x^\sigma T^{\mu\nu} = 0 \label{SEnew1}
	\end{equation}

	and, recalling that $x^\mu = X^\mu + \delta x^\mu$, \eqref{SEeqn1} gives
	
	\begin{equation}
		\int dV T^{\alpha\beta} = \frac{dX^\alpha}{dt}\int dV T^{\beta0} + \frac{d}{dt}\int dV \delta x^\alpha T^{\beta0} + \tensor{\Gamma}{^\beta_{\mu\nu}}\int dV \delta x^\alpha T^{\mu\nu}. \label{SEnew2}
	\end{equation}

	Then, integrating \eqref{SEeqn4} and using \eqref{SEnew1} and \eqref{SEnew2}, after some calculations (wherein again we use similar tricks to before) we arrive at
	
	\begin{equation}
		\int dV \delta x^\alpha T^{\beta\gamma} + \int dV \delta x^\beta T^{\alpha\gamma} = \frac{dX^\alpha}{dt}\int dV \delta x^\beta T^{\gamma0} + \frac{dX^\beta}{dt}\int dV \delta x^\alpha T^{\gamma0}. \label{SEnew3}
	\end{equation}

	These are the equations of motion for the particle up to the dipole moment. We need to make things a bit more understandable, and to do so we begin by defining
	
	\begin{equation}
		M^{\lambda\mu\nu} = -u^0\int dV \delta x^\lambda T^{\mu\nu},\label{M3}
	\end{equation}

	such that $M^{\lambda\mu\nu}$ is symmetric in its last 2 indices, but not all 3. Also, since the integral is evaluated at a constant $t$, $M^{0\mu\nu}$ is always 0. The spin tensor of the particle is closely related to this quantity, where the spin tensor is given by
	
	\begin{equation}
		S^{\alpha\beta} = \int dV \delta x^\alpha T^{\beta0} - \int dV \delta x^\beta T^{\alpha0} = -\frac{1}{u^0}(M^{\alpha\beta0}-M^{\beta\alpha0}). \label{spintoM}
	\end{equation}

	Using the definition in \eqref{M3}, as well as the fact that $\frac{dX^\alpha}{dt} = \frac{u^\alpha}{u^0}$, we have from \eqref{SEnew3} that 
	
	\begin{equation}
		u^0(M^{\alpha\beta\gamma}+M^{\beta\alpha\gamma}) = u^\alpha M^{\beta\gamma0} + u^\beta M^{\alpha\gamma0}.\label{3.7}
	\end{equation}

	Let us call this the ``$\alpha-\beta-\gamma$" version of this equation, based on the indices of the first $M$ tensor. We can also get cyclic perturbations of this equation, and by adding the $\gamma-\alpha-\beta$ equation and subtracting the $\beta-\gamma-\alpha$ one, we obtain 
	
	\begin{equation}
		2u^0 M^{\alpha\beta\gamma} = u^\alpha(M^{\beta\gamma0}+M^{\gamma\beta0} ) -u^0(u^\beta S^{\alpha\gamma} + u^\gamma S^{\alpha\beta}),
	\end{equation}

	where we also used \eqref{spintoM}. From \eqref{3.7} for the case where $\gamma=0$ and \eqref{spintoM} with $\beta=0$, we have that 
	
	\begin{equation}
		u^0(M^{\alpha\beta0}+M^{\beta\alpha0}) = u^\alpha M^{\beta00}+u^\beta M^{\alpha00} = -u^\alpha u^0 S^{\beta0} -u^\beta u^0 S^{\alpha0}.
	\end{equation}

	Putting all the pieces together, and using that the spin tensor is antisymmetric,  we have that
	
	\begin{equation}
		2M^{\alpha\beta\gamma} = \frac{u^\alpha}{u^0}(u^\beta S^{0\gamma} + u^\gamma S^{0\beta}) - (S^{\alpha\beta}u^\gamma + S^{\alpha\gamma}u^\beta),\label{MtoS}
	\end{equation}

	so that the $M^{\lambda\mu\nu}$ can all be expressed in terms of velocities and spin tensors. Now we can rewrite \eqref{SEnew2} using \eqref{Morigdef} and \eqref{M3} as 
	
	\begin{equation}
		M^{\alpha\beta} = \frac{u^\alpha}{u^0}M^{\beta0} - \frac{d}{d\tau}\left(\frac{M^{\alpha\beta0}}{u^0}\right)- \tensor{\Gamma}{^\beta_{\mu\nu}}M^{\alpha\mu\nu}.\label{calceqn1}
	\end{equation}

	Calculating the $\beta=0$ term and substituting it back in, we have 
	
	\begin{equation}
			M^{\alpha\beta} = \frac{u^\alpha}{u^0}\left(\frac{u^\beta}{u^0}M^{00}-\frac{d}{d\tau}\left(\frac{M^{\beta00}}{u^0}\right)- \tensor{\Gamma}{^0_{\mu\nu}}M^{\beta\mu\nu}\right)- \frac{d}{d\tau}\left(\frac{M^{\alpha\beta0}}{u^0}\right)- \tensor{\Gamma}{^\beta_{\mu\nu}}M^{\alpha\mu\nu}.
	\end{equation}

	So, taking $M^{\alpha\beta}-M^{\beta\alpha}=0$ owing to its symmetry, using the previous equation along with \eqref{MtoS} and the $\beta=0$ form of \eqref{calceqn1} and \eqref{spintoM}, we have the equation of motion for the spin tensor:
	
	\begin{equation}
		\frac{dS^{\alpha\beta}}{d\tau}+\frac{u^\alpha}{u^0}\frac{dS^{\beta0}}{d\tau} - \frac{u^\beta}{u^0}\frac{dS^{\alpha0}}{d\tau}+ (\tensor{\Gamma}{^\alpha_{\mu\nu}}-\frac{u^\alpha}{u^0}\tensor{\Gamma}{^0_{\mu\nu}})M^{\beta\mu\nu} - (\tensor{\Gamma}{^\beta_{\mu\nu}}-\frac{u^\beta}{u^0}\tensor{\Gamma}{^0_{\mu\nu}})M^{\alpha\mu\nu} = 0,\label{count1}
	\end{equation}

	where we can use \eqref{MtoS} to write the last two terms in terms of the spin tensor. We can also multiply \eqref{SEnew1} by $u^0$ and rewrite it as
	
	\begin{equation}
		\frac{d}{d\tau}\left(\frac{M^{\alpha0}}{u^0}\right)+\tensor{\Gamma}{^\alpha_{\mu\nu}}M^{\mu\nu} -\partial_\sigma\tensor{\Gamma}{^\alpha_{\mu\nu}}M^{\sigma\mu\nu}=0.\label{count2}
	\end{equation}
	
	At this point, we have 10 unknowns: $M^{00}$, 3 independent components of the velocity vector, and 6 independent components of the antisymmetric spin tensor. At first glance, we appear to have 6 equations from \eqref{count1} (owing to the antisymmetry in $\alpha\leftrightarrow\beta$) and 4 equations from \eqref{count2}, but unfortunately 3 of those in \eqref{count1} are trivial identities \cite{Papapetrou:1951pa}, and as such we will need supplementary conditions, to be chosen a bit later.\\
	
	We wish to write our equations of motion in covariant notation. Firstly, it can be shown \cite{Papapetrou:1951pa} that $S^{\alpha\beta}$ is a tensor, $\frac{1}{u^0}\left(M^{\alpha0}+\tensor{\Gamma}{^\alpha_{\mu\nu}}u^\mu S^{\nu0}\right)$ is a vector, and as such, 
	
	\begin{equation}
		m = \frac{u_\alpha}{u^0}\left(M^{\alpha0}+\tensor{\Gamma}{^\alpha_{\mu\nu}}u^\mu S^{\nu0}\right)\label{masseqn}
	\end{equation}
	
	is a scalar representing the mass of the particle, which has changed from the monopole particle which recall had $m = \frac{M^{00}}{(u^0)^2}$.\\
	
	Next, we define $\frac{Df}{D\tau} = u^\nu\nabla_\nu f$ and note that $u^\nu\partial_\nu f = \frac{df}{d\tau}$. Writing \eqref{count1} in terms of this covariant derivative, and using some symmetries of indices, we immediately have the first of the Mathisson-Papapetrou equations:
	
	\begin{equation}
		\frac{DS^{\alpha\beta}}{D\tau} = p^\alpha u^\beta - p^\beta u^\alpha,\label{MP1}
	\end{equation}

	wherein $p^\mu = \frac{M^{\mu0}}{u^0}-\frac{u^\nu}{u^0}\tensor{\Gamma}{^\mu_{\alpha\nu}}S^{0\alpha}$ is the momentum of our dipole particle.
	
	Using our covariant derivative now, we have from \eqref{count1} a Bianchi-type identity for the covariant derivative of the spin:
	
	\begin{equation}
		\frac{u^0}{u^0}\frac{DS^{\alpha\beta}}{D\tau} +\frac{u^\alpha}{u^0}\frac{DS^{\beta0}}{D\tau} +\frac{u^\beta}{u^0}\frac{DS^{0\alpha}}{D\tau}=0,\label{Bianchi}
	\end{equation}

	where $\frac{u^0}{u^0}$ is obviously unity but was left in for symmetry. Using this, along with the fact that $u^0S^{\alpha0} = -M^{\alpha00}$, the $\beta=0$ equation from \eqref{calceqn1} yields
	
	\begin{equation}
		M^{\alpha0} + \tensor{\Gamma}{^\alpha_{\mu\nu}}S^{\mu0}u^\nu = \frac{u^\alpha}{u^0}\left(M^{00}+\tensor{\Gamma}{^0_{\mu\nu}}S^{\mu0}u^\nu\right) + \frac{DS^{\alpha0}}{D\tau}.
	\end{equation}
	
	From this, and using the mass in \eqref{masseqn} as well as \eqref{Bianchi}, we find
	
	\begin{equation}
		\frac{M^{\alpha0}}{u^0} + \tensor{\Gamma}{^\alpha_{\mu\nu}}S^{\mu0}\frac{u^\nu}{u^0} = mu^\alpha +u_\beta\frac{DS^{\alpha\beta}}{D\tau},\label{partmomentum}
	\end{equation}
	
	which is equivalent to $p^\alpha$, as we saw above, and so we now have 2 expressions for the momentum of the particle. In usual flat space for a spinless particle, we would not have the second term on either side of this equation: it is interesting to note that the dipole nature of the particle causes a change to its momentum based on the coupling of curvature and spin.
	
	Inserting this result into \eqref{count2}, and using all the results derived until this point along with the relationship between the Riemann curvature tensor and the Christoffel symbols, we obtain the second Mathisson-Papapetrou (M-P) equation:
	
	\begin{equation}
		\frac{Dp^\alpha}{D\tau} = \frac{1}{2}\tensor{R}{^\alpha_{\beta\mu\nu}}u^\beta S^{\mu\nu}.\label{MP2}
	\end{equation}

	The final piece of information we need comes in the form of the Dixon-Tulczyjew supplementary condition. To begin, we can rewrite our spin tensor in terms of the particle's spin vector using \cite{Bini:2014poa}
	
	\begin{equation}
		S^\alpha = \frac{1}{2}\tensor{\eta}{_\beta^{\alpha\mu\nu}}u^\beta S_{\mu\nu},
	\end{equation}

	wherein $\eta_{\beta\alpha\mu\nu}=\sqrt{-g}\epsilon_{\beta\alpha\mu\nu}$ and $\epsilon_{\beta\alpha\mu\nu}$ is the 4D Levi-Civita tensor. The converse of this is that $S_{\mu\nu} = \eta_{\beta\alpha\mu\nu}u^\beta S^\alpha$.
	
	From this relation, we can write the magnitude of the spin vector $s$ in terms of the spin tensor:
	
	\begin{equation}
		s^2=S^\alpha S_\alpha = \frac{1}{2}S_{\mu\nu}S^{\mu\nu}.
	\end{equation}

	For the case of a neutrino, we know that $s^2=\braket{S^2}=\tilde{s}(\tilde{s}+1)=\frac{3}{4}$ (when $\hbar=1$) since neutrinos are spin-$\frac{1}{2}$ particles. This quantity $s^2$ is conserved, and as such we have that
	
	\begin{equation}
		\frac{D(s^2)}{D\tau} = 0 = \frac{D}{D\tau}\left(\frac{1}{2}S_{\mu\nu}S^{\mu\nu}\right) = S_{\mu\nu}\frac{DS^{\mu\nu}}{D\tau} = S_{\mu\nu}(p^\mu u^\nu - p^\nu u^\mu) = 2S_{\mu\nu}p^\mu u^\nu,
	\end{equation}

	where we used \eqref{MP1} and the antisymmetry of the spin tensor. This gives the Dixon-Tulczyjew supplementary condition: that $S^{\mu\nu}p_\nu=0$ \footnote{This could give an alternate condition, $S^{\mu\nu}u_\nu=0$. There are in fact many different supplementary conditions that can be considered, but for our purposes we will focus on the D-T condition.}. This gives the final 3 equations needed to solve our system.
	
	Thus, the full system of equations - known collectively now as the Mathisson-Papapetrou-Dixon-Tulczyjew equations - are
	
	\begin{equation}
		\frac{Dp^\alpha}{D\tau} = \frac{1}{2}\tensor{R}{^\alpha_{\beta\mu\nu}}u^\beta S^{\mu\nu},
	\end{equation}
	\begin{equation}
		\frac{DS^{\alpha\beta}}{D\tau} = p^\alpha u^\beta - p^\beta u^\alpha,
	\end{equation}
    \begin{equation}
		S^{\mu\nu}p_\nu=0.\label{MP3}
	\end{equation}
	
	\subsection{As a Result of QFT Considerations}
	
	Though classical and quantum ``spin" have a tenuous relationship, and the comparison between them is often misguided, the M-P equations seem to provide a scenario wherein the classical spin - that is, the movement of a particle around its centre of mass - and the quantum spin - a quantum number conserved as an intrinsic property of a particle - align. Though this derivation will lead to (almost) the same equations, there are some caveats, as we shall discuss.\\
	
	We begin with the Dirac equation in curved spacetime:
	
	\begin{equation}
		i\gamma^\mu\nabla_\mu\psi-\frac{m}{\hbar}\psi=0, \label{Diraceqn}
	\end{equation}
	
	where the covariant derivative acting on a spinor is related to the orthonormal tetrad field $\tensor{h}{_a^\alpha}$ via 
	
	\begin{equation}
		\nabla_\mu = \partial_\mu + \frac{1}{4}\left(\nabla_\mu\tensor{h}{_a^\alpha}\right)h_{\alpha b}\gamma^b\gamma^a.
	\end{equation}

	These orthonormal tetrad fields are related by $g_{\alpha\beta}\tensor{h}{_a^\alpha}\tensor{h}{_b^\beta} = \eta_{ab}$, with $\eta_{ab}$ the Minkowski metric. In this subsection, Greek indices will be raised and lowered using our spacetime metric, while the Minkowski metric applies to Latin indices. The factors of $\hbar$ will be kept in explicitly, as this is the small parameter that we will perform perturbations around. For the sake of following the derivation in \cite{Audretsch:1981wf}, we once again work in the $(+,-,-,-)$ signature.
	
	In order to analyse the classical limit of our quantum particle, we expand its wave function in a WKB expansion, keeping the first two terms:
	
	\begin{equation}
		\psi(x) = e^{iS(x)/\hbar} \sum_{n=0}^{\infty}(-i\hbar)^na_n(x) = e^{iS(x)/\hbar}(a_0(x)-i\hbar a_1(x)+\dots).
	\end{equation}

	Inserting this into \eqref{Diraceqn} and equating coefficients of powers of $\hbar$, we get for the $\hbar^{-1}$ and $\hbar^0$ equations
	
	\begin{equation}
		(\gamma^\mu\nabla_\mu S+m)a_0 = 0, \label{a0eqn}
	\end{equation}

	\begin{equation}
		(\gamma^\mu\nabla_\mu S+m)a_1 = -\gamma^\mu\nabla_\mu a_0. \label{a1eqn}
	\end{equation}

	In order for $a_0$ not to be just 0 (such that our wave function is not the trivial solution), we require that $\det(\gamma^\mu\nabla_\mu S+m)=0$, so we can write $\gamma^\mu\nabla_\mu S=-m$. Multiplying this by $\gamma^\nu\nabla_\nu S$ and using the anticommutation relation between gamma matrices
	
	\begin{equation}
		\gamma^{(\mu}\gamma^{\nu)}=g^{\mu\nu}, \label{anticommutator}
	\end{equation} 

	we have that $\nabla^\mu S\nabla_\mu S =m^2$. Owing to this, we define a momentum-like quantity $P_\mu = -\nabla_\mu S$, such that it satisfies the on-shell condition ($P^2=m^2$) and if the wave function has a plane-wave structure, then this is the momentum. Continuing this comparison, we define a velocity-like quantity $v_\mu=\frac{p_\mu}{m}$, which is normalised correctly: $v^\mu v_\mu=1$. It is difficult to apply this formalism to the case of a massless fermion, as there does not seem to be a candidate to act as a meaningful, correctly-normalised velocity. These $v^\mu$ are simply timelike normalised vectors orthogonal to the spacelike hypersurfaces of $S=$constant that form a congruence of geodesic worldlines through each $S=$constant. The covariant derivative of this congruence can be decomposed into
	
	\begin{equation}
		\nabla_\alpha v_\beta = \sigma_{\alpha\beta} + \omega_{\alpha\beta} + \frac{\theta}{3}k_{\alpha\beta},\label{decomposition}
	\end{equation}

	where $\frac{\theta}{3}k_{\alpha\beta} = \frac{\theta}{3}(g_{\alpha\beta}-v_\alpha v_\beta)$ is the expansion of the congruence\footnote{This $k_{\alpha\beta}$ is usually denoted $h_{\alpha\beta}$ but this could cause confusion with respect to the tetrad fields.}, $\sigma_{\alpha\beta}=\tensor{h}{_\alpha^\mu}\tensor{h}{_\beta^\nu}\nabla_{(\mu} v_{\nu)}-\frac{1}{3}\nabla_\mu v^\mu k_{\alpha\beta}$ its shear, and $\omega_{\alpha\beta} = \nabla_{[\mu} v_{\nu]}-\frac{D}{D\tau}(v_{[\alpha})v_{\beta]}$ is the vorticity, which is 0 since these are geodesics. So $\nabla_\alpha v_\beta = \sigma_{\alpha\beta} +\frac{\theta}{3}k_{\alpha\beta}$.\\
	
	Let us return to solving \eqref{a0eqn}. We make the ansatz that $a_0(x) = \beta_1(x)b_{01}(x) + \beta_2(x) b_{02}(x)$, and using that 
	
	\begin{equation}
		\gamma^0 = \begin{pmatrix}
			\mathds{1} & 0\\ 0 & -\mathds{1}
		\end{pmatrix} \ \ \ \ \ ; \ \ \ \ \ \gamma^i = \begin{pmatrix}
		0 & \sigma^i\\-\sigma^i & 0
	\end{pmatrix},
	\end{equation}

	where $\sigma^i$ are of course the Pauli matrices, calculating similarly to \eqref{refme}, we have the orthogonal solutions
	
	\begin{equation}
		b_{01} = \sqrt{\frac{E+m}{2m}}\begin{pmatrix}
			1\\0\\ \frac{k^3}{E+m} \\ \frac{k^1+ik^2}{E+m}
		\end{pmatrix} \ \ \ \ \ \ \ \ \ b_{02} = \sqrt{\frac{E+m}{2m}}\begin{pmatrix}
		0\\1\\ \frac{k^1-ik^2}{E+m} \\ -\frac{k^3}{E+m}
	\end{pmatrix},
	\end{equation}
	
	where we have defined $E=p^\mu \tensor{h}{_\mu^0}$ and $k^i=p^\mu \tensor{h}{_\mu^i}$.
	
	The homogeneous part of \eqref{a1eqn} is the same as \eqref{a0eqn}, and so $a_1$'s 3 solutions are $b_{01}$,$b_{02}$, and a final solution proportional to the inhomogeneity $\gamma^\mu\nabla_\mu a_0$. In order for this to be another solution and not merely a linear combination of the others, it must be orthogonal to them. Thus we have
	
	\begin{equation}
		\bar{b}_{0i}\gamma^\mu \nabla_\mu a_0 = 0,\label{solvecond}
	\end{equation}

	where $i=1,2$ and the bar denotes Pauli conjugation, $\bar{f}=f^\dagger\gamma^0$.\\
	
	To continue, we can select a particular tetrad. We do not lose any generality in doing so, as tetrads can be later rotated to any general tetrad to retreive a general result \cite{Audretsch:1981wf}. We choose our tetrad such that
	
	\begin{equation}
		\tensor{h}{_\mu^0} = v^\mu \ \ \ \ ; \ \ \ \ \nabla_\alpha\tensor{h}{_\mu^i} = 0.
	\end{equation}

	In this particular case, we have $\nabla_\mu = \partial_\mu$, $E=m$ and $k^i=0$. It is easy to see then that 
	
	\begin{equation}
		\bar{b}_{0i}\gamma^\mu b_{0j} = v^\mu \delta_{ij}, \label{orthogvector}
	\end{equation}

	since $\gamma^\mu={h}{^\mu_a}\gamma^a$. Similarly, in this particular tetrad, we have $(\nabla_\mu E){h}{^\mu_a} = 0$ as well as $(\nabla_\mu k^i){h}{^\mu_0} = 0$, but $(\nabla_\mu k^i){h}{^\mu_j} = m{h}{^\mu_i}(\sigma_{\mu\alpha}+\frac{\theta}{3}k_{\mu\alpha}){h}{^\alpha_j}$, where we used \eqref{decomposition}.
	
	Using this, as well as some properties of the shear (it is symmetric, traceless and orthogonal to $v^\alpha$), we arrive at 
	
	\begin{equation}
		\bar{b}_{0i}\gamma^\mu \nabla_\mu b_{0j} = \frac{\theta}{2}\delta_{ij}. \label{orthogderiv}
	\end{equation}

	Note that in this calculation, derivatives and products were calculated before taking the particular values for $E$ and $k^i$. Now we wish to see what \eqref{solvecond} becomes. Plugging in the ansatz for $a_0$ and using \eqref{orthogvector} and \eqref{orthogderiv}, we have that
	
	\begin{equation}
		\partial_\alpha\beta_iv^\alpha + \beta_i \bar{b}_{0i}\gamma^\mu\nabla_\mu b_{0i} + \beta_j \bar{b}_{0i}\gamma^\mu\nabla_\mu b_{0j}  = 0 \ \ \ \rightarrow \ \ \ \partial_\alpha\beta_iv^\alpha = -\frac{\theta}{2}\beta_i,
	\end{equation}

	where in the above $i\neq j$. Using that $(\nabla_\mu k^i){h}{^\mu_0} = 0$, we also have that $\nabla_\mu b_{0i} v^\mu =0$.
	
	We can now rewrite $a_0$ as $a_0=fb_0$, where $b_0$ is normalised ($\bar{b}_0b_0=1$) and $f^2=\beta_1^*\beta_1 + \beta_2^*\beta_2$. $f$ contains the expansion of the congruence. That is, 
	
	\begin{equation}
		\partial_\alpha fv^\alpha = -\frac{\theta}{2}f \ \ \ \ ; \ \ \ \ \nabla_\alpha b_0v^\alpha = 0. \label{orthogcond}
	\end{equation}

	The next part of the process involves performing a Gordon decomposition of the Dirac number current:
	
	\begin{equation}
		j^\mu = j_c^\mu+j_m^\mu, \label{Gdecomp}
	\end{equation}

	with $j^\mu = \bar{\psi}\gamma^\mu \psi$ the particle number current, $j_c^\mu = \frac{\hbar}{2mi}\left(\nabla^\mu\bar{\psi}\psi - \bar{\psi}\nabla^\mu\psi\right)$ acting analogously to a convection four-current and $j_m^\mu = \frac{\hbar}{2m}\nabla_\nu\left(\bar{\psi}\sigma^{\mu\nu}\psi\right)$ analogous to a magnetisation four-current. It is not difficult to show \eqref{Gdecomp} is true by use of the Dirac equation. This $\sigma^{\mu\nu}$ is not the shear of the worldline congruence, but rather is defined by $\sigma^{\mu\nu}=i\gamma^{[\mu}\gamma^{\nu]}$.
	
	We have $\nabla_\mu j^\mu=0$ by the Dirac equation, $\nabla_\mu j_m^\mu=0$ by symmetry arguments and thus $\nabla_\mu j_c^\mu=0$ by \eqref{Gdecomp}.
	
	Inserting our WKB expansion, we have up to order $\hbar$
	
	\begin{equation}
		j_c^\mu = f^2v^\mu -i\hbar(\bar{a}_0a_1-\bar{a}_1a_0)v^\mu-\frac{i\hbar}{2m}\left(\nabla^\mu\bar{b}_0b_0-\bar{b}_0\nabla^\mu b_0\right),
	\end{equation}
	\begin{equation}
		j_m^\mu = \frac{\hbar}{2m}\nabla_\nu\left(\bar{a}_0\sigma^{\mu\nu}a_0\right).
	\end{equation}

	Next we can define a vector $u^\mu = \frac{1}{\mathcal{N}}j_c^\mu$ as the normalised convection current. This also defines a congruence of timelike curves, and at zeroth order it is the same as $v^\mu$. This $u^\mu$ can be interpreted as the velocity of the particles. Since we just need $u^\mu$ to be normalised up to order $\hbar$, that is, $u^\mu u_\mu=1+\mathcal{O}(\hbar^2)$, using \eqref{orthogcond} we can write
	
	\begin{equation}
		u^\mu = v^\mu + \frac{\hbar}{2mi}\left(\nabla^\mu\bar{b}_0b_0-\bar{b}_0\nabla^\mu b_0\right).
	\end{equation}
	
	Similarly, we can define a tensor $S^{\alpha\beta}$ via
	
	\begin{equation}
		S^{\alpha\beta} = \frac{\bar{\psi}\sigma^{\alpha\beta}\psi}{\bar{\psi}\psi},
	\end{equation}
	
	which is simply $\bar{b}_0\sigma^{\alpha\beta}b_0$ at zeroth order. Since $\bar{\psi}\sigma^{\alpha\beta}\psi$ can be viewed as a spin density and $\bar{\psi}\psi$ as a number density, $S^{\alpha\beta}$ can be interpreted as the spin of a single particle. 
	
	Finally, we look at how the $u^\mu$ curves deviate from being geodesics. We find that, to order $\hbar$,
	
	\begin{equation}
		\frac{Du^\mu}{D\tau}=u^\nu \nabla_\nu u^\mu = \frac{\hbar}{2mi}v^\nu\nabla_\nu\left(\nabla_\mu\bar{b}_0b_0 - \bar{b}_0\nabla_\mu b_0\right).
	\end{equation}

	So, using that for spinors, covariant derivatives commute as \cite{Shapiro:2016pfm}
	
	\begin{equation}
		\nabla_{[\mu}\nabla_{\nu]}\psi = \frac{i}{4}R_{\mu\nu\alpha\beta}\sigma^{\alpha\beta}\psi,
	\end{equation}

	and once again using \eqref{orthogcond} we have
	
	\begin{equation}
		\frac{Dp^\mu}{D\tau}=\frac{\hbar}{2}\tensor{R}{^\mu_{\nu\alpha\beta}}u^\nu S^{\alpha\beta},
	\end{equation}

	where we have defined the momenta of the particles as $p^\mu=m u^\mu$. This is the exact expression we have in \eqref{MP2} with the $\hbar$ dependence explicit. Similarly, we find at lowest order
	
	\begin{equation}
		\frac{DS^{\mu\nu}}{D\tau} = v^\alpha\nabla_\alpha S^{\mu\nu} = v^\alpha\nabla_\alpha\left(\bar{b}_0\sigma^{\mu\nu}b_0\right) = 0,
	\end{equation}

	where we used \eqref{orthogcond} one final time. This is the same as \eqref{MP1} since we have already defined $p^\mu=mu^\mu$.\\
	
	What we have just seen is that, to order $\hbar$, the equations arising from analysing fermions in a curved spacetime are exactly the Mathisson-Papapetrou equations. The Dixon-Tulczyjew supplementary condition applies as well, as it is simply an expression of the conservation of the magnitude of the spin vector.
	
	This may seem perfect, but there are conceptual questions remaining. Is it really meaningful to define the spin tensor $S^{\mu\nu}$ and velocity $u^\mu$ in the way shown above? The velocity $u^\mu$ was taken to be the normalised convection current, which at zeroth order in $\hbar$ is the particle number current. In reality, this is a normalised flow rate of probability - its interpretation as the velocity of a single particle may be a step too far. Similarly defining the spin tensor of a particle via densities is a bit na\"ive, as this more accurately may represent the average spin per particle. Thus, while the matching of the classical and quantum M-P equations to within numerical factors is certainly astounding, some more elucidation on the physical meaning of some of the quantities in the quantum picture is needed.

	\subsection{Spin Effects During a Lensing Event}
	
	The M-P equations provide an interesting opportunity: since the magnitude of the neutrino's spin is known, given a metric, we may be able to deduce its mass in an observation. The data needed to do exactly this is the additional time delay that arises owing to the spin during a lensing event.\\
	
	To simplify our equations, we work in the approximation scheme wherein the particle's characteristic size, $\frac{s}{m}$, is much smaller than the characteristic length of the spacetime. For a Schwarzchild-like metric, this would be of the order $GM$. For the case of a neutrino in a Schwarzschild metric, this is obviously the case. Thus, the small quantity we will perform perturbation theory around is $\frac{s}{GMm}$, and we will work to linear order.
	
	To begin our expansion, consider \eqref{MP1} multiplied by $u_\beta$. This yields
	
	\begin{equation}
		p^\alpha = mu^\alpha - u_\beta\frac{DS^{\alpha\beta}}{D\tau},
	\end{equation}

	since we have returned to the $(-,+,+,+)$ signature so $m=-u_\mu p^\mu$. Contracting this with $p_\alpha$ and using \eqref{MP1} again gives
	
	\begin{equation}
		p_\alpha p^\alpha = -m^2 +\frac{DS^{\alpha\beta}}{D\tau}u_\beta \frac{D\tensor{S}{_\alpha^\gamma}}{D\tau}u_\gamma,
	\end{equation}

	so $p^2=-m^2+\mathcal{O}(s^2)$. Then, multiplying \eqref{MP1} by $-p_\beta$ yields
	
	\begin{equation}
		-p_\beta \frac{DS^{\alpha\beta}}{D\tau} = mp^\alpha+p^2u^\alpha \ \ \ \ \rightarrow \ \ \ \ S^{\alpha\beta}\frac{Dp_\beta}{D\tau}= mp^\alpha+p^2u^\alpha,\label{quadorder}
	\end{equation}

	where in the second part we used \eqref{MP3}. From \eqref{MP2}, $\frac{Dp_\beta}{D\tau}$ is $\mathcal{O}(s)$, so up to linear order in $s$, \eqref{quadorder} gives $0 = m(p^\alpha-mu^\alpha)$, and so in our linear approximation, we have $p^\alpha = mu^\alpha$.
	
	This immediately leads to 
	
	\begin{equation}
		\frac{DS^{\alpha\beta}}{D\tau} = 0,
	\end{equation}
	
	that is, to linear order, the spin tensor is parallel transported. From \eqref{MP2}, we have that 
	
	\begin{equation}
		\frac{D(mu^\alpha)}{D\tau} = \frac{Dm}{D\tau}u^\alpha + m\frac{Du^\alpha}{D\tau} = \frac{1}{2}\tensor{R}{^\alpha_{\beta\mu\nu}}u^\beta S^{\mu\nu}.\label{mandu}
	\end{equation}

	Multiplying this by $u_\alpha$ and using the antisymmetry of the Riemann tensor in its first 2 indices, we have 
	
	\begin{equation}
		-\frac{Dm}{D\tau} + mu_\alpha\frac{Du^\alpha}{D\tau} = 0.
	\end{equation}

	Finally, using that $mu_\alpha\frac{Du^\alpha}{D\tau} = m\frac{D}{D\tau}(u_\alpha u^\alpha) - mu_\alpha\frac{Du^\alpha}{D\tau} = - mu_\alpha\frac{Du^\alpha}{D\tau}$, clearly $mu_\alpha\frac{Du^\alpha}{D\tau}=0$ and we have
	
	\begin{equation}
		\frac{Dm}{D\tau}=0,
	\end{equation}

	so the mass of the particle is also conserved along its journey. Using this, \eqref{mandu}, and writing the spin vector as $S^\mu=sN^\mu$ where $N^\mu$ is simply a direction vector of the spin, we also have that 
	
	\begin{equation}
		m\frac{Du^\mu}{D\tau} = \frac{1}{2}\tensor{R}{^\mu_{\nu\alpha\beta}}u^\nu S^{\alpha\beta} = \frac{s}{2}\sqrt{-g}\epsilon_{\delta\gamma\lambda\rho}\tensor{R}{^\mu_{\nu\alpha\beta}}g^{\alpha\lambda}g^{\beta\rho}u^\nu u^\delta N^\gamma.\label{uitos}
	\end{equation}

	Following the lead of \cite{Bini:2014poa}, we define the function $\tensor{H}{^\mu_\gamma}$ as $\tensor{H}{^\mu_\gamma} = \frac{1}{2}\sqrt{-g}\epsilon_{\delta\gamma\lambda\rho}\tensor{R}{^\mu_{\nu\alpha\beta}}g^{\alpha\lambda}g^{\beta\rho}u^\nu u^\delta$ in order to neaten up equations.
	
	Returning to $\frac{DS^{\alpha\beta}}{D\tau} = 0$, in terms of the $S^\mu$ vector, this yields
	
	\begin{equation}
		u^\alpha\frac{DS^\beta}{D\tau} + S^\beta\frac{Du^\alpha}{D\tau} = 0,
	\end{equation}

	and from \eqref{uitos}, we see that $\frac{Du^\alpha}{D\tau}$ is of $\mathcal{O}(s)$. Thus, to linear order, we have that the direction of the spin vector is also parallel transported: $s\frac{DN^\beta}{D\tau}=0$. Therefore, we have 
	
	\begin{equation}
		\frac{Du^\mu}{D\tau} = \frac{s}{m}\tensor{H}{^\mu_\nu}N^\nu,\label{MPfinal1}
	\end{equation}
	\begin{equation}
		\frac{DN^\mu}{D\tau}=0.\label{MPfinal2}
	\end{equation}
	
	Clearly, taking the spinless limit returns the usual geodesic equation. But, as we saw in the previous subsection, taking the massless limit does not give a meaningful result: this formalism can only be applied to massive particles.\\
	We wish to consider a particle lensed by a massive object with mass $M$ on its way to us. The Schwarzschild metric is given by
	
	\begin{equation}
		ds^2=-(1-\frac{2M}{r})dt^2+(1-\frac{2M}{r})^{-1}dr^2+r^2d\Omega^2,
	\end{equation}
	
	where we have taken $G=c=1$ and $d\Omega^2=r^2d\theta^2+r^2\sin^2\theta d\phi^2$.
	
	More generally, we can analyse a Schwarzschild-like metric, which can apply to Reissner-Nordstrom and other cases. In this case, we can write our metric as
	
	\begin{equation}
		ds^2=-f(r)dt^2+\frac{1}{h(r)}dr^2+r^2d\Omega^2.
	\end{equation}
	
	We will calculate quantities for both the general and specific cases below.
	In this spacetime, the non-vanishing Christoffel symbols are
	
	\begin{equation}
		\Gamma^0_{01}=\Gamma^0_{10}=\frac{f'(r)}{2f(r)}=\frac{M}{r(r-2M)}
	\end{equation}
	\begin{equation}
		\Gamma^1_{00}=\frac{f'(r)h(r)}{2}=\frac{M(r-2M)}{r^3} \ \ \ \ \ \ \ \Gamma^1_{11}=-\frac{h'(r)}{2h(r)}=-\frac{M}{r(r-2M)} 
	\end{equation}
	\begin{equation}
		\Gamma^1_{22}=-rh(r)=-(r-2M) \ \ \ \ \ \ \ \Gamma^1_{33}=-r\sin^2\theta h(r)=-(r-2M)\sin^2\theta
	\end{equation}
	\begin{equation}
		\Gamma^2_{12}=\Gamma^2_{21}=\frac{1}{r} \ \ \ \ \ \ \ \Gamma^2_{33}=-\cos\theta\sin\theta
	\end{equation}
	\begin{equation}
		\Gamma^3_{13}=\Gamma^3_{31}=\frac{1}{r} \ \ \ \ \ \ \ \Gamma^3_{23}=\Gamma^3_{32}=\frac{\cos\theta}{\sin\theta},
	\end{equation}
	
	and the non-zero Riemann tensor elements are (along with all their permutations)
	
	\begin{equation}
		R^0_{101}=\frac{1}{4}\left[\left(\frac{f'(r)}{f(r)}\right)^2-2\frac{f''(r)}{f(r)}-\frac{f'(r)h'(r)}{f(r)h(r)}\right]=\frac{2M}{r^2(r-2M)} 
	\end{equation}
	\begin{equation}
		R^0_{202}=-\frac{rf'(r)h(r)}{2f(r)}=-\frac{M}{r}
	\end{equation}
	\begin{equation} 
		R^0_{303}=-\frac{r\sin^2\theta f'(r)h(r)}{2f(r)}=-\frac{M\sin^2\theta}{r}
	\end{equation}
	\begin{equation}
		R^1_{212}=-\frac{rh'(r)}{2}=-\frac{M}{r} 
	\end{equation}
	\begin{equation}
		R^1_{313}=-\frac{r\sin^2\theta h'(r)}{2}=-\frac{M\sin^2\theta}{r} 
	\end{equation}
	\begin{equation}
		R^2_{323}=\sin^2\theta(1-h(r))=\frac{2M\sin^2\theta}{r}.
	\end{equation}
	
	The next task is to calculate the $\tensor{H}{^\mu_\nu}$ components. For example, the $\tensor{H}{^t_\theta}$ component is
	\begin{multline}
		\tensor{H}{^t_\theta}=\frac{1}{2}\sqrt{-g}\epsilon_{\delta2\lambda\rho}\tensor{R}{^0_{\nu\alpha\beta}}g^{\alpha\lambda}g^{\beta\rho}U^\nu U^\delta \\ = \frac{1}{2}\sqrt{-g}\left[\epsilon_{1203}R^0_{303}g^{00}g^{33}U^3U^1 +\epsilon_{1230}R^0_{330}g^{00}g^{33}U^3U^1 +\epsilon_{3201}R^0_{101}g^{00}g^{11}U^1 U^3 +\epsilon_{3210}R^0_{110}g^{00}g^{11}U^1U^3\right].
	\end{multline}
	Using the symmetries of the Riemann tensor and permuting the Levi-Civita accordingly, we have
	\begin{multline}
		\tensor{H}{^t_\theta}=\sqrt{-g}U^1U^3(R^0_{303}g^{00}g^{33}-R^0_{101}g^{00}g^{11}) \\
		=-\frac{r^2}{4}\frac{dr}{d\tau}\frac{d\phi}{d\tau}\left[-2\frac{f'(r)h(r)}{f^2(r)r}+\frac{(f'(r))^2h(r)}{f^3(r)} -2\frac{f''(r)h(r)}{f^2(r)}-\frac{f'(r)h'(r)}{f^2(r)}\right]= \frac{3M}{r(1-\frac{2M}{r})}\frac{dr}{d\tau}\frac{d\phi}{d\tau}.
	\end{multline}
	
	Repeating this procedure for all components, we find the non-zero elements to be
	\begin{equation}
		\tensor{H}{^t_\theta}=-\frac{r^2}{4}\frac{dr}{d\tau}\frac{d\phi}{d\tau}\left[-2\frac{f'(r)h(r)}{f^2(r)r}+\frac{(f'(r))^2h(r)}{f^3(r)} -2\frac{f''(r)h(r)}{f^2(r)}-\frac{f'(r)h'(r)}{f^2(r)}\right]= \frac{3M}{r(1-\frac{2M}{r})}\frac{dr}{d\tau}\frac{d\phi}{d\tau},
	\end{equation}
	\begin{equation}
		\tensor{H}{^t_\phi}=-\frac{r^2}{4}\frac{dr}{d\tau}\frac{d\theta}{d\tau}\left[2\frac{f'(r)h(r)}{rf^2(r)}+\frac{h(r)}{f(r)}\left(\left(\frac{f'(r)}{f(r)}\right)^2-2\frac{f''(r)}{f(r)}-\frac{f'(r)h'(r)}{f(r)h(r)}\right)\right] =-\frac{3M}{r(1-\frac{2M}{r})}\frac{dr}{d\tau}\frac{d\theta}{d\tau},
	\end{equation}
	\begin{equation}
		\tensor{H}{^r_\theta}=\frac{r^2}{4}\frac{dt}{d\tau}\frac{d\phi}{d\tau} \left[2\frac{h(r)h'(r)}{r}+h^2(r)\left(\left(\frac{f'(r)}{f(r)}\right)^2-2\frac{f''(r)}{f(r)}-\frac{f'(r)h'(r)}{f(r)h(r)}\right)\right]
		=\frac{3M(1-\frac{2M}{r})}{r}\frac{dt}{d\tau}\frac{d\phi}{d\tau},
	\end{equation}
	\begin{equation}
		\tensor{H}{^r_\phi}=-\frac{r^2}{4}\frac{dt}{d\tau}\frac{d\theta}{d\tau} \left[2\frac{h(r)h'(r)}{r}+h^2(r)\left(\left(\frac{f'(r)}{f(r)}\right)^2-2\frac{f''(r)}{f(r)}-\frac{f'(r)h'(r)}{f(r)h(r)}\right)\right] =-\frac{3M(1-\frac{2M}{r})}{r}\frac{dt}{d\tau}\frac{d\theta}{d\tau}.
	\end{equation}
	
	We now need only to plug these into \eqref{MPfinal1}, and use our Christoffel symbols and \eqref{MPfinal2} to obtain the equations of motion. In the Schwarzschild case, we have
	
	\begin{equation}
		\frac{d^2t}{d\tau^2}+\frac{2M}{r(r-2M)}\frac{dt}{d\tau}\frac{dr}{d\tau}+\frac{3sM}{mr(1-\frac{2M}{r})}\frac{dr}{d\tau}\left(\frac{d\theta}{d\tau}N^\phi-\frac{d\phi}{d\tau}N^\theta\right)=0,
	\end{equation}
	\begin{multline}
		\frac{d^2r}{d\tau^2}+\frac{M(r-2M)}{r^3}\left(\frac{dt}{d\tau}\right)^2-\frac{M}{r(r-2M)}\left(\frac{dr}{d\tau}\right)^2-(r-2M)\left[\left(\frac{d\theta}{d\tau}\right)^2+\sin^2\theta\left(\frac{d\phi}{d\tau}\right)^2\right] \\
		+\frac{3sM(1-\frac{2M}{r})}{mr}\frac{dt}{d\tau}\left(\frac{d\theta}{d\tau}N^\phi-\frac{d\phi}{d\tau}N^\theta\right)=0,
	\end{multline}
	\begin{equation}
		\frac{d^2\phi}{d\tau^2}+\frac{2}{r}\frac{d\phi}{d\tau}\frac{dr}{d\tau}+\frac{2\cos\theta}{\sin\theta}\frac{d\phi}{d\tau}\frac{d\theta}{d\tau} +\frac{3sM}{mr^3\sin^2\theta}\frac{d\theta}{d\tau}\left(\frac{dt}{d\tau}N^r-\frac{dr}{d\tau}N^t\right)=0,
	\end{equation}
	\begin{equation}
		\frac{d^2\theta}{d\tau^2}++\frac{2}{r}\frac{d\theta}{d\tau}\frac{dr}{d\tau}-\cos\theta\sin\theta\left(\frac{d\phi}{d\tau}\right)^2 -\frac{3sM}{mr^3}\frac{d\phi}{d\tau}\left(\frac{dt}{d\tau}N^r-\frac{dr}{d\tau}N^t\right)=0,
	\end{equation}
	\begin{equation}
		\frac{dN^\theta}{d\tau}-\cos\theta\sin\theta\frac{d\phi}{d\tau}N^\phi+\frac{1}{r}\frac{dr}{d\tau}N^\theta+\frac{1}{r}\frac{d\theta}{d\tau}N^r=0,
	\end{equation}
	\begin{equation}
		\frac{dN^\phi}{d\tau}+\frac{1}{r}\frac{dr}{d\tau}N^\phi+\frac{1}{r}\frac{d\phi}{d\tau}N^r+\frac{\cos\theta}{\sin\theta}\frac{d\theta}{d\tau}N^\phi +\frac{\cos\theta}{\sin\theta}\frac{d\phi}{d\tau}N^\theta=0,
	\end{equation}
	\begin{equation}
		\frac{dN^r}{d\tau}+\frac{M(r-2M)}{r^3}\frac{dt}{d\tau}N^t-\frac{M}{r(r-2M)}\frac{dr}{d\tau}N^r-(r-2M)\left[\frac{d\theta}{d\tau}N^\theta+\sin^2\theta\frac{d\phi}{d\tau}N^\phi\right]=0,
	\end{equation}
	\begin{equation}
		\frac{dN^t}{d\tau}+\frac{M}{r(r-2M)}\left[\frac{dr}{d\tau}N^t+\frac{dt}{d\tau}N^r\right]=0.
	\end{equation}
	
	Luckily, these equations can be simplified (and actually for consistency, must be). Recall that we are only working to 1st order in $\frac{s}{mM}$, as this is our small parameter. \\
	
	Let us consider the case of our spinning particle being lensed by the Schwarzschild object. In this case, when there is no spin, we can take $\theta=\frac{\pi}{2}$ and $\frac{d\theta}{d\tau}=\frac{d^2\theta}{d\tau^2}=0$ - that is, the particle moves in the equatorial plane only. However, from our equations of motion it is clear that the spin induces some motion in the $\theta$-direction. Since the source is the spin, these deviations from the zeroth-order case must be small. Thus, we can take $\theta=\frac{\pi}{2}+\tilde{\theta}$ and $\frac{d\theta}{d\tau}=\frac{d\tilde{\theta}}{d\tau}$, where both $\tilde{\theta}$ and $\frac{d\tilde{\theta}}{d\tau}$ are of order $\frac{s}{mM}$.\\
	
	Expanding trigonometric quantities, we see that to order $s$, $\cos\theta\sin\theta\approxeq\frac{\cos\theta}{\sin\theta}\approxeq-\tilde{\theta}$ and that $\sin^2\theta\approxeq1$. \\
	
	
	
	
	We are interested in the time delay induced by the spin correction, and so we require an equation for $\frac{dt}{dr}$. In line with this goal, only 4 of the 8 equations of motion need to be solved. At order $s$, these are, in the specific Schwarzschild case,

	\begin{equation}
		\frac{d^2t}{d\tau^2}+\frac{2M}{r(r-2M)}\frac{dt}{d\tau}\frac{dr}{d\tau}-\frac{3sM}{mr(1-\frac{2M}{r})}\frac{dr}{d\tau}\frac{d\phi}{d\tau}N^\theta=0,
		\label{dt-1}
	\end{equation}
	\begin{equation}
		\frac{d^2r}{d\tau^2}+\frac{M(r-2M)}{r^3}\left(\frac{dt}{d\tau}\right)^2-\frac{M}{r(r-2M)}\left(\frac{dr}{d\tau}\right)^2-(r-2M)\left(\frac{d\phi}{d\tau}\right)^2 - \frac{3sM(1-\frac{2M}{r})}{mr}\frac{dt}{d\tau}\frac{d\phi}{d\tau}N^\theta=0,
		\label{dr-1}
	\end{equation}
	\begin{equation}
		\frac{d^2\phi}{d\tau^2}+\frac{2}{r}\frac{d\phi}{d\tau}\frac{dr}{d\tau}=0,
	\end{equation}
	\begin{equation}
		\frac{dN^\theta}{d\tau}+\frac{1}{r}\frac{dr}{d\tau}N^\theta=0.
	\end{equation}
	
	Let us now solve these for the Schwarzschild case. We begin with the latter two:
	
	\begin{equation}
		\frac{d^2\phi}{d\tau^2}+\frac{2}{r}\frac{d\phi}{d\tau}\frac{dr}{d\tau}=0\quad\Rightarrow\quad\frac{d}{d\tau}\left(\ln\left(\frac{d\phi}{d\tau}r^2\right)\right)=0
	\end{equation}
	which implies a conserved quantity, $J$, the total angular momentum per unit mass of the particle:
	\begin{equation}
		\frac{d\phi}{d\tau}r^2=J\quad \Leftrightarrow\quad \frac{d\phi}{d\tau}=\frac{J}{r^2}.\label{Jconserv}
	\end{equation}
	
	Similarly,
	
	\begin{equation}
		\frac{dN^\theta}{d\tau}+\frac{1}{r}\frac{dr}{d\tau}N^\theta=0\quad\Rightarrow\quad\frac{d}{d\tau}(rN^\theta)=0
	\end{equation}
	and so we have another conserved quantity, $K$, and
	\begin{equation}
		N^\theta=\frac{K}{r}\,.\label{Kconserv}
	\end{equation}

	Clearly, since $N^\mu$ is normalised, $K$ is a constant between -1 and 1\footnote{Importantly, this $K$ is constant for each individual particle - many particles emitted from the same source can each have a different value of $K$.}. This leaves 2 equations to be solved:
	
	\begin{equation}
		\frac{d^2t}{d\tau^2}+\frac{2M}{r^2(1-\frac{2M}{r})}\frac{dt}{d\tau}\frac{dr}{d\tau}-\frac{3sK}{Mm}\frac{J}{r^2}\left(\frac{M}{r}\right)^2\frac{1}{(1-\frac{2M}{r})}\frac{dr}{d\tau}=0,
	\end{equation}
	
	\begin{equation}
		\frac{d^2r}{d\tau^2}+\frac{M(1-\frac{2M}{r})}{r^2}\left(\frac{dt}{d\tau}\right)^2 -\frac{M}{r^2(1-\frac{2M}{r})}\left(\frac{dr}{d\tau}\right)^2-\frac{J^2(1-\frac{2M}{r})}{r^3}- \frac{3sK}{Mm}\frac{J}{r^2}\left(\frac{M}{r}\right)^2(1-\frac{2M}{r})\frac{dt}{d\tau}=0.
	\end{equation}
	
	Even before solving these equations, some observations can be made. Firstly, since the effect of the spin on the time delay depends on $K$, the spin-induced time delay $\Delta t_s$ will have a range of values, depending on the initial spin of the particle when it is emitted from its source. In fact, depending on whether the particle's spin is ``up" (completely in the $\theta$-direction) or ``down" (completely in the $-\theta$-direction), there will be a range of values with range $2\Delta t_s$. If this range is not observed, it would be a smoking gun sign of polarisation - emitted particles (in this case, neutrinos) from the source are only emitted with spin in a certain direction. Note that this direction does not have to only be in the $\pm\theta$-directions, but could also be somewhere between them - that is, $K$ has some intermediate value between -1 and 1. 
	
	Secondly, since $\frac{s}{mM}$ is a small parameter, and in the weak-field limit so is $\frac{M}{r}$, the term that appears owing to the spin is extremely small. In fact, it is reasonable to assume, given the miniscule size of a neutrino, that $\frac{s}{mM}$ is much smaller than $\frac{M}{r}$ and in some cases (depending on the mass of the lensing object and the distance of closest approach) may even be smaller than $\left(\frac{M}{r}\right)^2$ or even $\left(\frac{M}{r}\right)^3$. Normally, the time delay effect during lensing is only calculated up to linear order in $\frac{M}{r}$. In order to meaningfully include the correction that the spin induces, we would need to include all corrections in the weak-field approximation up to $\left(\frac{M}{r}\right)^6$. This, in turn, would require exceptionally accurate measurements of the time delay in order to confirm $\Delta t_s$ and be able to utilise it. For example, for a particle skimming the sun wherein the time delay is of the order of 10 days, we would need to measure the time delay precisely within $10^{-25}$ seconds - about 5 orders more precisely than has yet been achieved in perfect laboratory conditions \cite{fastlight}. \\
	
	In this chapter, we have shown how a particle with spin should couple to the curvature of the spacetime it is in, both from a classical- and quantum-mechanical viewpoint. We then saw how we might be able to use this effect to find an observable time delay that depends on the mass of the neutrino, and in doing so, deduce the incoming neutrino's mass. Unfortunately, it seems that for the near future, the accuracy of time-delay measurements is extremely far from what would be required to glean any information using this technique, and we leave that task for hyper-advanced civilizations of the future.
	
	\newpage
	\section{Conclusions}
	
	Neutrinos certainly are exciting particles with a lot of unutilised potential. Their ability to teach us about our universe seems unmatched, and is only sadly limited by the technology of today.
	
	The first half of this work covered some of the fundamental physics needed to understand our current theories of neutrinos and our cosmic history. Though certainly only scratching the surface of the vast fields of cosmology and particle physics, these chapters attempted to illuminate particular pockets of physics relevant to the final trio of chapters, in order for the experimental setups described to be understood.
	
	Using the formalism of general relativity along with some empirical observations of our cosmos, we were able to obtain the universe's spacetime metric and understand the evolution of our universe and the constituents within it. Following that, switching to the language of quantum field theory, we studied neutrinos and how they fit into the standard model: which other particles they interact with, how they obtain their masses, and why they have no electric charge. It was also fascinating to understand how neutrinos are able to change flavour while propagating, and see how experiments allow us to fit the parameters of our model to reality.\\
	
	Combining these ideas and delving into particle cosmology, we predicted the C$\nu$B's existence, along with some testable predictions of what characteristics it should have. Refining these expectations, we now had a clear picture of the relic neutrinos in our universe, and deviations from these expectations would be a clear sign of exciting new physics. To detect these relic neutrinos, we discussed why inverse beta decay on tritium nuclei is our most promising candidate, while looking at a few other neutrino detectors.\\
	
	As we saw next, the upcoming planned PTOLEMY experiment is cause for great anticipation. Utilising tritium nuclei, the experiment will be the first probe into the pre-CMB universe: an enormous leap. It is important to remember, however, that this is simply a first step in relic neutrino astronomy. While this ancient information will certainly assist in elucidating some enigmas surrounding neutrinos and our early universe, we cannot expect too much. As already discussed, any deviations from the expected capture rate have a myriad of explanations, including experimental errors, lack of precision in other quantities (such as the PMNS matrix), and many different ideas for changes to the standard model. The primary hurdle, it would seem, is the need to improve the energy resolution of the detector.\\
	
	Shifting the focus to gravitational lensing, we then (after covering some fundamental concepts and calculations) detailed 2 more theoretical experiments that could be revealing. The first, involving supernova neutrinos, could be useful in gaining a more accurate description of supernovae themselves. In fact, this experiment was unique in that it was not limited by experimental setup or lack of information, but rather by the low probability of a perfect event occurring soon, wherein a large star explodes on exactly the opposite side of the galaxy.
	
	The other idea that used lensing involved the lensing of the relic C$\nu$B neutrinos in order to investigate the evolution of lensing objects - such as stars, galaxies, or black holes. Though the idea of using neutrino detections to essentially watch these astrophysical objects' lives unfold before us is fantastic, this experiment is certainly also the furthest from our reach. While the next (first) generation of C$\nu$B detectors will struggle to detect even a few relic neutrinos a year with no directional information, to achieve the lofty aim of watching galactic evolution using neutrinos we would need the technology to detect massive numbers of cosmic neutrinos, as well as extremely accurate information regarding their angle of incidence. \\
	
	The final technique explored involved looking at higher-order corrections owing to the neutrino's spin while being lensed in order to constrain - or even deduce - its mass. While fascinating links were made between classical and quantum physics, in the end it seems that the spin-curvature correction is certainly too small to be detected soon, owing to the inaccuracies of our current measurements and the incredible precision needed. \\
	
	In conclusion, despite being held back by the technology of today, there is no doubt a glowing future for neutrino astronomy - particularly relic neutrino astronomy. With some patience, over the coming years and with multiple generations of neutrino detectors, the information garnered about our universe and the standard model will certainly be unrivalled.

\end{document}